\documentclass[useAMS,usenatbib,usegraphicx,pdftex]{mn2e}
\usepackage{graphicx}
\usepackage{amsmath,amssymb}
\usepackage{pslatex}
\usepackage{float}
\usepackage{bm}
\usepackage[shortcuts]{extdash}
\bibpunct{(}{)}{;}{a}{}{,}

\title[A Catalogue of Extragalactic Non-nuclear X-ray Sources]{A new, clean catalogue of extragalactic non-nuclear X-ray sources in nearby galaxies}
\author[H. P. Earnshaw, et al.]{\parbox{\textwidth}{H. P. Earnshaw$^{1,2}$\thanks{E-mail:
hpearn@caltech.edu}, T. P. Roberts$^{1}$, M. J. Middleton$^{3}$, D. J. Walton$^{4}$, S. Mateos$^{5}$}\\
\\
\parbox{\textwidth}{
$^1$Centre for Extragalactic Astronomy, Department of Physics, Durham University, South Road, Durham, DH1 3LE, UK\\
$^2$Cahill Center for Astronomy and Astrophysics, California Institute of Technology, Pasadena, CA 91125, USA\\
$^3$Physics \& Astronomy, University of Southampton, Southampton, Hampshire SO17 1BJ, UK\\
$^4$Institute of Astronomy, University of Cambridge, Madingley Road, Cambridge CB3 0HA, UK\\
$^5$Instituto de F\'{i}sica de Cantabria (CSIC-Universidad de Cantabria), E-39005, Santander, Spain}}

\begin{document}

\date{}

\pagerange{\pageref{firstpage}--\pageref{lastpage}} \pubyear{}

\maketitle

\label{firstpage}

\begin{abstract}
	
We have created a new, clean catalogue of extragalactic non-nuclear X-ray sources by correlating the 3XMM-DR4 data release of the {\it XMM-Newton} Serendipitous Source Catalogue with the Third Reference Catalogue of Bright Galaxies and the Catalogue of Neighbouring Galaxies, using an improved version of the method presented in \citet{walton11}. Our catalogue contains 1,314 sources, of which 384 are candidate ultraluminous X-ray sources (ULXs). The resulting catalogue improves upon previous catalogues in its handling of spurious detections by taking into account {\it XMM-Newton} quality flags. We estimate the contamination of ULXs by background sources to be 24 per cent. We define a `complete' subsample as those ULXs in galaxies for which the sensitivity limit is below $10^{39}$\,erg\,s$^{-1}$ and use it to examine the hardness ratio properties between ULX and non-ULX sources, and ULXs in different classes of host galaxy. We find that ULXs have a similar hardness ratio distribution to lower-luminosity sources, consistent with previous studies. We also find that ULXs in spiral and elliptical host galaxies have similar distributions to each other independent of host galaxy morphology, however our results do support previous indications that the population of ULXs is more luminous in star-forming host galaxies than in non-star-forming galaxies. Our catalogue contains further interesting subpopulations for future study, including Eddington Threshold sources and highly variable ULXs. We also examine the highest-luminosity ($L_{\rm X} > 5\times10^{40}$\,erg\,s$^{-1}$) ULXs in our catalogue in search of intermediate-mass black hole candidates, and find nine new possible candidates. 

\end{abstract}


\section{Introduction}
\label{sec:intro}

Over the past decade, ultraluminous X-ray sources (ULXs) have proved to be a fruitful laboratory for challenging and refining our understanding of extreme accretion onto compact objects (see \citealt{kaaret17} for a recent review). Defined as non-nuclear extragalactic X-ray point sources with X-ray luminosity $L_{\rm X} > 10^{39}$\,erg\,s$^{-1}$, these objects are significant because they possess luminosities above those expected for a typical stellar-mass ($\sim10$\,M$_{\odot}$) black hole (BH) accreting at its Eddington limit. The presence of such objects has two possible implications: either these sources are intermediate-mass black holes (IMBHs; $10^2<M_{\rm BH}<10^5$\,M$_{\odot}$; \citealt{colbert99}) accreting at sub-Eddington rates and likely in similar accretion states to those we observe in stellar-mass BH binaries (e.g. \citealt{servillat11}), or they are stellar-mass (typically 1--20\,M$_{\odot}$) compact objects undergoing a non-standard form of accretion -- either undergoing relativistic beaming (e.g. \citealt{kording02}) sufficient to make a sub-Eddington source appear to have a luminosity above Eddington, or accreting in a genuinely super-Eddington mode with mild geometric beaming (e.g. \citealt{poutanen07}).

The formation of a population of IMBHs to match the ULX population we observe presents a number of problems. IMBHs cannot simply result as the end-point of a typical massive star's lifetime as X-ray binaries do -- they require particular formation scenarios such as the collapse of early-Universe population III stars \citep{madau01}, runaway mergers in globular clusters (e.g. \citealt{zwart02,guerkan04,vesperini10}) or the direct collapse of gas inside extremely metal-poor galaxies in the early Universe (e.g. \citealt{loeb94,bromm03}). While these scenarios could explain individual objects, they are problematic when it comes to explaining the entire ULX population. This applies especially to star-forming galaxies, where unrealistic production rates of IMBHs are required to match observations \citep{king04}. Additionally, the best quality ULX data from {\it XMM-Newton} (e.g. \citealt{stobbart06,gladstone09}) followed by high-energy observations of ULXs using {\it NuSTAR} (e.g. \citealt{bachetti13, walton15}) have established that the majority of ULXs do not have the X-ray spectral and timing properties of sources in sub-Eddington accretion states. Rather, they appear to be accreting in an ultraluminous regime (e.g. \citealt{sutton13b}), in which the accretion disc becomes supercritical and the observed X-ray properties may be dominated by a massive, radiatively-driven outflowing wind. We note that evidence for strong winds has now been seen in a number of ULXs \citep{pinto16,pinto17,walton16,kosec18}. Reprocessing in this wind introduces a soft excess to the spectrum, and the wind itself potentially scatters and/or obscures the hard central emission depending upon the mass accretion rate and the angle of inclination with respect to the observer \citep{middleton15a}. 

While the high luminosity of these objects meant that the assumption that they are BHs was not an unreasonable one, the recent discovery of pulsations from a small number of ULXs (e.g. \citealt{bachetti14,fuerst16,israel17a,israel17b,carpano18}) has established that at least some fraction of the ULX population are highly super-Eddington neutron stars (NSs). Exactly how big of a fraction remains an open question -- not all NS ULXs are expected to exhibit pulsations (e.g. \citealt{king17}), and the discovery of a ULX with a cyclotron line but no pulsations \citep{brightman18} provides evidence of the existence of NS ULXs beyond those observed as pulsars. Studies considering the broadband spectra of ULXs suggest that all high-quality ULX spectra could be consistent with a NS ULX model \citep{koliopanos17,pintore17,walton18}, indicating that the fraction of ULXs that are NSs could conceivably be very high.

Although the properties of the majority of ULXs can be explained by super-Eddington accretion onto stellar-mass BHs or NSs, the most luminous of the population, hyper-luminous X-ray sources (HLXs; $L_{\rm X}>10^{41}$\,erg\,s$^{-1}$), are more challenging to explain even with highly super-Eddington accretion (although known super-Eddington sources such as NGC~5907~ULX-1, a highly super-Eddington NS, have been observed to reach such luminosities; e.g. \citealt{israel17a}). Therefore the HLX regime is often used to look for genuine IMBH candidates, with one such object being ESO~243-49~HLX-1 (HLX-1, \citealt{farrell09}). It is also possible to find plausible IMBH candidates in the main ULX population if they show strong evidence of accreting in a sub-Eddington rather than a super-Eddington state. Examples include a radio detection of steady jet emission (e.g. NGC~2273-3c, \citealt{mezcua15} -- although likely super-Eddington sources have also demonstrated radio emission e.g. Ho~II~X-1, \citealt{cseh15}), strong band-limited noise and a hard spectrum indicating a source accreting in the low/hard state (e.g. M51~ULX-7; \citealt{earnshaw16}), or changes in spectral state similar to those seen in sub-Eddington X-ray binaries (as also seen in HLX-1, \citealt{servillat11}). Together with the ambiguity between BH and NS ULXs, the presence of IMBH candidates in the ULX population shows it to be a complex and heterogeneous one.

While closely studying the properties of individual well-known ULXs can provide us with detailed insight into the accretion mechanisms of this population (e.g. \citealt{mukherjee15,walton15b,luangtip16,walton17}), this is only possible for a small number of sources for which, through a combination of high luminosity, relative proximity, and lengthy observing campaigns, there is a wealth of high-quality data available. Insight into the properties of the ULX population as a whole requires the creation of large samples of ULXs, which has been made possible through the collected observations of X-ray space telescopes, especially {\it ROSAT}, {\it Chandra} and {\it XMM-Newton} \citep{roberts00, colbert02, swartz04, liu05a, liu05b, swartz11, walton11}. These samples have allowed us to investigate the characteristics of this population as a whole -- for example, it appears to be the case that ULXs are more numerous and luminous in star-forming galaxies than in non-star-forming galaxies \citep{swartz04,liu05b}, which would imply different ULX populations depending upon the nature of their host environment. The association of one population with star-forming galaxies makes them likely to be predominantly located in high-mass X-ray binaries (HMXBs) with a high-mass, short-lived stellar companion such as an OB-supergiant acting as the fuel supply. Those ULXs in elliptical galaxies and non-star-forming regions of spiral galaxies are more likely to be low-mass X-ray binaries (LMXBs) with smaller companion stars, and begin accreting long after star formation has ended. These sources could be similar in nature to the Galactic sources SS~433 and GRS~1915+105 respectively, a HMXB and LMXB that exhibit super-Eddington accretion \citep{king02}. This notion of two populations of ULXs is tentatively supported by the differing luminosity functions of ULXs found in spiral and elliptical host galaxies, with a shallower power-law slope for those in spiral (i.e. star-forming) galaxies indicating the presence of a greater number of higher luminosity ULXs (e.g. \citealt{walton11}).

Aside from their extreme luminosities, ULXs do not differ a great deal from lower-luminosity X-ray sources, possessing much the same bulk X-ray properties as non-ULXs in terms of spectral shape and colour over the soft energy range of {\it XMM-Newton} and {\it Chandra} \citep{swartz04}. However, different subgroups of objects within the general X-ray source population such as HMXBs and LMXBs, as well as supernova remnants and supersoft sources, are suggested to have different distributions in X-ray colour \citep{prestwich03}. Subgroup differences may also be evident in the ULX population, where it has been suggested that the hard and soft ultraluminous regimes can be distinguished by X-ray colour \citep{pintore14}.

Below the luminosity regime of ULXs, extragalactic X-ray sources are more difficult to detect, with detailed studies of the sub-ULX population limited to the Milky Way and the Local Group prior to the {\it XMM-Newton} and {\it Chandra} missions (e.g. \citealt{fabbiano89}). Extending study of the X-ray populations of other galaxies to a larger volume allows the inclusion of a greater variety of environments in our understanding of the X-ray populations in the nearby Universe, and this has begun to be enabled by current missions. So far, deep observations with the {\it Chandra} mission have enabled source catalogues to be created for other galaxies (e.g. \citealt{terashima04,mineo13,long14}), from which X-ray luminosity functions can be produced. However, observations by {\it XMM-Newton} deliver a greater amount of data than equivalent observations using {\it Chandra} and provide enough data for a large number of galaxies to enable study of the X-ray colours, and in some cases the spectral and timing properties, of their sub-ULX population (e.g. \citealt{jenkins04,akyuz13}). Coming to a greater understanding of the lower-luminosity X-ray populations of other galaxies is important since these sources are more common than ULXs and provide insight into the distribution and behaviour of sub-Eddington accreting sources in different galaxy environments. 

Samples of bright extragalactic X-ray sources are also useful for identifying extreme or otherwise unusual sources for further study. The most obvious example is the search for HLXs that could potentially be viable IMBH candidates (e.g. \citealt{sutton12}, examining high-luminosity ULXs in the catalogue presented in \citealt{walton11}). However, the ability to select sources based upon other properties, such as spectral shape and variability properties, allows other routes for locating objects of interest. In this respect, the {\it XMM-Newton} Serendipitous Source Catalogue is an excellent resource for constructing a ULX sample, containing the fluxes in various energy bands and hardness ratios between bands for every detected source. In addition, it benefits from the large collecting area, good sensitivity and wide field of view of the mission which makes the telescope effective at making serendipitous detections. It also contains a number of quality indicators which assist in reducing spurious detections.

In this paper we create a new, clean sample of extragalactic non-nuclear X-ray sources using the 3XMM-DR4 data release of the {\it XMM-Newton} Serendipitous Source Catalogue. We describe our sample creation in Section~\ref{sec:data}, and examine and discuss the bulk properties of our sample and some significant subsets in Section~\ref{sec:results}. We present our conclusions in Section~\ref{sec:conc}.

\begin{figure*}
	\begin{center}
		\includegraphics[width=78mm]{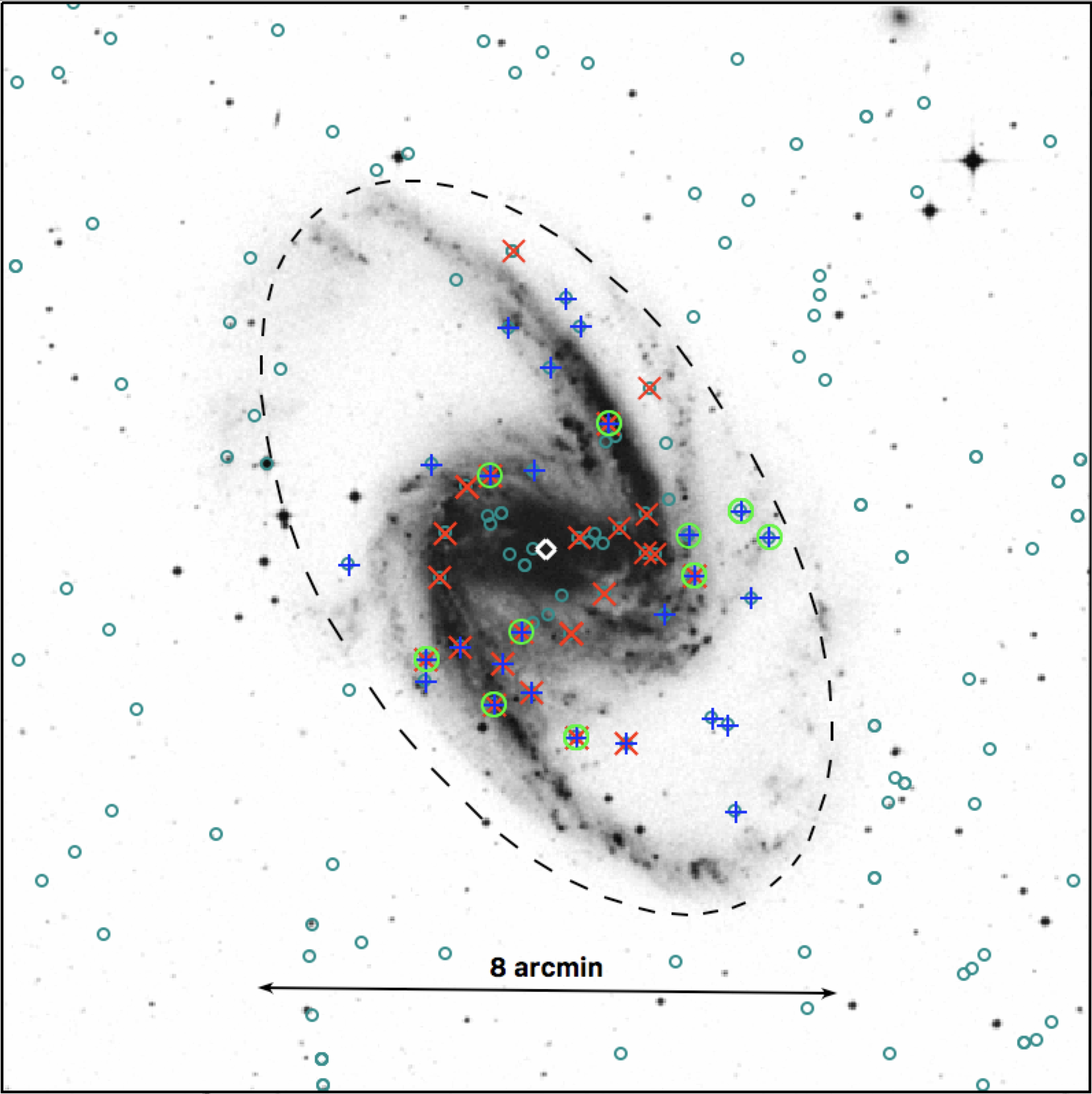}
		\hspace{1mm}
		\includegraphics[width=78mm]{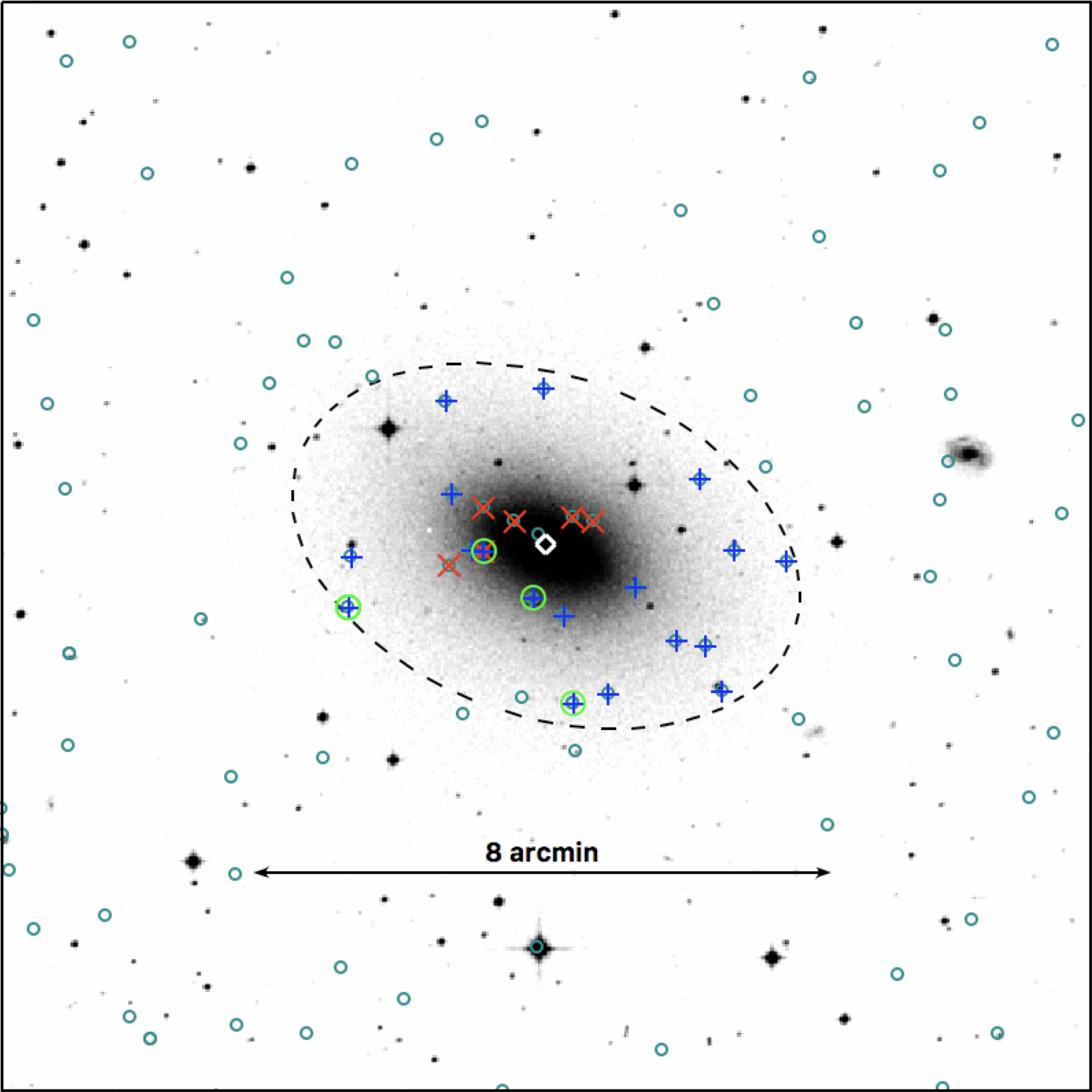}
	\end{center}
	\caption{Examples of the catalogue selection method for NGC~1365 ({\it left}) and NGC~4697 ({\it right}). Sources are overlaid on SAO-DSS optical images of the galaxies. All sources from 3XMM-DR4 are marked with small dark cyan circles (including those later removed for being extended or having low detection significance). The D25 ellipses of the galaxies are marked with a dashed black line, and the significant detections of point sources falling inside them are marked either with a blue cross for a source with no major detection quality flags, or with a red cross for a flagged detection (see Section~\ref{sec:flag}). For sources with multiple detections, it is possible for one detection to have a major quality flag and another not to. The central AGN is marked with a white diamond and ULXs are marked with large green circles.} 
	\label{fig:method}
\end{figure*}

\section{Data and sample selection}
\label{sec:data}

We produced our catalogue of ULXs using the 3XMM-DR4 data release of the {\it XMM-Newton} Serendipitous Source Catalogue \citep{rosen15,rosen16}. Images were taken using the European Photon Imaging Camera (EPIC) instrument of {\it XMM-Newton} between the dates of 3 February 2000 and 8 December 2012, with a coverage of $\sim2\%$ of the total sky. The survey catalogue contains 531,261 detections of 372,728 unique sources, a $\sim50\%$ increase from the 2XMM-DR1 data release used for the predecessor to this catalogue (see below), and features improvements in the instrument calibration, data processing algorithms and the Science Analysis Software available for data reduction. It is therefore an ideal resource for creating a large sample of ULXs that can be used for studies of these rare and extreme objects, as well as for obtaining large numbers of lower luminosity objects (we note that further 3XMM data has been released since this project's inception, providing material for a follow-up expansion of this sample, with the latest data release being 3XMM-DR7 in 2017; \citealt{rosen16}). 

The 2XMM-DR1 release of the {\it XMM-Newton} Serendipitous Source Catalogue was used by \citet{walton11}, henceforth referred to as W11, to create a catalogue of ULXs. Our method is based heavily on that used in W11, applied to the 3XMM-DR4 release but containing a number of improvements. We also extended the catalogue to lower-luminosity extragalactic X-ray sources. This section describes the process of catalogue creation and the basic properties of the final sample, but we also recommend that the reader also refer to W11 for further details. 

\subsection{Sample creation}
\label{sec:samp}

We began by cross-correlating 3XMM-DR4 with a list of galaxies from the Third Reference Catalogue of Bright Galaxies (RC3; \citealt{devaucouleurs91}), as in W11. RC3 contains 23,022 individual galaxies within a distance of 500\,Mpc, which we supplemented with the Catalogue of Neighbouring Galaxies (CNG; \citealt{karachentsev04}), which contains 451 galaxies with distance $D \lesssim 10$\,Mpc, in order to build a more complete picture of the local population of X-ray binaries in other galaxies. We added the 366 galaxies not already included in RC3 to our list, then updated the central position coordinates and recessional velocities of all galaxies with their values according to the NASA Extragalactic Database (NED)\footnote{http://nedwww.ipac.caltech.edu/}. A small number of nearby galaxies extend over a greater area of the sky than the 30\,arcminute {\it XMM-Newton} field of view, which means that individual observations of these objects are incomplete in terms of detecting the full source population. Additionally, we expect that the X-ray source population of these galaxies is heavily contaminated by background sources, given their low distance and large area on the sky. Therefore, since we prioritise a clean sample in this investigation, we removed from our list all galaxies with a $D_{25}$ isophotal major axis greater than 25\,arcminutes -- these 13 galaxies are Andromeda, the Large and Small Magellanic Clouds, M33, M54, M81, M101, NGC~55, NGC~253, NGC~5128, Draco Dwarf, Sculptor Dwarf, and Sextans Dwarf Spheroidal. While this removes a large number of known ULXs and X-ray binaries from our sample, the X-ray populations of these galaxies are relatively well-studied elsewhere (e.g. \citealt{jenkins05, stiele11, williams15}).

In order to obtain the most accurate distances for the galaxies, we first used data from the NED Distances Database (NED-D\footnote{http://ned.ipac.caltech.edu/Library/Distances/}; \citealt{steer17}). If a galaxy had distance measurements within NED-D obtained from Cepheid variables, the tip of the red giant branch or Type Ia supernovae (in that order of precedence), we took the mean of the distances calculated using that method to obtain the best galaxy distance. The distances of remaining galaxies with recessional velocity $cz < 1000\rm~km~s^{-1}$ were obtained from the Catalogue of Nearby Galaxies (NBG; \citealt{tully88}). Any galaxies with $cz < 1000\rm~km~s^{-1}$ and no NED-D or NBG distance were discarded, as the velocities of nearby galaxies will be dominated by peculiar motion and an accurate distance cannot be obtained using the recessional velocity alone. The velocity of galaxies with $cz > 1000\rm~km~s^{-1}$ and no NED-D or NBG distance we considered to be dominated by the Hubble flow and thus their distances were calculated using Hubble's law, with $H_0 = 75\rm~km~s^{-1}$ for consistency with \citet{tully88}. Finally, we ran a correlation of this galaxy list with the 3XMM-DR4 summary of observations to find all galaxies that fall within the sky coverage of 3XMM-DR4, which gave us a final list of 1,868 3XMM-DR4 field galaxies.

Since we wanted to investigate potential differences in the ULX population in different galactic environments, we grouped the galaxies by type. We define elliptical galaxies as galaxies with Hubble stage $T < 0$ in RC3 or CNG, spiral galaxies as those with $0 \leq T \leq 9$, and irregular galaxies as those with $T > 9$. (This is slightly different from W11, who define galaxies with T = 0 as ellipticals.)

This list of field galaxies was matched with all significantly-detected point sources in the 3XMM-DR4 catalogue, defined as those objects with extent $<6$\,arcseconds, to exclude extended sources, and a maximum likelihood of detection $>8$ (equivalent to a $3.5\sigma$ detection) over the full energy range of EPIC. As in W11, we performed this matching using the {\sc topcat}\footnote{Tool for Operations on Catalogues and Tables; http://www.star.bris.ac.uk/mbt/topcat/} software to find all point sources that fell (to within their combined $1\sigma$ statistical and systematic position errors) inside the $D_{25}$ isophotal ellipse of each galaxy. Where there was no position angle data for the galaxy, we performed a circular match within the minor axis of the galaxy. This left us with an initial sample of 2,712 X-ray point sources. For an example of the execution of this method, see Fig.~\ref{fig:method}.

The luminosity of each detection was calculated using the EPIC flux over all energy bands (0.2--12\,keV), and the calculated distance to the host galaxy as described above. The error on the luminosity was derived from the error on the flux, as while we expect that the error will be dominated by the uncertainty in the distance measurement, this is likely dominated by unknown systematics and so is not well quantified. We defined ULXs as those sources with $L_{\rm X} \geq 10^{39}$\,erg\,s$^{-1}$ or luminosity within 1$\sigma$ of this value, however we also retain a much larger number of lower luminosity sources in our sample as an extension of the catalogue. We note that these are all apparent luminosities as we do not make any corrections for absorption.

\subsection{Removal of known contaminants}
\label{sec:cont1}

A large number of contaminants are still present in the sample at this stage. The majority of contaminants are the active galactic nuclei (AGNs) of the host galaxies. To ensure that our sample only contains the non-nuclear objects, these host AGNs must be removed. While many AGNs will be much more luminous even than most ULXs, a cut cannot be made on the basis of luminosity alone, since ULX luminosities can overlap with those of low luminosity AGNs (LLAGNs), which may have luminosities as low as $\sim10^{38}$\,erg\,s$^{-1}$ (e.g. \citealt{ghosh08}, \citealt{zhang09}). Instead, we removed possible AGNs based on their separation from the centre of the host galaxy as defined by NED. 

We defined a minimum separation $r_{\rm min}$ as the separation between the object's source position and the galaxy centre, minus three times the source position error, a slightly more conservative metric than that used by W11. We set aside all objects with $L_{\rm X} > 10^{42}$\,erg\,s$^{-1}$ into a separate sample of sources which we are confident are AGNs. We found that $>95\%$ of this AGN sample had $r_{\rm min} < 3$\,arcseconds. Therefore we removed all objects with $r_{\rm min} < 3$\,arcseconds, which we found gave a good balance of excluding probable AGNs and retaining as many candidate ULXs as possible. Most sources removed by this cut had an absolute separation from their host galaxy centre of $<5$\,arcseconds. We do expect to lose some genuine ULXs with this cut, and it has previously been found by \citet{swartz04} that the frequency of ULXs increases towards the centre of galaxies (although we note that Swartz et al. also exclude the central 5\,arcseconds from their own analysis), however genuine ULXs at this low separation will inevitably be confused with any AGNs at {\it XMM-Newton}'s angular resolution. This cut removed $\sim30\%$ of the sample in total, leaving 1,886 sources. The remaining $\sim5\%$ of probable bright AGNs are incidentally removed at various later stages of contamination removal.

To remove any remaining known contaminants, we cross-correlated the remaining objects with the \citet{veron10} catalogue of quasars to remove background QSOs, and the Tycho-2 catalogue \citep{tycho2} to remove foreground stars. We then cross-correlated all remaining objects with NED and SIMBAD\footnote{http://simbad.u-strasbg.fr/simbad/, \citet{wenger00}}, and removed all supernovae and background QSOs. Supernova remnants (SNRs) were retained as several well-known ULXs are classified as such by NED\footnote{One example of this is the ULX NGC~6946~X-1, which is a ULX coincident with the optical emission-line nebula MF16 and thus misidentified as a SNR although the nebula was created by the ULX \citep{roberts03}.}. These cross-correlations were all performed within a reasonably conservative radius of 10\,arcseconds and removed a total of 55 sources (a further $\sim3\%$ of the sample). At this point, being confident that known background contaminants have been removed, we resolved any duplicated detections caused by the overlapping of the $D_{25}$ isophotes between different galaxies by assuming that the source lies in the foreground galaxy, as the foreground galaxy is likely to cause a significant amount of absorption to anything lying behind it. In the case of systems at an equal distance, we assumed the source to belong to the galaxy that it is closest to the centre of. The sample at this stage contains 3,069 detections of 1,831 sources. A further handful of contaminants were removed after a close examination of the brightest sources in the sample, performed after the cut of flagged detections (see Section~\ref{sec:flag}).

\subsection{Flagged detections}
\label{sec:flag}

While we have made a number of minor refinements to W11's method in previous stages, the primary difference between our selection method and that of W11 is our consideration of quality flags assigned to detections by the {\it XMM-Newton} pipeline. The {\it XMM-Newton} flagging system highlights problematic detections, however we want to strike a balance between only including good quality data in the sample and retaining as many sources as possible for a large sample size. In this section we give a summary of the flagging system and our justification for the exclusion of sources marked with some types of flag. A detailed description of the various flags and their meanings can be found in the 2XMM and 3XMM User Guides to the Catalogue\footnote{http://xmmssc-www.star.le.ac.uk/Catalogue/2XMM/UserGuide\_xmmcat.html, http://xmmssc-www.star.le.ac.uk/Catalogue/3XMM-DR4/UserGuide\_xmmcat.html}.

There are twelve quality flags that can be assigned to a detection, and a further summary flag field that takes values between 0 and 4 depending upon the state of the quality flags. Flags 1, 2, 3 and 9 are automatically triggered when a source has low detector coverage, is near another source, is within extended emission, or is near the bright corner of the EPIC-MOS1 detector respectively, and therefore may have some problems with its recorded parameters. If any of these flags are true, the summary flag is given the value 1. Flags 4, 5 and 6 are automatically triggered in various circumstances that indicate that the source detection is possibly spurious, and if at least one of these flags is set to true, flag 7 is set as well. Flag 8 is triggered if the source is on the bright EPIC-MOS1 corner or a low-gain column on the EPIC-pn detector. If either flag 7 or flag 8 are true, the summary flag is increased to 2. 

The remaining flags are handled differently between the 2XMM and 3XMM pipelines and depend on the data release the source is reported in. In 2XMM, flag 10 is unused, and flag 11 is manually triggered on visual inspection of the observation if the source lies within a region where a spurious detection is likely -- for example, regions of bright extended emission, out-of-time events or `spiderleg' artefacts caused by scattering from the mirror module support struts -- with flag 12 indicating the bright source which is often the cause for these regions. For detections with flag 11 set to true, the summary flag is increased to 3. If flags 7 or 8 are also true, the summary flag is instead increased to 4. In 3XMM, flag 10 is put to use as an automatically triggered flag in the case of out-of-time events, although if only flag 10 is set, then the summary flag is only given the value 1. Visual inspection of the observations continue to be used to set flag 11, but rather than flag 12 exempting bright sources, flag 11 simply isn't set in these cases. 

\begin{figure}
\begin{center}
\includegraphics[width=82mm]{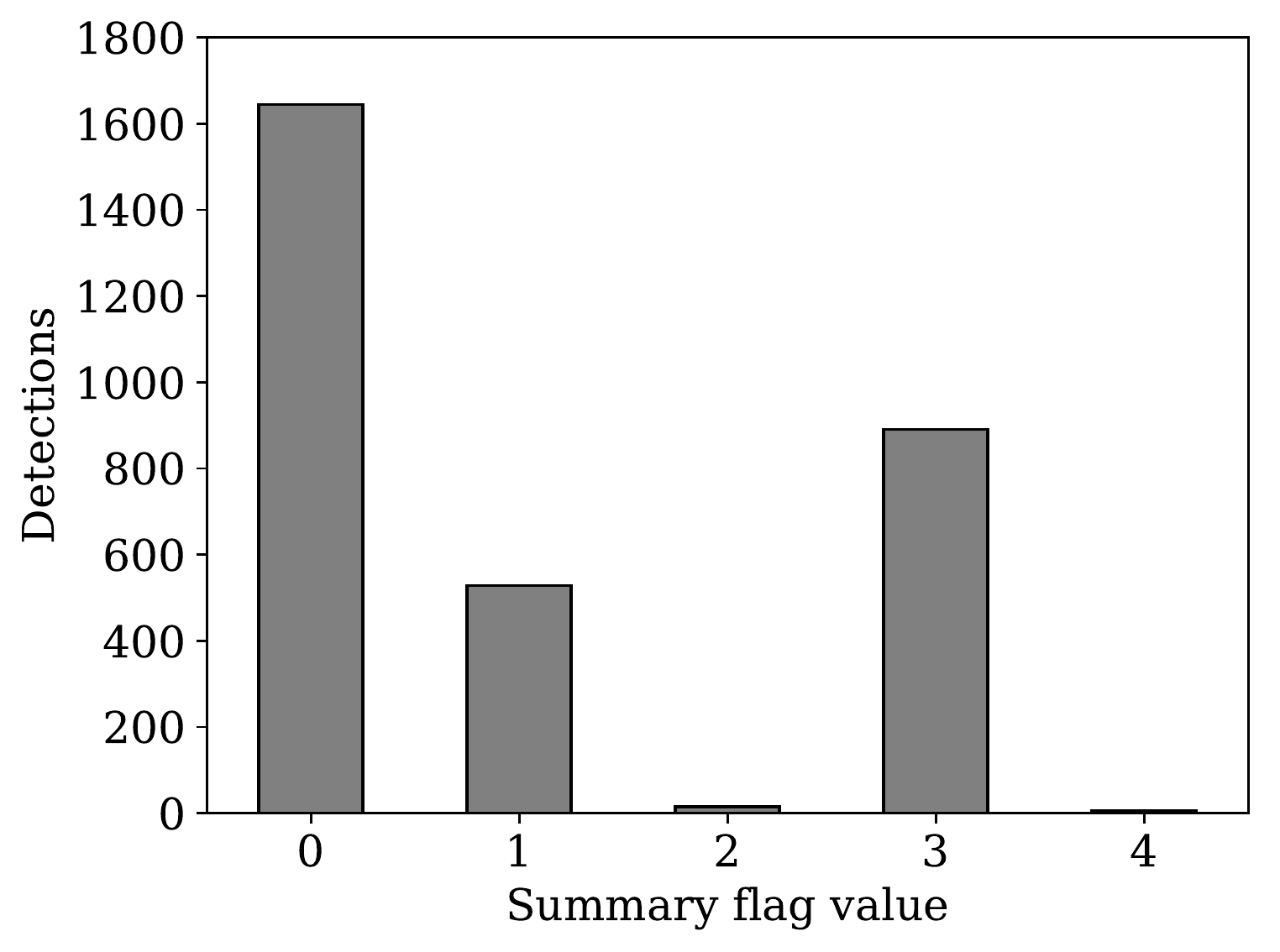}
\end{center}
\vspace{-3mm}
\caption{Bar chart of the occurrence of summary flag values throughout the sample at the point after contaminants have been removed in Section~\ref{sec:cont1}.} 
\label{fig:flags}
\end{figure}

\begin{figure}
\begin{center}
\vspace{-5mm}
\includegraphics[width=82mm]{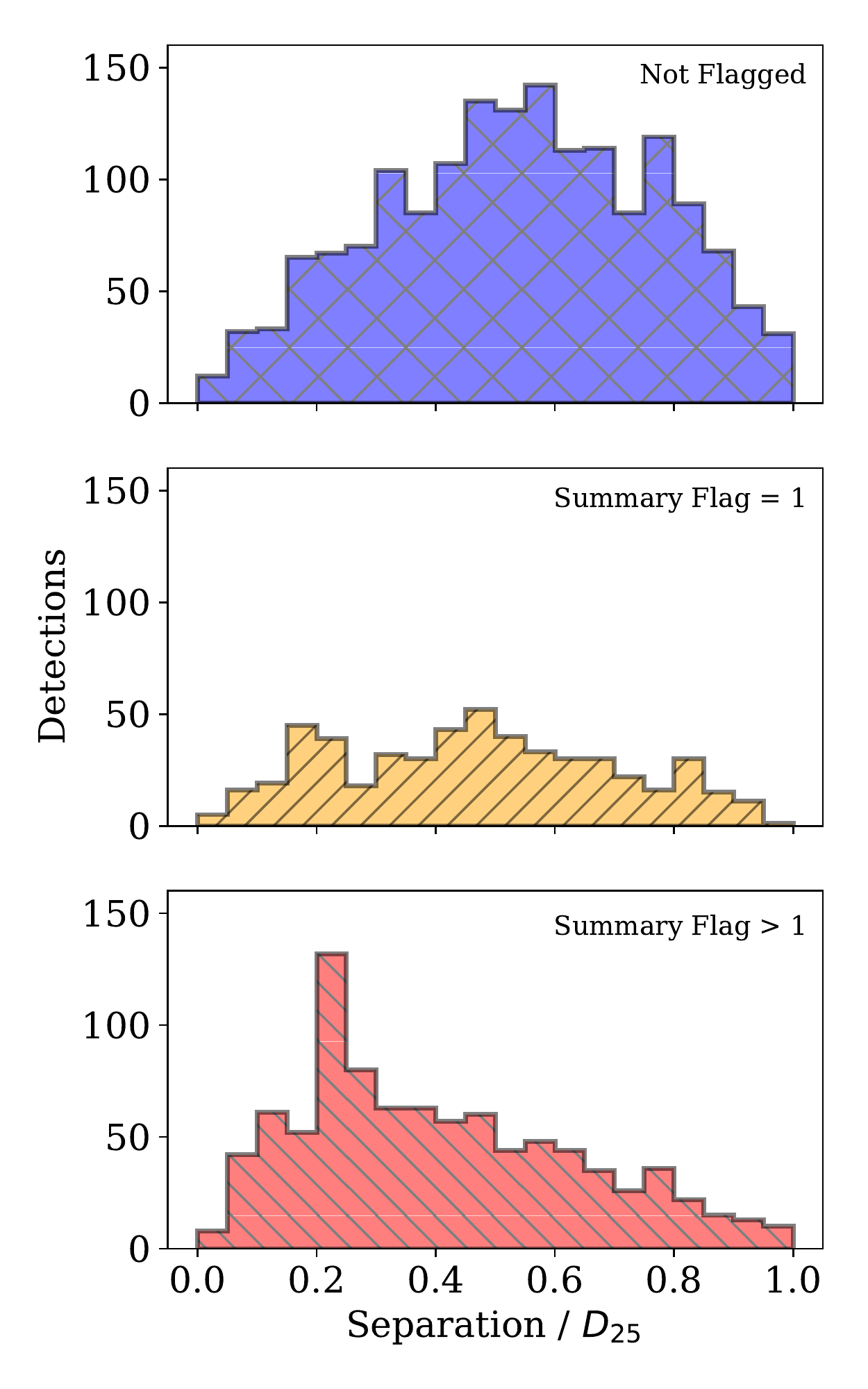}
\end{center}
\vspace{-5mm}
\caption{Histogram of flag occurrence by the separation of the detection from the centre of its host galaxy as a fraction of the semi-major axis. Unflagged sources are shown in blue, sources with summary flag equal to 1 are shown in yellow, and all other flagged sources are shown in red.} 
\label{fig:flagsep}
\end{figure}

\begin{figure}
\begin{center}
\includegraphics[width=82mm]{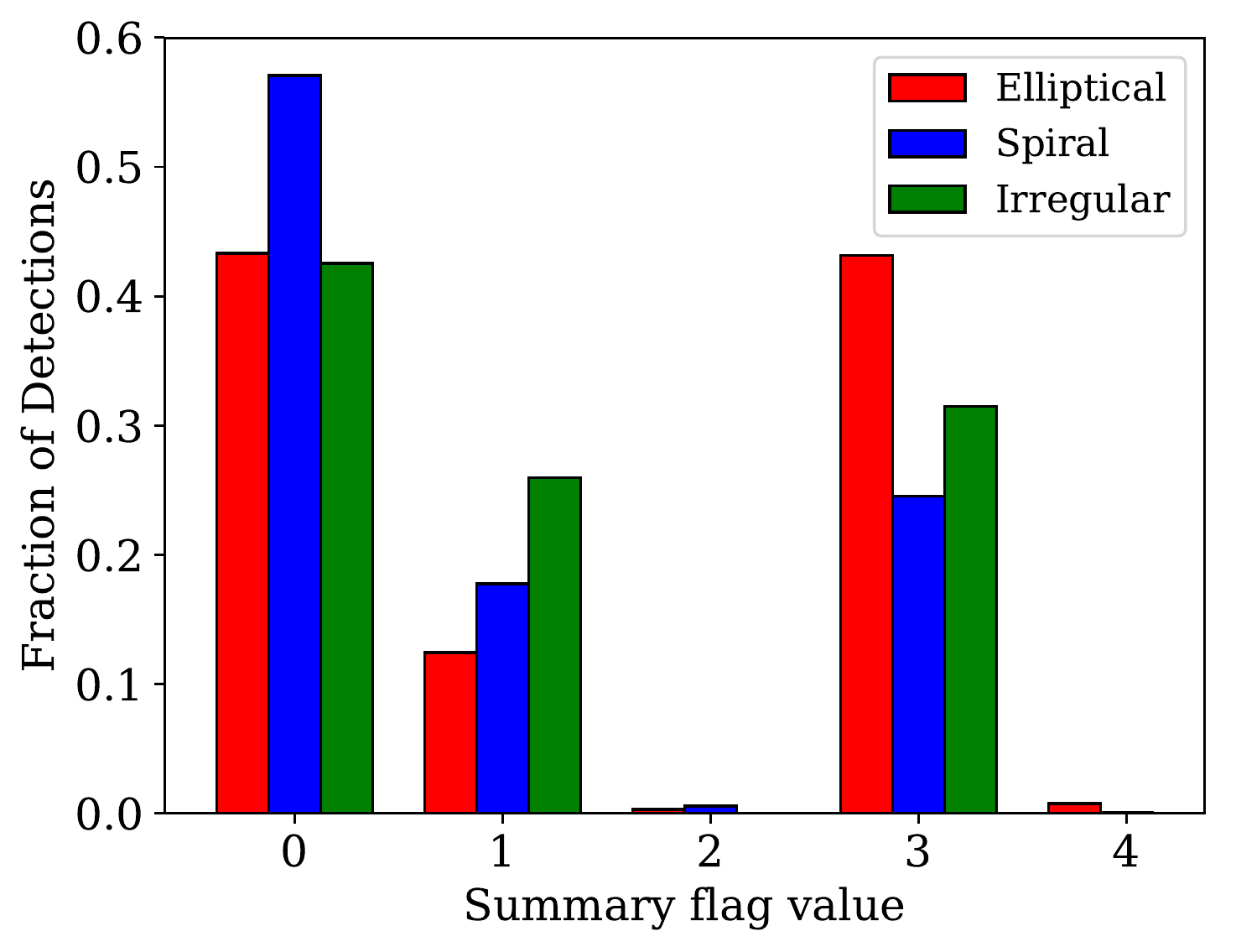}
\end{center}
\vspace{-3mm}
\caption{Bar chart of the fractional occurrence of summary flag values by host galaxy type. Values for elliptical galaxies are plotted in red, spiral galaxies in blue and irregular galaxies in green.} 
\label{fig:flaggals}
\end{figure}

Approximately 45\% of all detections have at least one quality warning flag attached to them and thus have a summary flag value greater than zero. Of the flagged detections, about 38\% have a summary flag of 1, and 61\% have a summary flag of 3, with only a handful having a summary value of 2 or 4 (see Fig.~\ref{fig:flags}). Therefore the main drivers of a detection being flagged are either minor problems with its source parameters, or it lying in a region where spurious detections are likely. We decided that removing all flagged detections would decrease the size of our sample by too much as we still want to retain as many sources as possible, so we kept sources with a summary flag of 1. Sources with a summary flag of 3, however, have a reasonable possibility of being spurious even without flags 7 and 8, which only consider limited scenarios that may cause a detection to be spurious, and there is no easy way of identifying which sources may be genuine without further visual inspection of each observation. Additionally, even those sources that are genuine but embedded within bright extended emission will have their source fluxes erroneously increased and their spectra distorted by the presence of contaminating soft emission. Therefore we removed all detections with a summary flag of 2, 3 or 4. 

Sources with quality flag 10 set to true are likely to be artefacts or out-of-time events rather than genuine detections, although on inspection of some of these sources we found that the genuine source of the out-of-time events was also often flagged in this way. Therefore we removed only those sources with quality flag 10 set to true that had a count rate $<0.05$\,ct\,s$^{-1}$, which keeps the majority of genuine sources on out-of-time columns and removes the majority of actual out-of-time events. 

\begin{figure}
\begin{center}
\includegraphics[width=85mm]{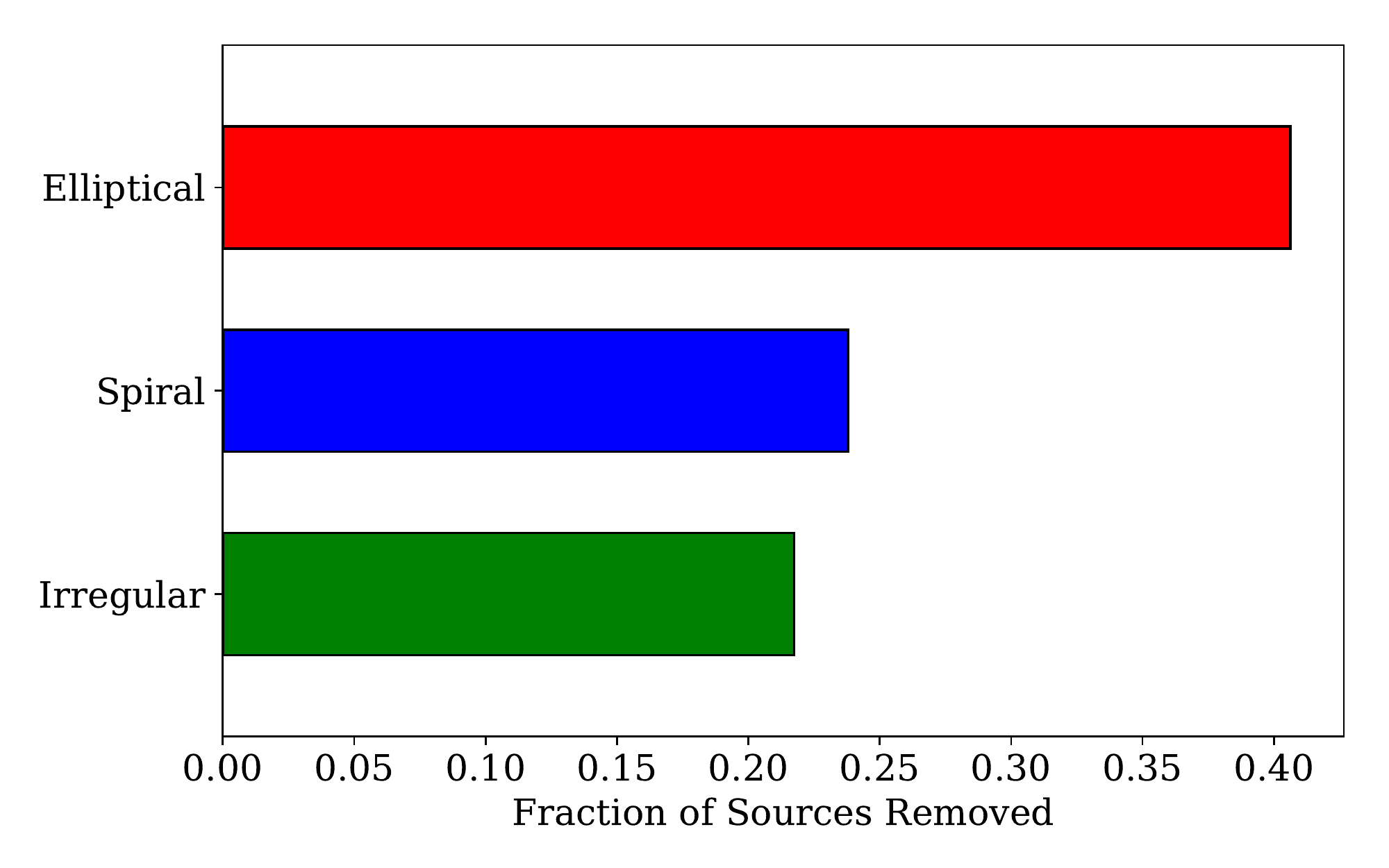}
\end{center}
\vspace{-3mm}
\caption{Bar chart of the fraction of sources removed completely from the sample by the flag cut, by host galaxy type.} 
\label{fig:flagrem}
\end{figure}

We acknowledge that removing these flagged detections introduces some bias in the sample based upon the environment of the sources. Sources with a summary flag of 3 are more likely to be located near the centre of galaxies, since that is where either bright extended emission or spiderleg artefacts caused by bright AGNs are located. We show the bias against more centrally-located sources in Fig.~\ref{fig:flagsep}, where it is clear that sources with a summary flag value greater than 1 are more likely to be centrally located than sources with no flag or a minor flag, which have a mostly flat distribution with fractional distance from the galaxy centre. There is also some bias against sources located in large, bright elliptical galaxies due to their extended X-ray emission, which can be seen in Fig.~\ref{fig:flaggals} in which elliptical galaxies are slightly over-represented for summary flag values greater than 1. This bias is confirmed when we examine the number of sources removed completely from the sample (rather than only having a fraction of their detections removed), and break them down by galaxy type -- around 40\% of sources in elliptical galaxies are removed, compared with about 24\% of sources in spirals and 22\% of sources in irregular galaxies (Fig.~\ref{fig:flagrem}).

However, we expect there to be fewer ULXs in elliptical environments than in star-forming environments from previous studies \citep{swartz04,liu05b}, and upon inspection of some objects discarded for this reason we conclude that many of the real sources are more likely to be background AGNs than genuine ULXs due to possessing optical or infrared counterparts that may indicate that the source is actually in a background galaxy. Therefore we are confident that removing those detections with a summary flag greater than 1 strikes a reasonable balance between creating a clean and reliable sample and still retaining a fairly large sample size. After applying this flag cut, the sample contains 2,149 detections of 1,322 sources. At this point, we performed a closer examination of some of the more extreme objects of the sample (for example, objects with high luminosity or very hard spectra; see Section~\ref{sec:bright}) and we removed all detections of eight sources which coincided with the AGN of their host galaxy but had a separation of $r_{\rm min} > 3$\,arcseconds due to an inaccurate position, or coincided with background objects of known redshift. 

\subsection{Sample properties}
\label{sec:prop}

\begin{figure}
	\begin{center}
		\hspace{-5mm}\includegraphics[width=88mm]{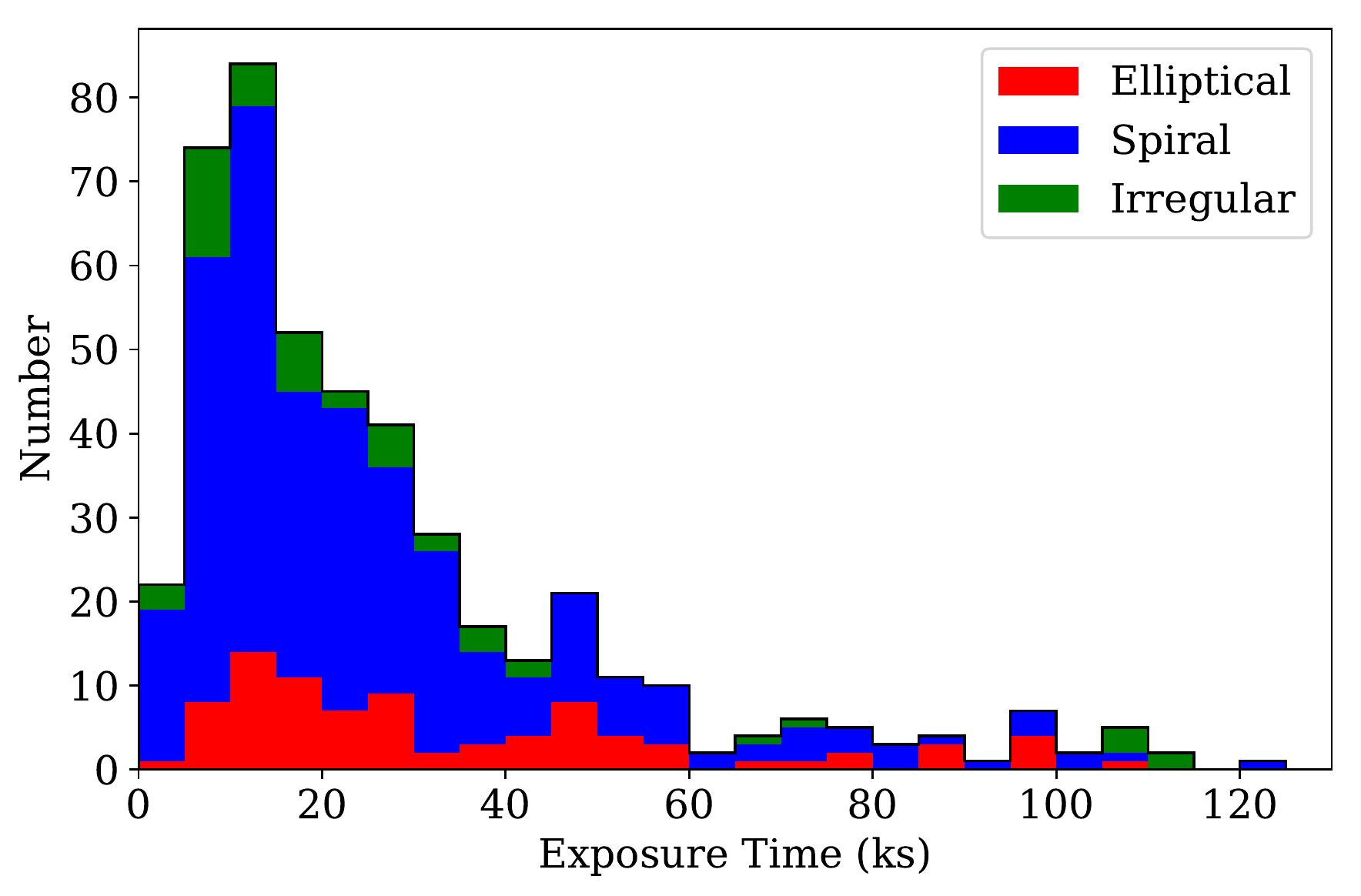}
	\end{center}
	\vspace{-3mm}
	\caption{Stacked histogram by galaxy type of the distribution of exposure times for the observations that went into creating our sample.} 
	\label{fig:exptimes}
\end{figure}

\begin{figure}
	\begin{center}
		\hspace{-5mm}\includegraphics[width=88mm]{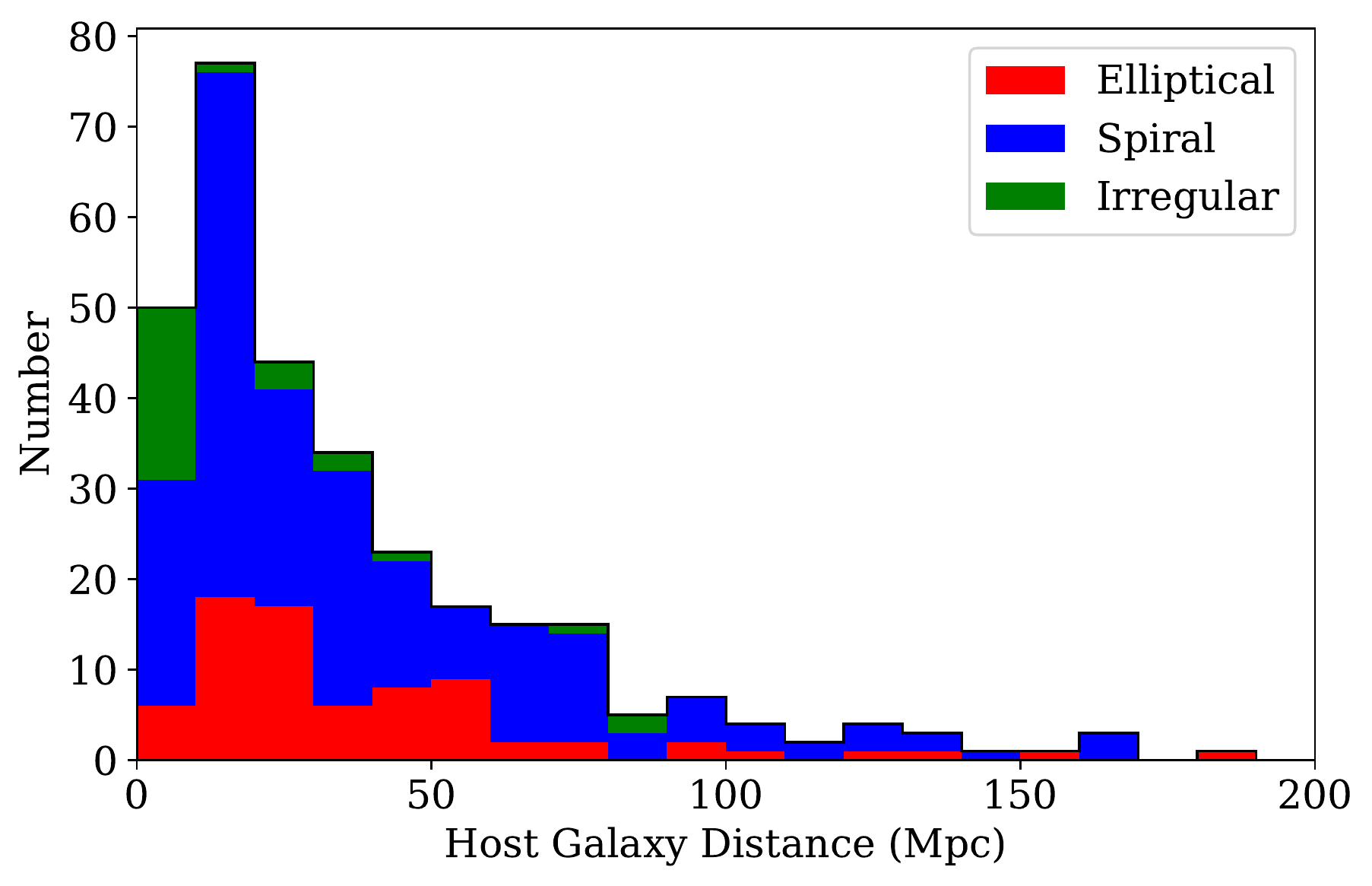}
	\end{center}
	\vspace{-3mm}
	\caption{Stacked histogram by galaxy type of the distribution of distances for the host galaxies of our sample.} 
	\label{fig:galdist}
\end{figure}

\begin{figure}
	\begin{center}
		\hspace{-5mm}\includegraphics[width=88mm]{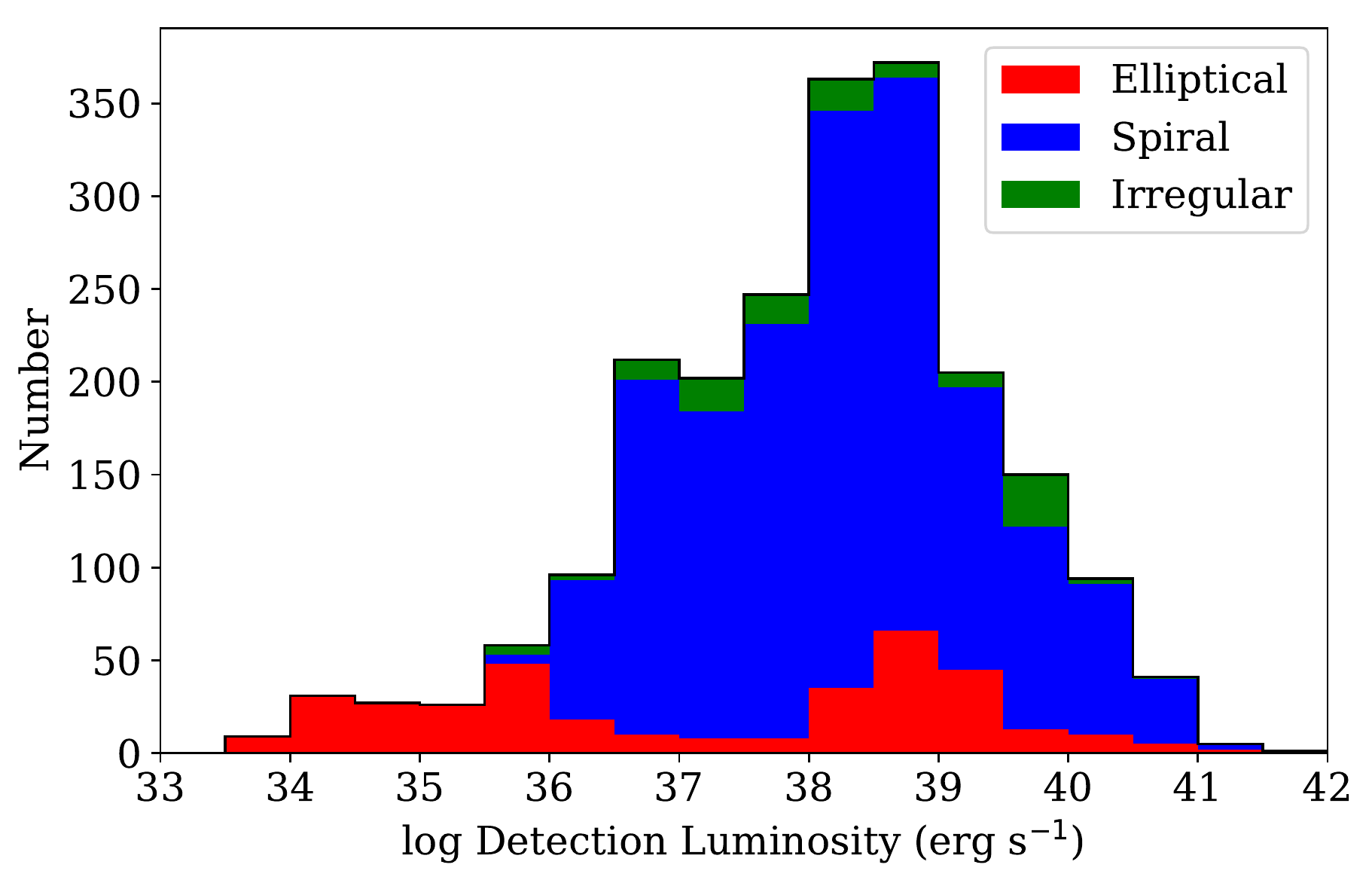}
	\end{center}
	\vspace{-3mm}
	\caption{Stacked histogram by galaxy type of the distribution of luminosities for the detections in our sample.} 
	\label{fig:luminosityhist}
\end{figure}

The final sample is available in digital format alongside this paper (see Appendix A) and contains 2,139 detections of 1,314 sources, located within 305 host galaxies. Of these sources, 384 are candidate ULXs, defined as having at least one detection with $L_{\rm X} > 10^{39}$\,erg\,s$^{-1}$ or else having $L_{\rm X}$ within $1\sigma$ of $10^{39}$\,erg\,s$^{-1}$. These data were obtained from 459 unique {\it XMM-Newton} observations with a median duration of $\sim20$\,ks (although observations are concentrated below the median, the exposure time distribution has a tail that extends to over 120\,ks -- see Fig.~\ref{fig:exptimes}). The median distance for the host galaxies of the sources in our sample is 23.3\,Mpc, with the distribution of distances similarly concentrated at relatively low distances with a long tail out to very distant galaxies (see Fig.~\ref{fig:galdist}). Given the sensitivity of {\it XMM-Newton}, for a typical 20\,ks observation of a galaxy 20\,Mpc away, we can expect to detect sources down to $\sim10^{38}$\,erg\,s$^{-1}$, depending on the spectral shape. Indeed, while there are several hundred ULXs, the majority of detections we include are of sources with luminosities in the range $10^{37} < L_{\rm X} < 10^{39}$\,erg\,s$^{-1}$ (see Fig.~\ref{fig:luminosityhist}). This makes the catalogue a good resource for sampling X-ray populations of interest with luminosities below the ULX regime, such as the Eddington threshold (see Section~\ref{sec:threshold}).

We present sample numbers, both altogether and by galaxy type, in Table~\ref{tab:numbers}, both for the full-luminosity sample and for those sources identified as ULXs. For ULXs, we also include the detections of those sources that do not themselves reach ULX luminosities, since many ULXs are highly variable in flux between observations. Therefore, not all detections of a ULX are required to have $L_{\rm X} > 10^{39}$\,erg\,s$^{-1}$ (or to be within 1$\sigma$) for the source to be classified as such. 

There is a common association of ULXs with star-forming regions that might suggest we would expect to find a greater number of ULXs in spiral or irregular (mostly star-forming) galaxies than in elliptical (mostly non-star-forming) galaxies. Indeed, we find that the majority of extragalactic non-nuclear X-ray sources we detect are found in spiral or irregular galaxies regardless of luminosity, with $\sim18\%$ of all spiral galaxies and $\sim35\%$ of all irregular galaxies containing at least one detectable non-nuclear X-ray source, compared with $\sim12\%$ of elliptical galaxies. Similarly, while fewer galaxies contain a ULX, we find more in spirals and irregulars: $\sim15\%$ of field galaxies contain at least one ULX in both cases, compared with $\sim9\%$ of ellipticals. This appears to be consistent with the association of ULXs with star-forming environments. We do note our previously-identified bias against bright elliptical galaxies in flagging, which will reduce the number of sources in elliptical galaxies that make it into our sample (see Section~\ref{sec:flag}), however when we correct for the proportion of flagged detections removed, as well as for the reduced number of elliptical galaxies in the field to begin with, we still find that spiral and irregular galaxies contain greater numbers of non-nuclear X-ray sources than elliptical galaxies do. All the same, whether an existing source is detected within a galaxy is dependent on the distance of the galaxy and how for long it is observed. Therefore it is more meaningful to make these comparisons with galaxies in which we are confident that all sources of a given luminosity are detected, to which end we create a complete subsample (see Section~\ref{sec:comp}).

\begin{table*}
	\begin{minipage}{100mm}
		\caption{Catalogue numbers for all sources and for ULXs, broken down by galaxy type.} \label{tab:numbers}
		\begin{center}
			\begin{tabular}{lcccc}
				\hline
				& Full sample & Spiral & Elliptical & Irregular \\
				\hline
				Field galaxies 				& 1,868 & 1,133 & 650 & 84 \\
				- with a nuclear source removed & 832 & 483 & 326 & 22 \\\\
				Detections 					& 2,139 & 1,659 & 362 & 118  \\ 
				Sources 						& 1,314 & 942 & 300 & 72 \\
				- with multiple detections 	& 351 & 300 & 35 & 16 \\
				Host galaxies 					& 305 & 201 & 75 & 29 \\
				- containing multiple sources 	& 145 & 99 & 32 & 14 \\\\
				ULX detections 				& 606 & 466 & 98 & 42 \\ 
				ULXs 							& 384 & 279 & 85 & 20 \\
				- with multiple detections 	& 92 & 71 & 13 & 8 \\ 
				ULX host galaxies 				& 241 & 168 & 58 & 15 \\
				- containing multiple ULXs 	& 76 & 61 & 11 & 4 \\
				\hline
			\end{tabular}
		\end{center}
	\end{minipage}
\end{table*}

\begin{table*}
	\begin{minipage}{104mm}
		\caption{The field galaxy and host galaxy properties for the overall sample, broken down by galaxy type.} \label{tab:galaxies}
		\vspace{-2mm}
		\begin{center}
			\begin{tabular}{cccccc}
				\hline 
				Field galaxies & All & Spiral & Elliptical & Irregular \\ 
				\hline 
				$\tilde{d}^a$ & $70.3^{+22.1}_{-19.1}$ & $73.9^{+20.9}_{-18.2}$ & $68.0^{+22.4}_{-16.8}$ & $27.9^{+10.8}_{-18.4}$ \\ 
				$\tilde{M_{\rm B}}^b$ & $-20.4\pm0.5$ & $-20.3\pm0.5$ & $-20.5\pm0.5$ & $-17.5^{+0.8}_{-1.1}$ \\ 
				\hline 
				\hline 
				Source host galaxies & All & Spiral & Elliptical & Irregular \\ 
				\hline 
				$\tilde{d}^a$ & $23.2^{+16.5}_{-5.6}$ & $24.5^{+16.5}_{-6.6}$ & $25.0^{+19.5}_{-4.7}$ & $4.4^{+10.4}_{-0.8}$ \\ 
				$\tilde{M_{\rm B}}^b$ & $-20.3^{+0.6}_{-0.4}$ & $-20.4^{+0.5}_{-0.4}$ & $-20.6^{+0.5}_{-0.4}$ & $-17.1^{+0.5}_{-0.7}$ \\ 
				$\tilde{{\rm N}}^c$ & $1^{+1}_{-0}$ & $1^{+1}_{-0}$ & $1^{+1}_{-0}$ & $1^{+1}_{-0}$ \\ 
				$\bar{{\rm N}}^d$ & $4.3\pm0.5$ & $4.7\pm0.7$ & $4.0\pm0.8$ & $2.5\pm0.4$ \\ 
				\hline 
				\hline 
				ULX host galaxies & All & Spiral & Elliptical & Irregular \\ 
				\hline 
				$\tilde{d}^a$ & $32.2^{+16.3}_{-11.4}$ & $32.4^{+17.1}_{-13.6}$ & $35.5^{+14.7}_{-11.4}$ & $21.8^{+15.3}_{-12.7}$ \\ 
				$\tilde{M_{\rm B}}^b$ & $-20.5\pm0.5$ & $-20.5^{+0.5}_{-0.4}$ & $-20.8^{+0.4}_{-0.5}$ & $-18.0^{+0.7}_{-0.8}$ \\ 
				$\tilde{{\rm N}}^c$ & $1\pm0$ & $1^{+1}_{-0}$ & $1\pm0$ & $1\pm0$ \\ 
				$\bar{{\rm N}}^d$ & $1.59\pm0.08$ & $1.7\pm0.1$ & $1.5\pm0.2$ & $1.3\pm0.2$ \\ 
				\hline
			\end{tabular}
		\end{center}
		\vspace{-1mm}
		$^a$Median galaxy distance in Mpc, along with the inter-tercile range. \\
		$^b$Median absolute magnitude in the $B$ system, corrected for Galactic and internal extinction, along with the inter-tercile range.\\
		$^{c,d}$Median/mean number of sources/ULXs located in the host galaxy, along with the inter-tercile range/standard error on the mean.
	\end{minipage}
\end{table*}

We present a selection of average host galaxy properties in Table~\ref{tab:galaxies}, both for the entire list of field galaxies, those galaxies containing any good detections of X-ray sources, and ULX host galaxies. ULX host galaxies are on average more distant than the overall population of those containing non-nuclear X-ray sources, illustrating the relative scarcity of ULXs and also that they can be detected at larger distances. In general though, galaxies containing non-nuclear X-ray sources have a lower average distance than field galaxies because of the limitations of {\it XMM-Newton's} sensitivity and spatial resolution -- at higher distances it is not only hard to detect any but the very brightest sources, but it is also more likely that a non-nuclear X-ray source within a galaxy will be indistinguishable from an AGN due to the positional uncertainty extending over the central 3\,arcsecond region that we use to define AGNs, and thus the source is filtered out of our catalogue.

In order to determine how many of the ULX candidates that we identify are new, we matched our sources with the following previous ULX catalogues: \citet{colbert02}, \citet{swartz04}, \citet{liu05a}, \citet{liu05b}, \citet{swartz11} and W11. These previous catalogues use data from {\it ROSAT}, {\it Chandra} and {\it XMM-Newton}, and various slightly different methods between them. The largest, W11, contains 475 ULX candidates. We find in our sample 161 new candidate ULXs which do not appear in any of these catalogues. Additionally we find 201 of the sources identified in the W11 catalogue in our own as well (the remaining sources are previously identified ULXs that do not appear in W11). 

The discrepancy in numbers between our catalogue and W11's, with 274 W11 ULXs not found as ULXs in our catalogue, is mostly due to two main factors. The first factor is the changes between the 2XMM and 3XMM pipelines to produce the initial Serendipitous Source Catalogue. This leads to some of the W11 objects being classed as extended or not significantly detected (about 6\% of the discrepancy), or slight changes in the calculated flux causing the derived luminosity to be below the ULX threshold ($\sim16\%$ of the discrepancy; this may also be due to differences in the distance that we assign to each galaxy) -- note that in this latter case the sources are still found in our catalogue, just not classed as ULXs. The second major factor is our removal of detections with certain quality flags (see Section~\ref{sec:flag}), which accounts for $\sim60\%$ of the discrepancy. The remaining difference is accounted for by galaxies we discarded due to an inadequate distance measurement, and sources discarded in our contamination removal stage. As for the sources found in our catalogue but not in W11, approximately half are new to DR4 and half are due to similar reasons to the above.


While our sample is smaller than the largest previous catalogue (although still large), due to our handling of flags and contaminants we are confident that it is cleaner than previous catalogues and also contains a large number of ULXs that have not yet been considered in most ULX population studies. We present the catalogue as a digital accompaniment to this paper -- details of its data fields are briefly discussed in Appendix~\ref{app:catalogue}.

\subsection{Complete subsamples}
\label{sec:comp}

In order to perform studies of the ULX population of our sample of extragalactic X-ray sources without bias towards brighter sources, we require a complete subsample -- that is, a sample made up of all ULXs within galaxies for which we are confident that all ULXs they possess have been detected. This is to ensure that the ULXs we examine are representative of the population as a whole. The method we use is presented in detail in W11, so we provide only a brief summary here.

We begin by using 3XMM sensitivity maps for {\it XMM-Newton} (see \citealt{carrera07} and \citealt{mateos08} for a description of the method of creating these) to find the minimum flux at a certain position on the detector for each observation and each EPIC camera, $f_{\rm obs}$, such that a source could be detected with maximum likelihood $>10$ across the energy range 0.5--12\,keV. We then calculate the minimum ULX flux in the same band, $f_{\rm ULX}$, emitted by a source with luminosity $L_{\rm X} = 10^{39}$\,erg\,s$^{-1}$ and an average ULX spectral shape which we define as an absorbed power-law spectrum with $\langle N_{\rm H}\rangle = 2.4\times10^{21}$\,cm$^{-2}$ and $\langle\Gamma\rangle = 2.2$, as determined from \citet{gladstone09}. We compare $f_{\rm obs}$ and $f_{\rm ULX}$ for each observation of the galaxy across its entire area, excluding chip gaps and a 7.5\,arcsec circular region excluded to account for our filtering out likely AGNs, and for each {\it XMM-Newton} EPIC camera. An observation of a galaxy is determined complete if $f_{\rm ULX} > f_{\rm obs}$ over the entire galaxy for at least one of the detectors and thus we can be confident that all ULXs in the galaxy have been detected at least once and will therefore appear in the catalogue. We also remove from the complete subsample any galaxies for which we do not have position angle data and thus only search for sources within the minor axis of the galaxy footprint, since we may be excluding genuine ULXs that lie outside of this region of their host galaxy. 

For the majority of galaxies we consider, the excluded central region does not account for a large area of the galaxy, being $<6\%$ of the total area for $>90\%$ of the galaxies. Therefore it is likely that our 'complete' subsample is still incomplete to a few per cent when it comes to excluded sources in the central region of each host galaxy, although we are unable to distinguish those sources from AGNs using {\it XMM-Newton} at this radius.

\begin{table*}
	\hspace{-5mm}
	\begin{minipage}{180mm}
		\caption{Catalogue numbers and galaxy properties for the complete subsamples, broken down by galaxy type.} \label{tab:complete}
		\vspace{-2mm}
		\begin{center}
			\begin{tabular}{@{~}c@{~}c@{~}c@{~}c@{~}cc@{~}c@{~}c@{~}cc@{~}c@{~}c@{~}c@{~}}
				\hline
				 & \multicolumn{12}{c}{Sample complete to $L_{\rm X,cutoff}$:} \\
				 & \multicolumn{4}{c}{$10^{40}$\,erg\,s$^{-1}$} & \multicolumn{4}{c}{$10^{39}$\,erg\,s$^{-1}$} & \multicolumn{4}{c}{$10^{38}$\,erg\,s$^{-1}$} \\
			     & All & Spiral & Elliptical & Irregular & All & Spiral & Elliptical & Irregular & All & Spiral & Elliptical & Irregular\\
				\hline
				Field galaxies & 694 & 409 & 253 & 32 & 248 & 153 & 74 & 21 & 42 & 23 & 9 & 10  \\
				$\tilde{d}$ & $41.5^{+16.5}_{-13.1}$ & $40.0^{+17.8}_{-13.8}$ & $48.3^{+11.4}_{-16.5}$ & $23.0^{+8.5}_{-12.2}$ & $19.7^{+3.0}_{-3.1}$ & $19.0\pm2.6$ & $20.9^{+3.6}_{-2.9}$ & $9.1^{+12.6}_{-5.1}$ & 			$4.3^{+3.7}_{-0.7}$ & $7.5^{+1.7}_{-3.2}$ & $0.7^{+1.7}_{-0.0}$ & $3.5^{+0.3}_{-0.4}$  \\ 
				Max. $d$ & 187.9 & 187.9 & 132.0 & 97.9 & 56.6 & 47.0 & 56.6 & 37.4 & 17.5 & 15.4 & 17.5 & 9.1 \\
				$\tilde{M_{\rm B}}$ & $-20.3^{+0.5}_{-0.4}$ & $-20.3^{+0.6}_{-0.4}$ & $-20.3^{+0.4}_{-0.5}$ & $-17.7^{+0.6}_{-0.4}$ & $-19.8^{+0.7}_{-0.6}$ & $-20.0\pm0.5$ & $-19.9\pm0.6$ & $-17.4^{+0.6}_{-0.5}$ & $-17.9^{+1.2}_{-1.4}$ & $-19.5^{+0.6}_{-0.2}$ & $-15.6^{+0.9}_{-0.2}$ & $-17.0^{+0.4}_{-0.5}$ \\
				\hline
				\hline
				 & \multicolumn{4}{c}{$10^{40}$\,erg\,s$^{-1}$} & \multicolumn{4}{c}{$10^{39}$\,erg\,s$^{-1}$} & \multicolumn{4}{c}{$10^{38}$\,erg\,s$^{-1}$}\\
				 & All & Spiral & Elliptical & Irregular & All & Spiral & Elliptical & Irregular & All & Spiral & Elliptical & Irregular\\
				\hline
				Host galaxies$^a$ & 52 & 43 & 5 & 4 & 93 & 63 & 22 & 8 & 28 & 22 & 2 & 4 \\
				$\tilde{d}$ & $42.3^{+14.2}_{-8.4}$ & $41.5^{+15.8}_{-4.8}$ & $45.0^{+3.9}_{-7.5}$ & $56.9\pm29.9$ & $18.0^{+2.8}_{-2.0}$ & $17.3^{+1.9}_{-1.8}$ & $23.0^{+0.8}_{-3.7}$ & $12.8^{+7.3}_{-3.9}$ & $7.5^{+1.6}_{-3.2}$ & $7.5^{+1.7}_{-3.1}$ & $9.1\pm2.9$ & $6.5\pm2.4$ \\ 
				Max. $d$ & 120.7 & 120.7 & 56.4 & 89.7 & 42.3 & 42.3 & 37.5 & 36.9 & 17.5 & 15.3 & 17.5 & 9.1  \\
				$\tilde{M_{\rm B}}$ & $-20.5^{+0.3}_{-0.5}$ & $-20.5\pm0.3$ & $-20.5^{+0.5}_{-0.6}$ & $-20.0^{+1.2}_{-1.1}$ & $-20.3^{+0.6}_{-0.3}$ & $-20.3^{+0.6}_{-0.2}$ & $-20.8^{+0.4}_{-0.3}$ & $-17.9^{+0.4}_{-0.3}$ & $-19.1^{+1.0}_{-0.5}$ & $-19.5^{+0.6}_{-0.2}$ & $-17.2\pm0.5$ & $-17.5^{+0.3}_{-0.2}$ \\ \\
				Sources$^b$ & 57 & 48 & 5 & 4 & 190 & 133 & 45 & 12 & 103 & 89 & 2 & 12 \\  
				$\tilde{{\rm N}}^b$ & $1\pm0$ & $1\pm0$ & $1\pm0$ & $1\pm0$ & $1^{+1}_{-0}$ & $2^{+0}_{-1}$ & $1^{+1}_{-0}$ & $1^{+1}_{-0}$ & $3^{+2}_{-1}$ & $3^{+3}_{-1}$ & $1\pm0$ & $2.5\pm0.5$ \\ 
				$\bar{{\rm N}}^b$ & $1.10\pm0.05$ & $1.12\pm0.06$ & $1\pm0$ & $1\pm0$ & $2.0\pm0.2$ & $2.1\pm0.2$ & $2.0\pm0.5$ & $1.5\pm0.2$ & $4.1\pm0.7$ & $4.5\pm0.8$ & $1\pm0$ & $3.2\pm0.8$ \\ 
				\hline
			\end{tabular}
		\end{center}
		Properties as given in Table~\ref{tab:galaxies} except where specified. \\
		$^a$The number of host galaxies containing at least one source with a detection above or within 1$\sigma$ of the cut-off luminosity for each complete subsample. \\
		$^b$The number of X-ray sources with at least one detection above or within 1$\sigma$ of the cut-off luminosity for each complete subsample, and its median/mean.
	\end{minipage}
\end{table*}

The complete subsample of ULXs contains 287 detections of 190 ULXs from 248 galaxies complete to $10^{39}$\,erg\,s$^{-1}$, 148 of which contain at least one non-nuclear X-ray source (although sources with luminosities below $10^{39}$\,erg\,s$^{-1}$ are not guaranteed to be detected) and 93 of which host at least one ULX -- that is, approximately 1 in 3 galaxies host a ULX. Sub-divided by galaxy type, we find that $\sim40\%$ of spiral and irregular galaxies contain a ULX, compared with $\sim30\%$ of ellipticals. Complete galaxies are of course at much lower distances than the galaxy list as a whole, with all complete galaxies within 56.6\,Mpc and at a median distance of 19.7\,Mpc. 

We repeated this process for cut-off luminosities of $10^{40}$ and $10^{38}$\,erg\,s$^{-1}$ for comparison. We give the numbers of galaxies broken down by galaxy type for each complete subsample in Table~\ref{tab:complete}, as well as the number of sources above the cut-off luminosity of each subsample, along with galaxy properties. Since the number of irregular galaxies and sources within them is so small, we place ULXs in irregular galaxies into the spirals group for all further analysis, as we would expect both irregular and spiral galaxies to be primarily star-forming environments. We refer to this group containing ULXs in both spiral and irregular galaxies purely as spiral-hosted for convenience. 

\subsection{Quantifying unknown contamination}
\label{sec:cont2}

While we have taken care to remove known contaminants (see Section~\ref{sec:cont1}), some fraction of the remaining sources will be made up of unknown contaminants, in this case background quasars that have not been identified as such. While these cannot be removed from the sample, we can attempt to quantify the extent of the remaining contamination by estimating the number of background sources we expect to appear in the catalogue. This method is also very similar to that used in W11, so we again briefly summarise it here and direct the reader to W11 for further details.

We took a typical quasar spectrum to be an absorbed power-law with effective photon index $\Gamma = 1.59$ \citep{piconcelli03,mateos08} and with a column density equal to the Galactic column density in the direction of the host galaxy under consideration. We defined a limiting flux $f_{\rm lim}$ as the highest one of two fluxes: the minimum ULX flux $f_{\rm ULX}$, calculated using the quasar spectrum, and the minimum detection sensitivity flux $f_{\rm obs}$, as defined in Section~\ref{sec:comp}. Using the \citet{moretti03} distribution\footnote{We choose to use the \citet{moretti03} distribution over more recent ones partly for consistency with W11, and partly because the distribution spans a large range in flux which makes it particularly useful for our purposes.} of background sources that should be resolved at a flux sensitivity $S$, i.e. $N(>S)$, we converted the $f_{\rm lim}$ maps to maps of the number of background contaminants expected to be observed for each pixel in the hard band (2--12\,keV) -- this was found to be more reliable than using a soft band in W11, due to minimising the effects of additional absorption by the apparent host galaxy. The total expected number of contaminants observed above the ULX luminosity is the sum of the background contaminants across the area of each galaxy, excluding chip gaps and the inner AGN region. This was calculated for each EPIC camera. 

We subtracted the known background contaminants (i.e. those sources identified as background AGN/QSOs in Section~\ref{sec:cont1}) from this number, then compared it with the number of sources above $10^{39}$\,erg\,s$^{-1}$ detected in our sample with maximum likelihood $>10$ in the hard band. We calculated a fractional background contamination for each EPIC camera and produced a weighted average, using counting statistics to determine uncertainties. We find an estimated background contamination of $23.9\pm3.4$ per cent across the entire sample of ULXs. Divided into galaxy types, ellipticals suffer a far larger amount of contamination, with a fractional contamination of $43.4\pm9.0$ per cent compared to $17.5\pm3.5$ per cent for spirals -- this is perhaps unsurprising given the expectation that a greater proportion of genuine ULXs will be found in star-forming galaxies than in elliptical galaxies. With fewer bright X-ray sources detected within the galaxies themselves, the contamination percentage will naturally be higher. The small numbers of sources involved also cause the errors on the elliptical contamination percentage to be relatively large.

\begin{table}
		\caption{The estimated remaining contamination percentage above three cut-off luminosities, as a weighted average over the three EPIC cameras.} \label{tab:contamination}
		\vspace{-2mm}
		\begin{center}
			\begin{tabular}{lccc}
				\hline
				$L_{\rm X,cutoff}$ & $10^{40}$\,erg\,s$^{-1}$ & $10^{39}$\,erg\,s$^{-1}$ & $10^{38}$\,erg\,s$^{-1}$ \\
				\hline
				Contamination (\%) & $12.0\pm2.2$ & $23.9\pm3.4$ & $22.3\pm2.7$ \\
				- Spirals & $5.5\pm1.4$ & $17.5\pm3.5$ & $19.0\pm2.8$ \\ 
				- Ellipticals & $43\pm13$ & $43.4\pm9.0$ & $35.6\pm7.3$ \\
				\hline
			\end{tabular}
		\end{center}
\end{table}

We repeated this estimation of the background contamination for sources detected above $10^{38}$ and $10^{40}$\,erg\,s$^{-1}$, and give the contamination percentages for these in Table~\ref{tab:contamination}. At the highest luminosities, a proportionally greater number of contaminants are removed by hand due to individual inspection of bright sources (see Section~\ref{sec:bright}), so the expected remaining contamination percentage is relatively low. 

This contamination estimate was performed before and separately to the removal of detections based upon {\it XMM-Newton} quality flags (see Section~\ref{sec:flag}). While further contaminants due to badly constrained fluxes and camera artefacts were removed at that step, we expect that the fractional contamination due to unknown background objects remained similar after the flag cut. It should be noted that even though we use the hard band to reduce absorption effects, these percentages are upper limits (particularly in the case of dusty spiral galaxies) as we do not take absorption intrinsic to the host galaxy into account -- background sources will still be subject to this absorption and thus the number of contaminants detected will be lower than the number predicted. 

\subsection{Catalogue limitations}
\label{sec:limit}

Given the similarities of our method to that of W11, our catalogue shares some of its limitations, including the presence of contamination due to background sources (although we improve on removing contaminants due to artefacts or badly constrained source flux), incompleteness due to the limited number of galaxies observed by {\it XMM-Newton} and the limited observation depth, and loss of some point sources near the centre of their host galaxy due to our AGN cut. In addition to this, we also introduce a further radial bias by filtering on {\it XMM-Newton} quality flags (see Section~\ref{sec:flag}), and lose some sources from the edges of galaxies where we do not have position angle data and only search within the minor axis of the galaxy footprint.

We also note that in a number of cases, the angular resolution of {\it XMM-Newton} is not sufficient to resolve sources that are close together in space. In some cases, a source appears in our catalogue as a particularly bright ULX, which can be resolved using {\it Chandra} into separate sources which may or may not be ULXs themselves (e.g. a luminous ULX in NGC~2276, \citealt{sutton12}). It is also possible that some clusters of unresolved sources, or simply sources embedded in a bright extended star-forming region, may have been identified as extended and thus filtered out of our sample.

Since the publication of 3XMM-DR4 in 2013 there have been a number of subsequent data releases of the 3XMM catalogue, the latest being 3XMM-DR7 in 2017. Each data release brings with it an additional quantity of data, as well as corrections to errors identified in previous versions of the catalogue. In particular, the 3XMM-DR5 data release came with the identification of problems in the 3XMM-DR4 release, including a number of corrupted event lists from mismatched mosaic mode sub-pointings (although no observations included in our sample were affected by this), and erroneously over-estimated errors in some detection positions, fluxes and hardness ratios \citep{rosen16}. This affects the overall derived values of these quantities as they are an error-weighted combination of values from the EPIC-MOS and EPIC-pn cameras. Additionally, erroneously large position errors may cause sources near the centre of galaxies to be incorrectly identified as AGN. Therefore, while 3XMM-DR4 was the most up-to-date catalogue at the beginning of this project, it is no longer the largest and most reliable dataset available today. We anticipate producing future iterations of this catalogue from more recent releases of the 3XMM catalogue.

Finally, we note that a handful of famous ULXs do not make it into our catalogue, mainly due to the galaxy-matching stage. For example, HLX-1 (e.g. \citealt{farrell09}) is absent from our sample because its host galaxy, ESO 243-49, does not appear in the RC3 or CNG galaxy catalogues, and Ho~IX~X-1 (e.g. \citealt{laparola01}) does not appear because no position angle data is available for its host galaxy in these catalogues, meaning that our matching algorithm only matched within the minor axis of the galaxy, which the source lies outside. Occurrences such as these mean that other ULXs could have been missed due to incompleteness in our galaxy data.

\section{Example catalogue applications: analysis \& discussion}
\label{sec:results}

A large, clean sample of ULX candidates provides an ideal resource for studying the bulk properties of the ULX population and identifying trends, as well as for picking out unusual objects that warrant further investigation. The {\it XMM-Newton} source catalogue that forms the basis of our sample contains a large number of science fields populated through its analysis pipeline, including count rates and fluxes in five different energy bands (and combinations thereof) and hardness ratios between these bands. These properties are useful for preliminary characterisation of the ULX population using only products from the {\it XMM-Newton} pipeline without further time-intensive analysis. In this section, we present an overview of the hardness ratios, luminosity and variability of the ULXs in our sample, and investigate whether these properties show any dependence upon the nature of the host galaxies of these sources.

\begin{figure}
\begin{center}
\vspace{-1mm}
\includegraphics[width=75mm,trim={0 0 0 1.2cm},clip=true]{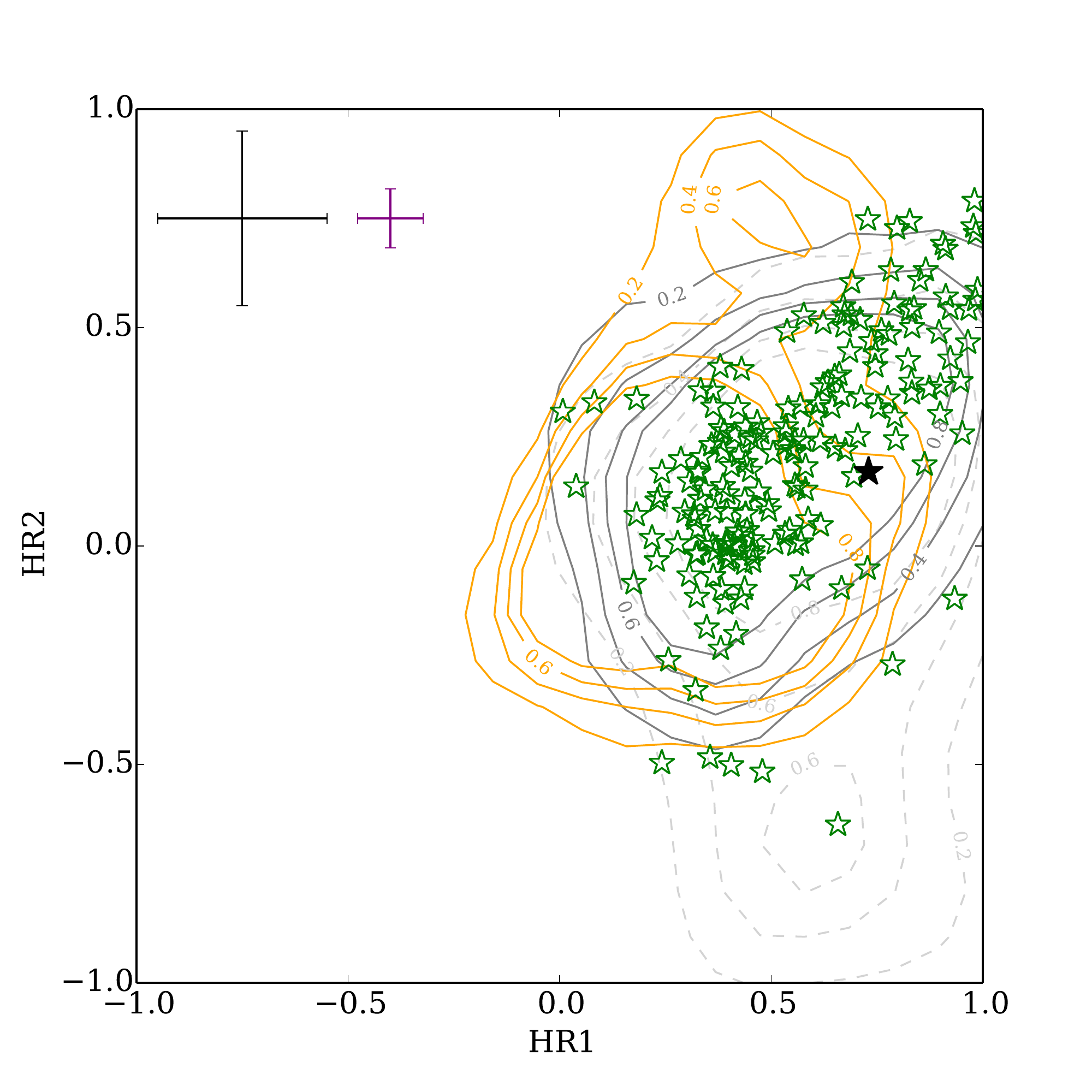}\\
\vspace{-3mm}
\includegraphics[width=75mm,trim={0 0 0 1.2cm},clip=true]{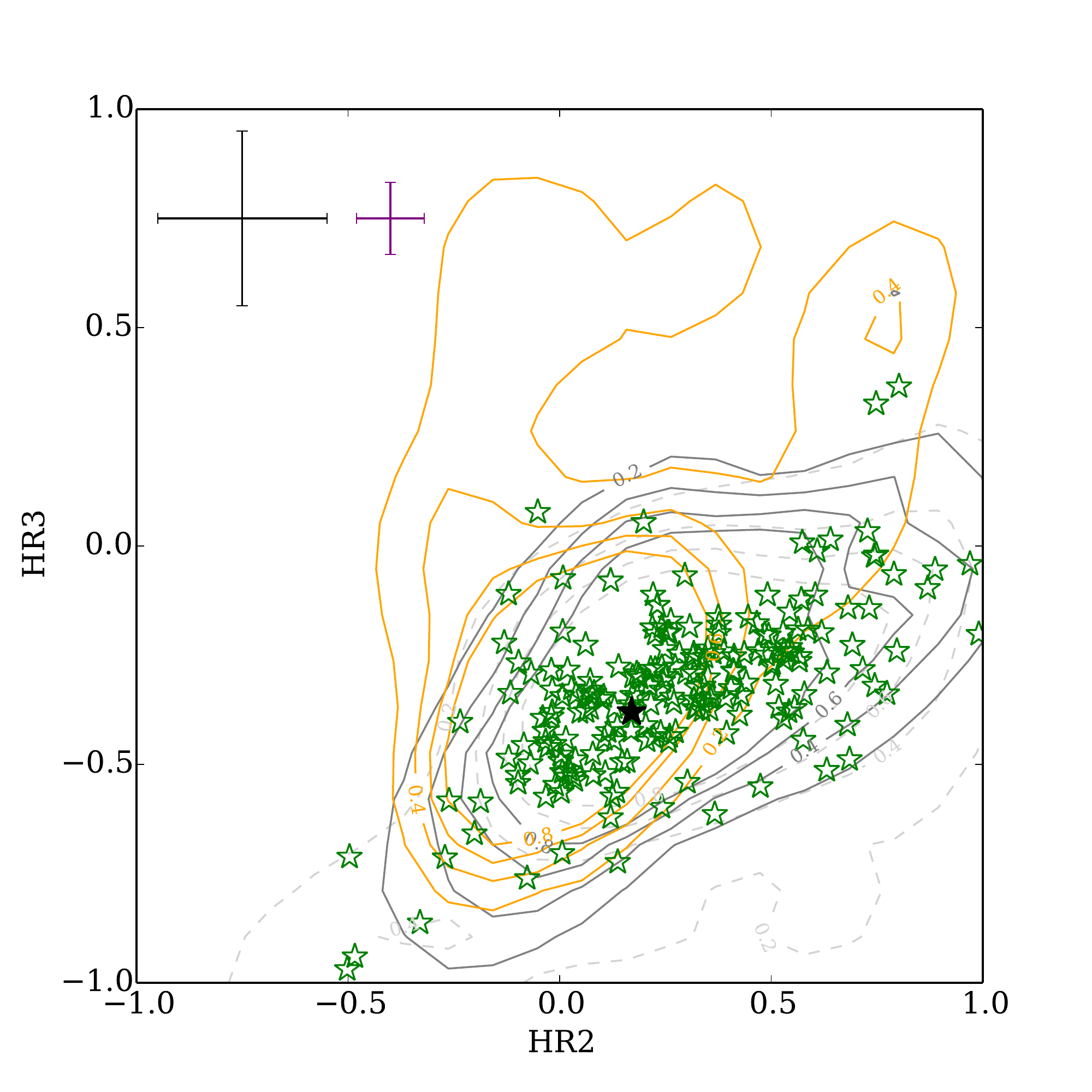}\\
\vspace{-3mm}
\includegraphics[width=75mm,trim={0 0 0 1.2cm},clip=true]{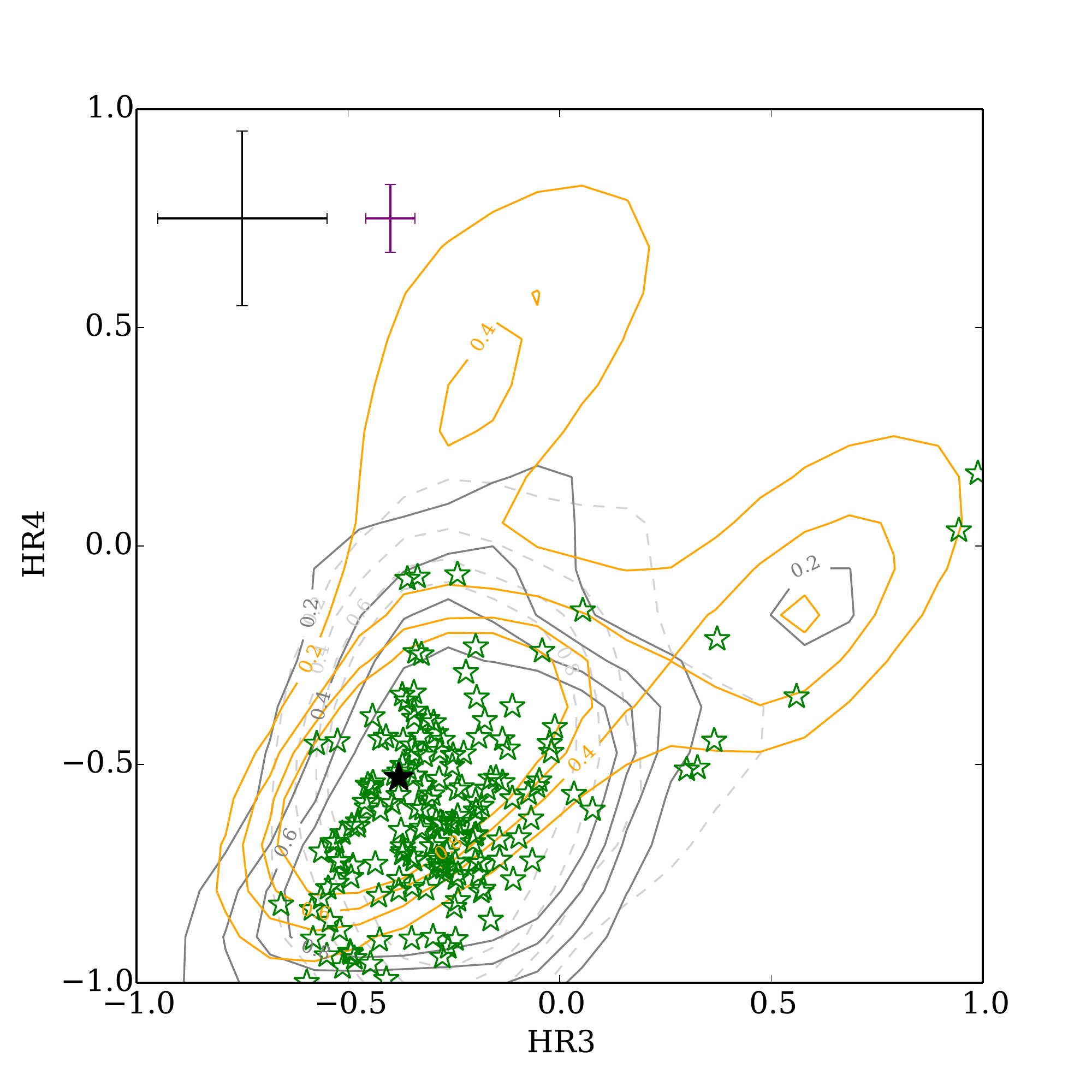}
\end{center}
\vspace{-6mm}
\caption{Hardness ratio plots for our complete ULX sample (green stars) compared with the distribution of extragalactic X-ray sources with $10^{38} \leq L_{\rm X} < 10^{39}$\,erg\,s$^{-1}$ complete to $10^{38}$\,erg\,s$^{-1}$ (grey contours) and AGNs (yellow contours). The distribution of lower luminosity sources ($L_{\rm X} < 10^{38}$\,erg\,s$^{-1}$) is shown in dashed light grey contours, although we note that this is not a complete sample. All sets of contours represent probability density values of 0.2, 0.4, 0.6 and 0.8 of the source distribution. The position of the typical ULX spectrum used in Sections~\ref{sec:comp} and~\ref{sec:cont2} is marked by a filled black star. Average $1\sigma$ error bars are shown in purple, with the maximum errors ($\pm0.2$) shown in black.} 
\label{fig:hrall}
\end{figure}

\subsection{Properties of the complete ULX subsample}
\label{sec:hr}

In this section we examine the hardness ratio (HR; i.e. X-ray colour index) and luminosity distributions of the complete subsample of ULXs (see Section~\ref{sec:comp}), for which we can be confident that we have detected all ULXs within their host galaxies and not simply the brightest from each galaxy, and thus have a representative sample of the ULX population.

The {\it XMM-Newton} pipeline records source count rates in five different energy bands: 0.2--0.5, 0.5--1.0, 1.0--2.0, 2.0--4.5 and 4.5--12.0\,keV. From these five bands, four HRs are defined by:

\vspace{-2mm}
\begin{equation}
{\rm HRn} = \frac{{\rm RATE}\_b - {\rm RATE}\_a}{{\rm RATE}\_b + {\rm RATE}\_a},
\end{equation}

where $a$ and $b$ are adjacent energy bands, with $a$ the lower-energy band. Where the count rate of one of the bands is zero, the HR takes an upper or lower limit of 1 or $-1$ respectively, therefore we only consider HRs where $-1 < {\rm HR} < 1$ in our analysis (between 2 and 7 per cent of detections are excluded on the basis of extreme HR values, depending on which HRs are under consideration). We also exclude those HRs with error $>0.2$ (that is, the $1\sigma$ errors as calculated in the 3XMM pipeline and provided in the {\it XMM-Newton} Serendipitous Source Catalogue, although these may be underestimated at low count rates; \citealt{park06}), as it causes their position on HR plots to be very unconstrained. This excludes 20--50 per cent of detections, again depending on the HRs in question. This excludes some sources from the plots where one of the component HRs has been excluded for one of these reasons, therefore the exact sample of sources shown varies between the three plots.

We first compare the population of ULXs as a whole with three other populations: the sources with $10^{38} \leq L_{\rm X} < 10^{39}$\,erg\,s$^{-1}$ in the sample complete to $10^{38}$\,erg\,s$^{-1}$, all sources with luminosities lower than $10^{38}$\,erg\,s$^{-1}$, and the bright AGNs first filtered out in Section~\ref{sec:cont1}. We do this in a set of HR-HR plots to determine how the overall spectral shape of ULXs compares with other sources (Fig.~\ref{fig:hrall}). On the whole, the ULXs have a very similar HR-HR distribution to lower-luminosity X-ray sources on a visual inspection, matching the underlying distribution of lower-luminosity sources almost exactly except for regions of the lowest-luminosity distribution containing particularly soft sources. This further supports findings in \citet{swartz04} showing that the general X-ray properties of ULXs over the energy range of {\it XMM-Newton} are statistically indistinguishable from the lower-luminosity population and the interpretation that the majority of ULXs are the highest luminosity objects within the stellar-mass BH population of their host galaxies. We also plotted where the typical ULX spectrum we used to determine the complete sample (see Section~\ref{sec:comp}; \citealt{gladstone09}) falls on this plot -- while it does not fall in the centre of the general distribution of ULXs, we can be confident that it is representative of a typical ULX spectrum. 

AGNs, however, have a very obviously different distribution of HRs, tending to have low HR2 values, indicating a steep spectrum at low energies, and are more likely to have high HR3 and HR4 values, indicating a hard high-energy tail. However, ULXs cannot be distinguished from AGNs by HRs alone, with the AGN population overlapping significantly with the majority of the ULX population. (We do note, though, that the AGN sample is not complete and representative in the same way that the ULX sample is.)

While these differences in the distribution are evident on a visual inspection alone, we also performed a multivariate non-parametric Cramer test \citep{baringhaus04} on each two-HR distribution, comparing the complete ULX population both with the complete $10^{38}$--$10^{39}$\,erg\,s$^{-1}$ population and the AGN sample. In all three cases, we can reject the null hypothesis that the ULX population and the AGN sample have the same underlying distribution to greater than 3$\sigma$ confidence. Conversely, in all cases the ULX population and the $10^{38}$--$10^{39}$\,erg\,s$^{-1}$ population are consistent with sharing the same underlying distribution.

The majority of ULXs lie within a reasonably tight locus with $0.25 < {\rm HR1} < 1$, $-0.25 < {\rm HR2} < 0.5$, $-0.5 < {\rm HR3} < 0$ and $-1 < {\rm HR4} < -0.25$. The exceptions are a small number of unusually soft ULXs with low values of HR2, overlapping with the distribution of much lower luminosity sources, and a small branch of harder sources with high HR3 values, which appear to overlap better with the AGN distribution than they do with the lower-luminosity sources -- while the $10^{38} \leq L_{\rm X} < 10^{39}$\,erg\,s$^{-1}$ source distribution also shows some evidence of overlapping with this regime, the small quantity of sources in these regions is consistent with the predicted background contamination. On examination of the ULXs with ${\rm HR3} > 0.2$, a small fraction appeared to have unidentified optical or near-infrared counterparts that may suggest the presence of a background contaminant, although obtaining a redshift for these sources or searching for a counterpart for the remainder of these high-HR3 ULXs is beyond the scope of this study. If further investigation of these sources reveals them to be background AGNs, then HR3 values may be an easy method to remove some of the remaining AGN contaminants from ULX samples. 

We further investigate the distribution of ULXs in HR space by dividing the sample into two groups based upon the nature of their host galaxy. ULXs in elliptical galaxies form one group, and ULXs in spiral and irregular galaxies form a second (which we call spiral-hosted for simplicity). 

\begin{figure}
\begin{center}
	\vspace{-2mm}
	\includegraphics[width=75mm,trim={0 0 0 1.2cm},clip=true]{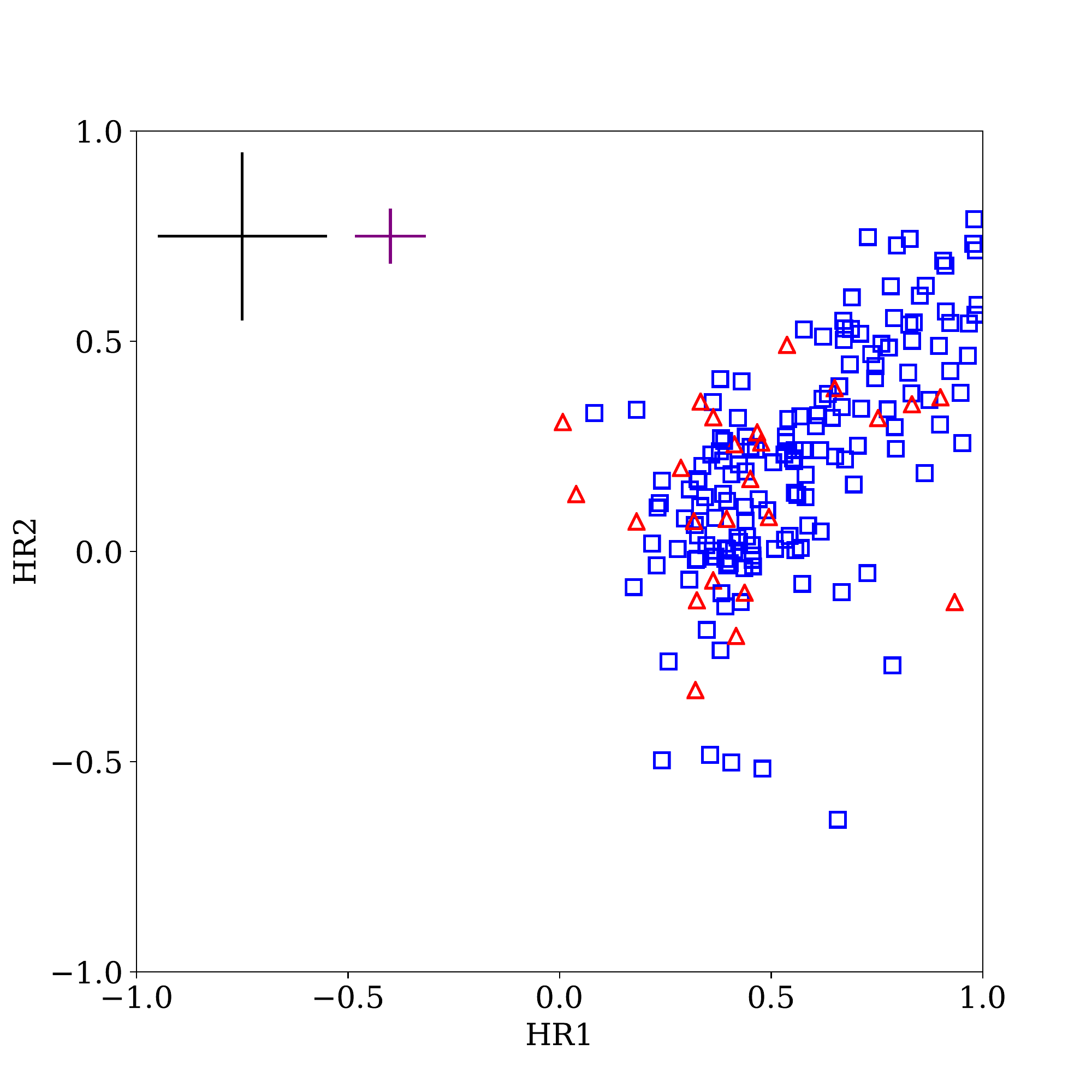}\\
	\vspace{-3mm}
	\includegraphics[width=75mm,trim={0 0 0 1.2cm},clip=true]{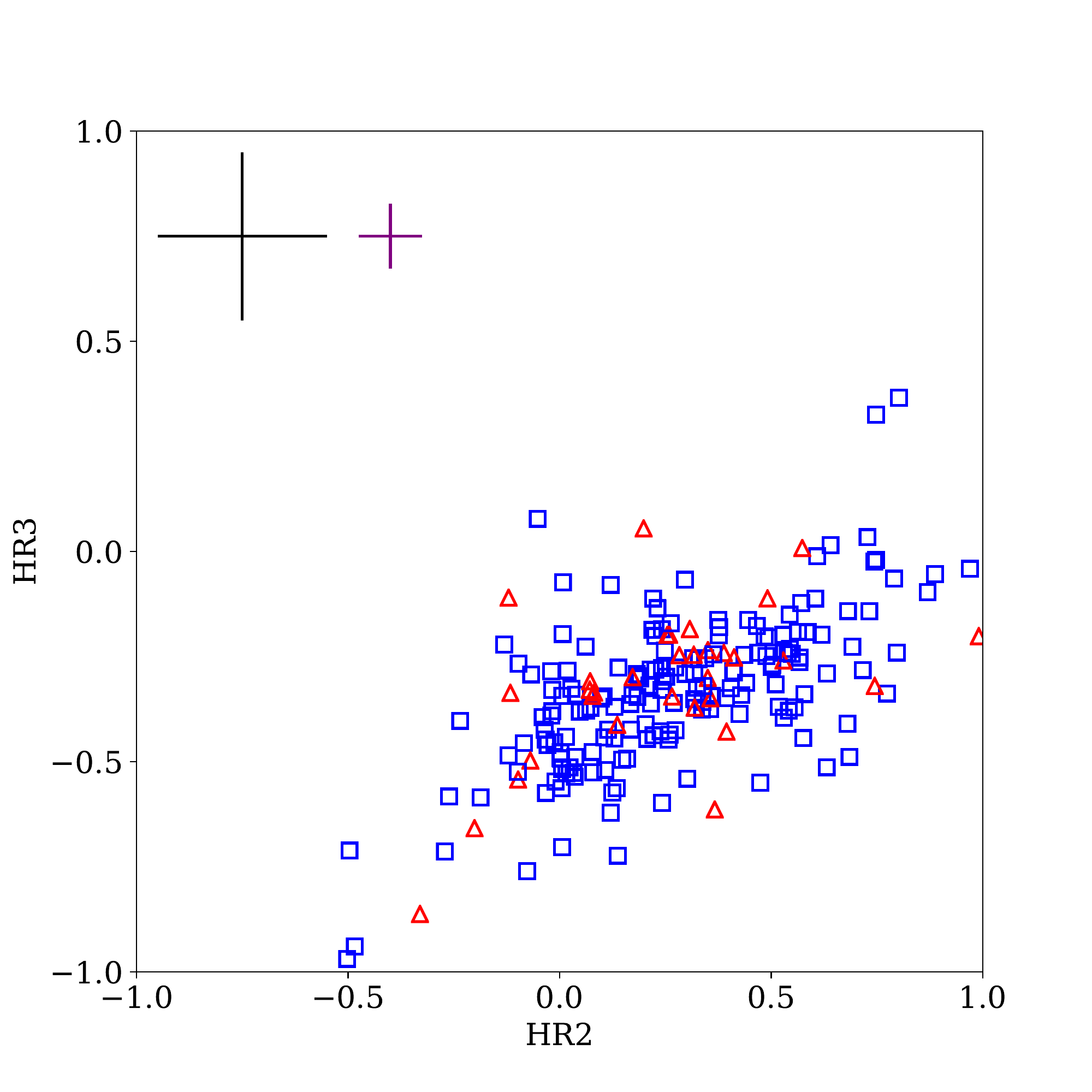}\\
	\vspace{-3mm}
	\includegraphics[width=75mm,trim={0 0 0 1.2cm},clip=true]{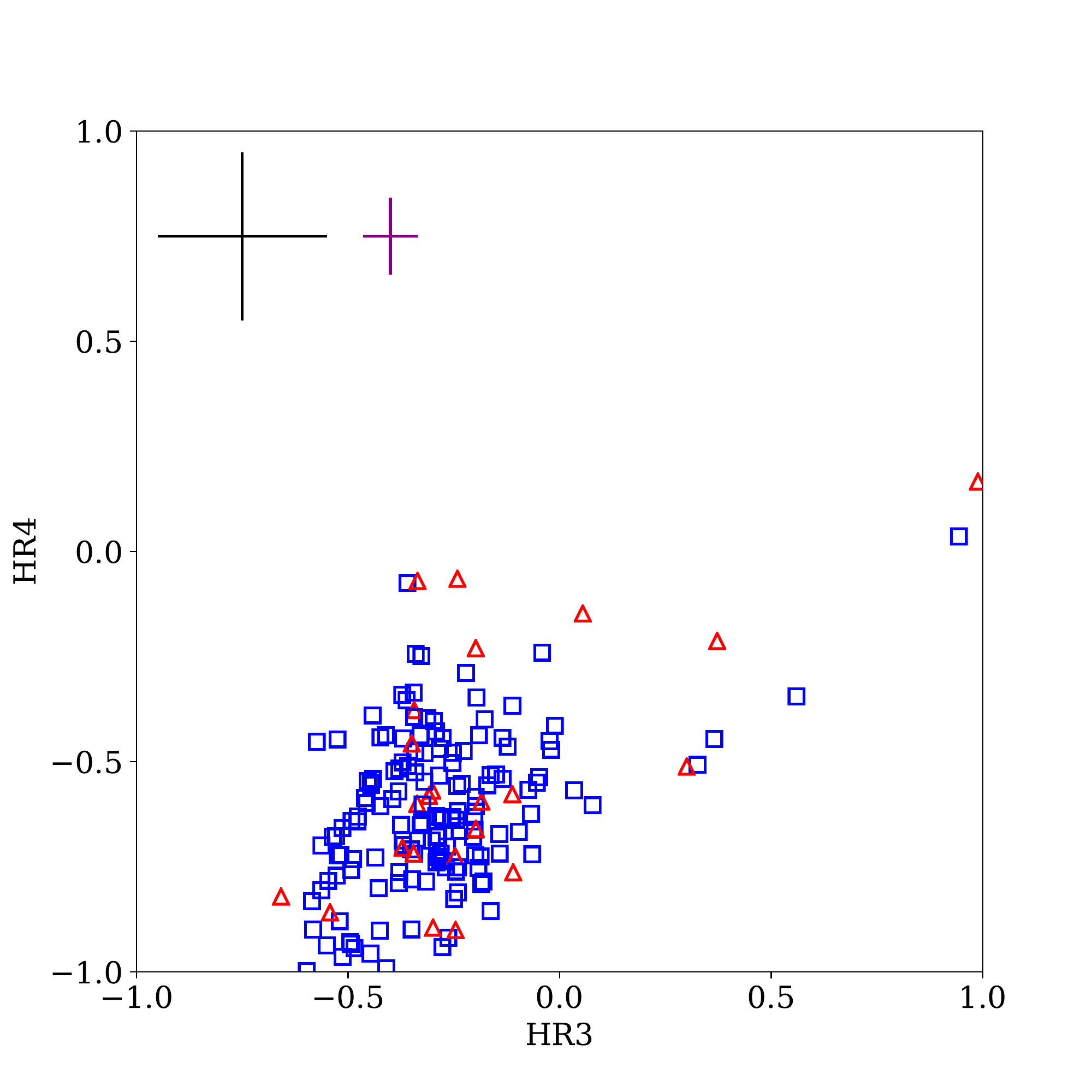}
\end{center}
\vspace{-6mm}
\caption{Hardness ratio plots for our ULX sample divided into sources located in elliptical host galaxies (red triangles) and those located in spiral or irregular galaxies (blue squares). Example error bars are as in Fig.~\ref{fig:hrall}.} 
\label{fig:hrgal}
\end{figure}

\begin{figure*}
\begin{center}
\includegraphics[width=80mm]{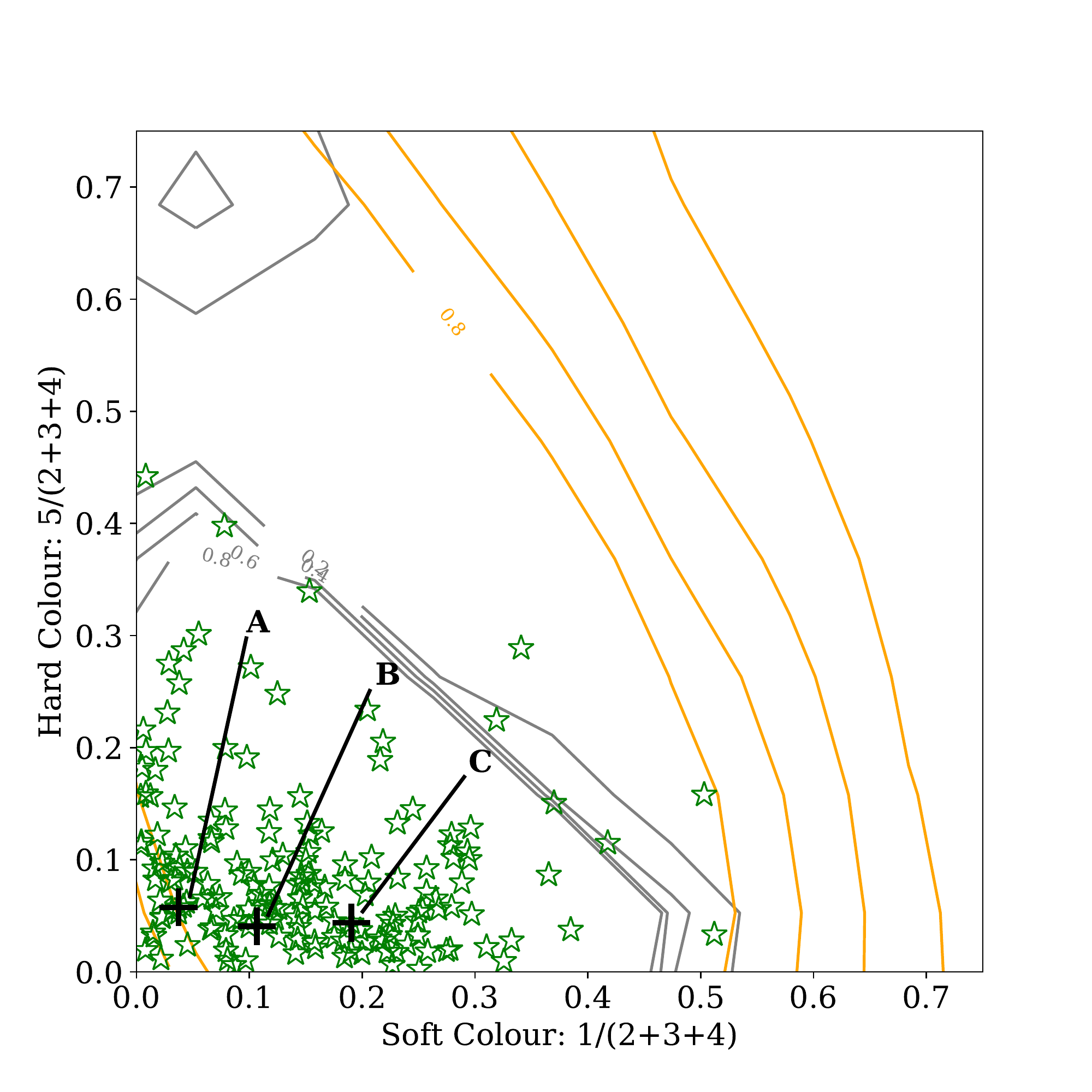}
\includegraphics[width=80mm]{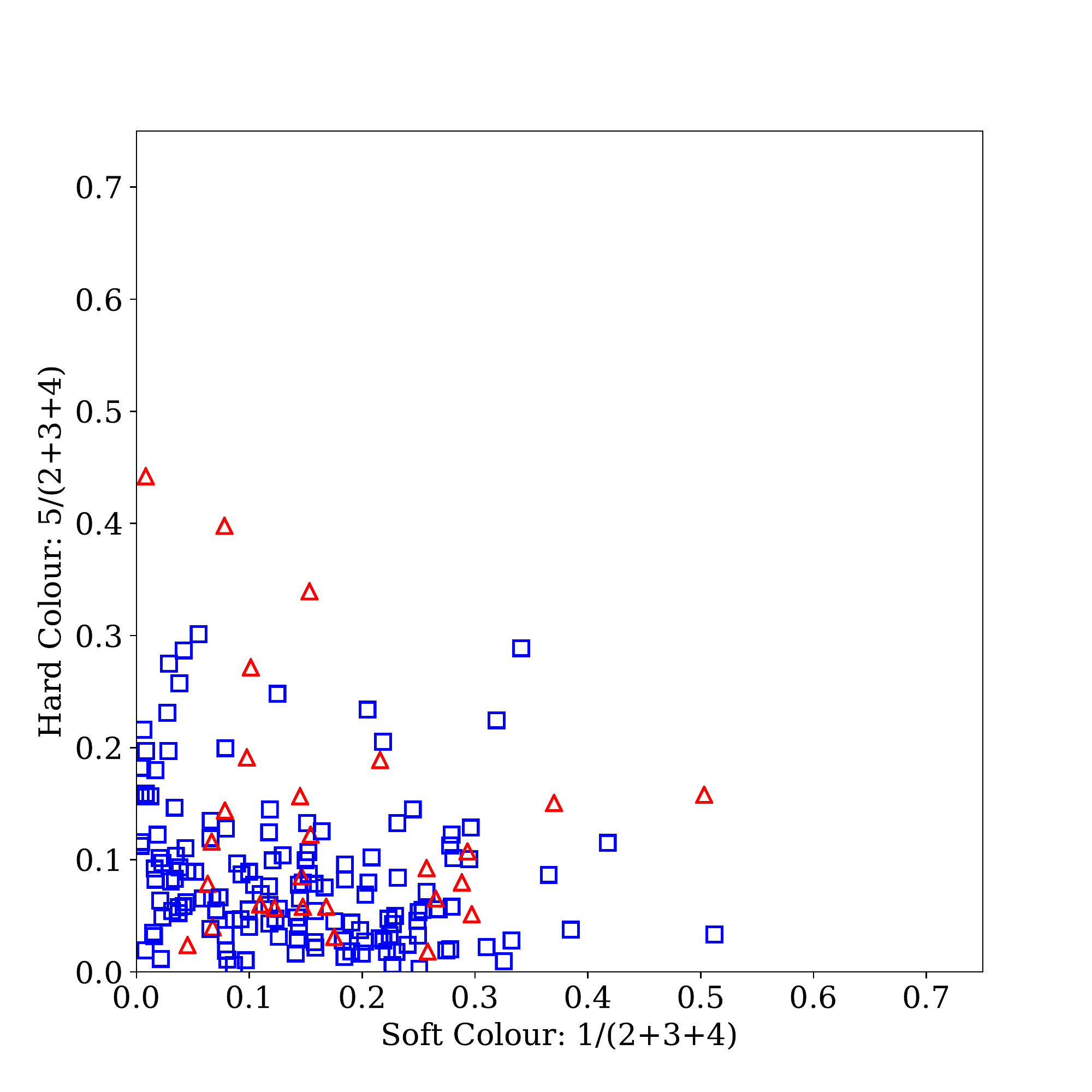}
\end{center}
\caption{Colour-colour plots based on those introduced in \citet{pintore14}, with the energy bands of {\it XMM-Newton} numbered from 1 to 5, for the complete ULX subsample compared with lower luminosity extragalactic X-ray sources and with AGN as in Fig.~\ref{fig:hrall} (left), and with the ULX sample divided into elliptical and spiral/irregular galaxies as in Fig.~\ref{fig:hrgal} (right). Three sources from \citet{sutton13b} are marked with black crosses on the left plot. A = NGC~2403~X-1 (detection classified as a broadened disc in \citealt{sutton13b}), B = NGC~6946~X-1 (detection classified as hard ultraluminous), C = NGC~4559~ULX2 (detection classified as soft ultraluminous).} 
\label{fig:pintore}
\end{figure*}

\begin{figure*}
	\begin{center}
		\includegraphics[width=80mm]{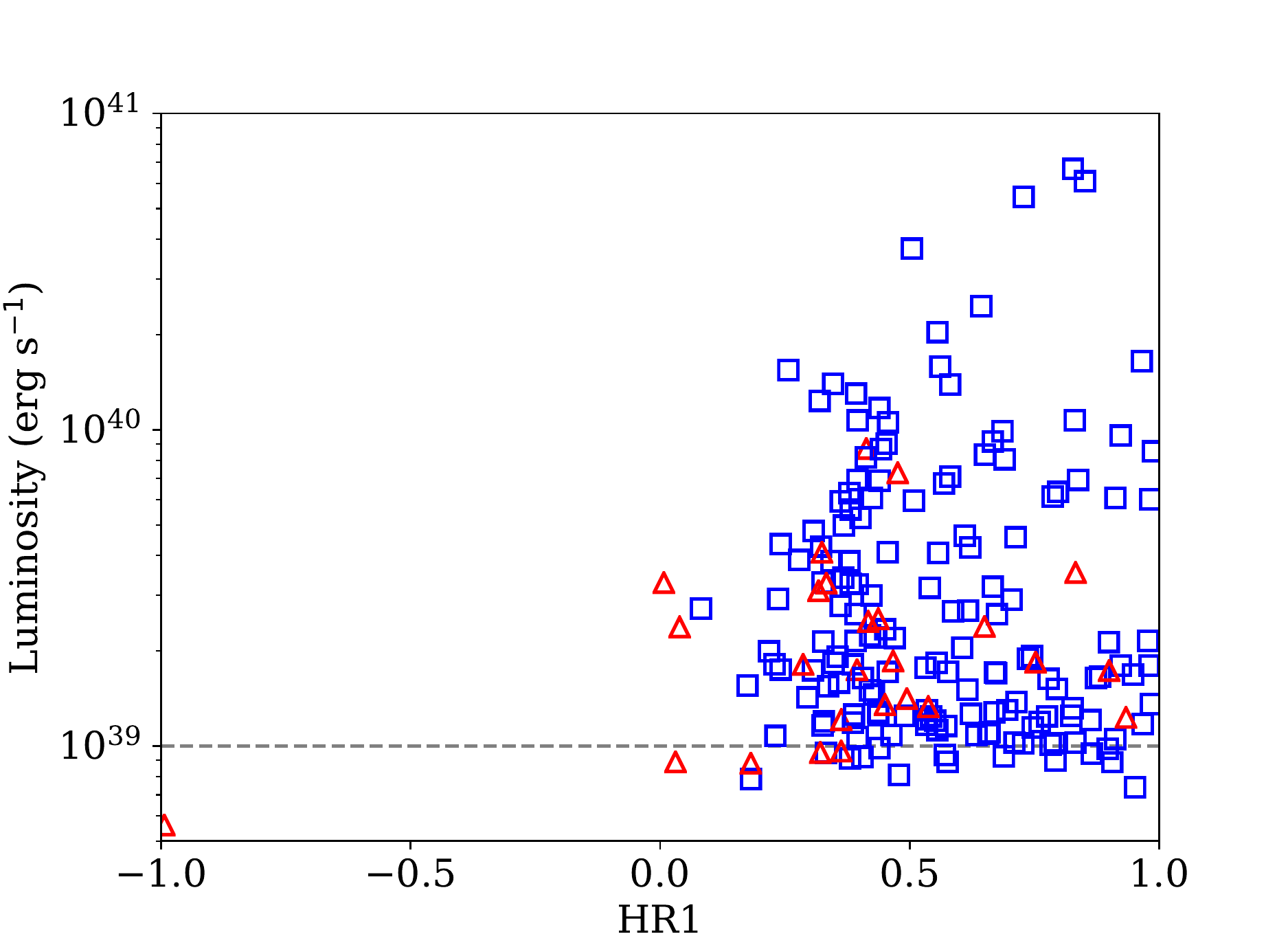}
		\includegraphics[width=80mm]{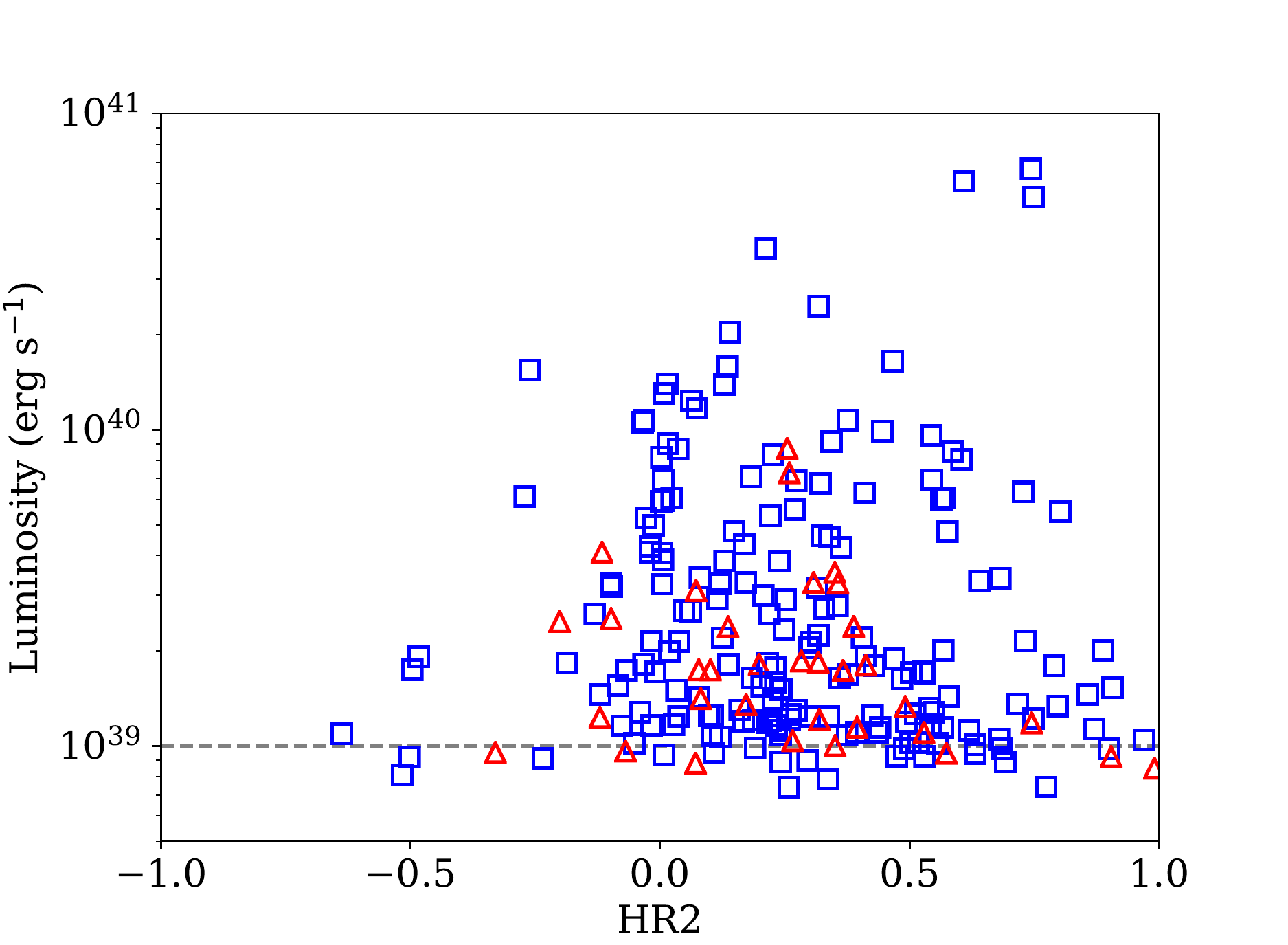}\\
		\includegraphics[width=80mm]{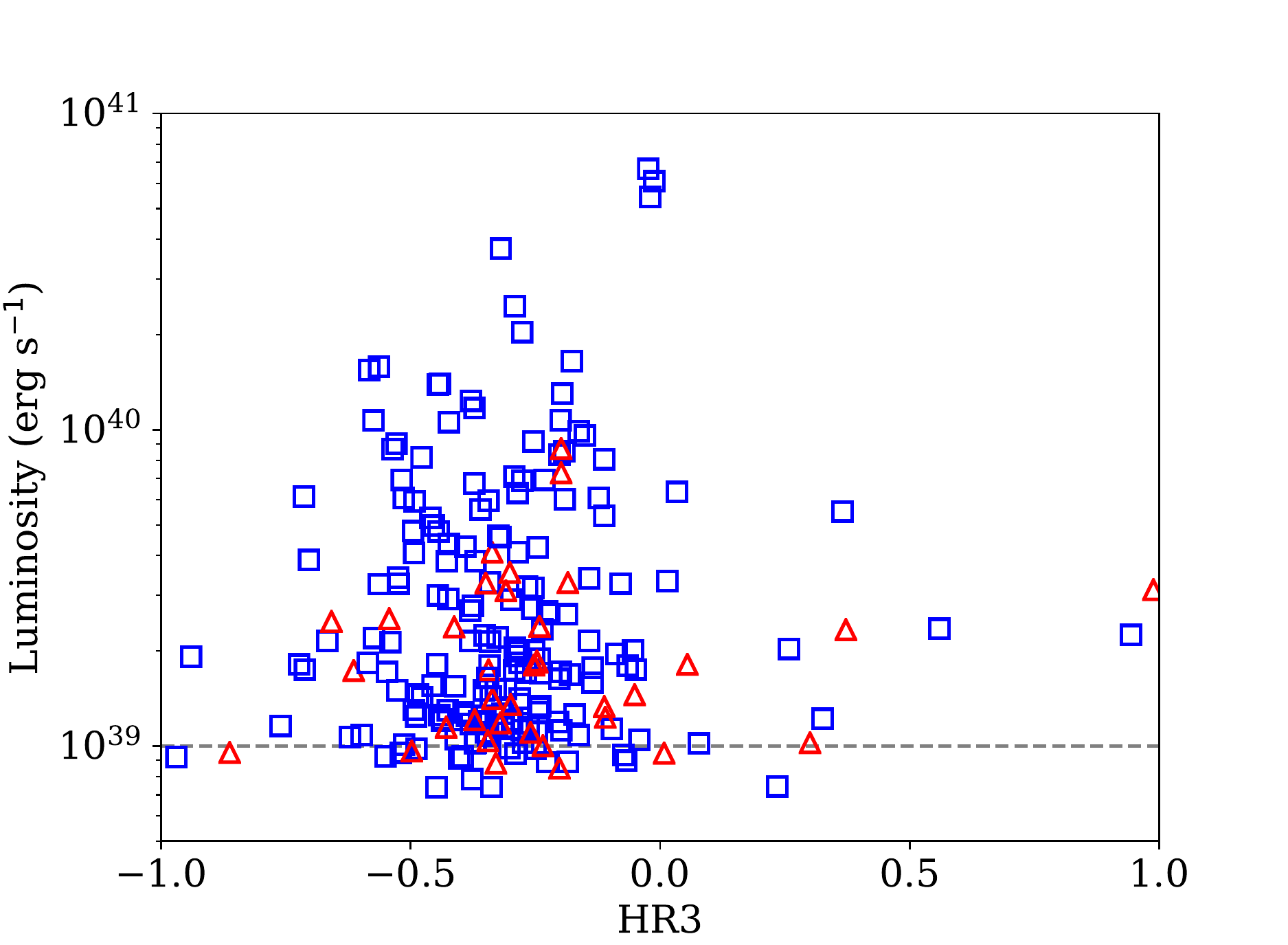}
		\includegraphics[width=80mm]{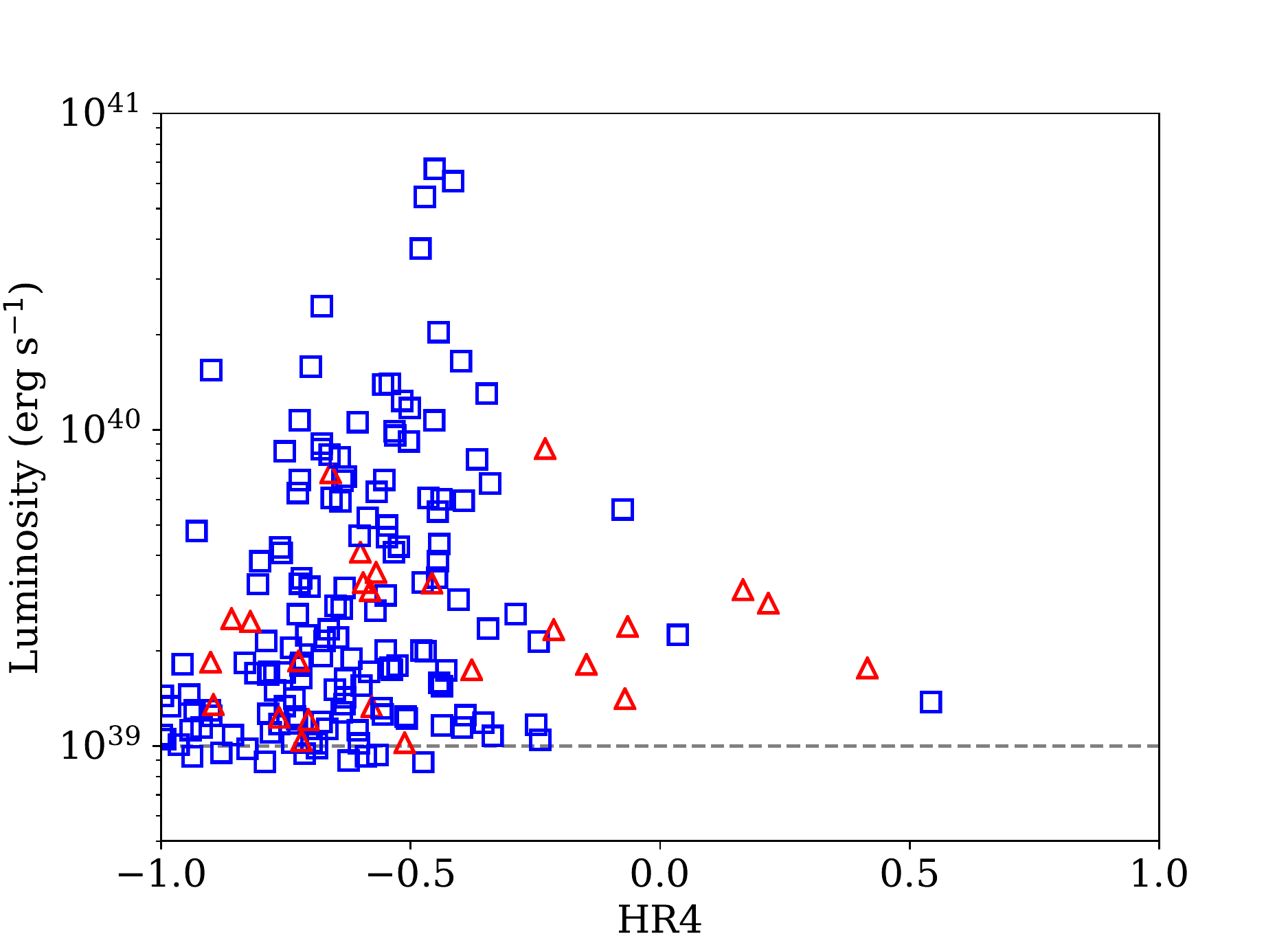}
	\end{center}
	\caption{The hardness ratios plotted against luminosity for our ULX sample divided into elliptical and spiral/irregular galaxies as in Fig.~\ref{fig:hrgal}.}
	\label{fig:hrlum}
\end{figure*}

When we divide the ULX sample into elliptical- and spiral-hosted sources in this way (Fig.~\ref{fig:hrgal}) and examine the HR1/HR2 plot, we find a high-HR2 hard tail made up exclusively of ULXs located in spiral galaxies. The plot requirement is for both HR1 and HR2 to have errors $<0.2$ -- if we only consider HR2 errors when examining this hard tail, we find that there are also five ULXs in elliptical galaxies in this region (see the HR2/HR3 plot and also Fig.~\ref{fig:hrlum}). This sub-population is still dominated by spiral-hosted sources, but since the number of ULXs in elliptical galaxies is smaller than the number of ULXs in other environments, we need to test whether this is simply due to having a smaller sample size. Therefore we randomly selected unique ULXs equal to the number of elliptical-hosted ULXs from the spiral ULX population and checked to see how many from that selection had ${\rm HR2} > 0.5$, and whether that number was consistently greater than five. We repeated this 10,000 times and found that the null hypothesis probability of both populations having the same underlying distribution in HR2 was $p = 0.08$. Therefore we are only confident at the $\sim1.7\sigma$ level that this high-HR2 population is particular to the spiral galaxy environment. 


The ULXs with particularly low values of HR2 are all located in spiral galaxies, with two located within a spiral arm and two on the outskirts of their host galaxy, and all have a reasonably low luminosity ($\sim10^{39}$\,erg\,s$^{-1}$). However, given the small number of sources, there is no particular indication that these sources are particular to spiral galaxies. On the other hand, the high-HR3 tail of ULXs lying in the AGN-dominated region of the HR diagrams is populated by sources in both spiral and elliptical galaxies, and has no particular environmental dependence.

\citet{pintore14}, henceforth P14, also examined X-ray colours to attempt to characterise the variability behaviour of ULXs between observations when fitted with a two-component model. They produced a soft X-ray vs. hard X-ray two-colour plot that appeared to indicate that the ULX population lay in a sequence from a high hard colour and low soft colour to a low hard colour and high soft colour, and could be divided loosely into two groups. The first, hard group was made up of less luminous and more absorbed sources, with the normalisation of the soft component and the parameters of the hard component driving the spectral changes. The second contained more luminous and less absorbed sources, with variations in $N_{\rm H}$ and the normalisation of the soft component being more significant than the hard component in driving spectral changes. The first group contained sources in the broadened disc and hard ultraluminous accretion regimes as classified by \citet{sutton13b}, and the second contained sources in the soft ultraluminous regime -- it was suggested that X-ray colour, and the soft colour in particular, could be used to distinguish between the two groups in cases of lower-quality data.

In order to test whether this is indeed the case, we make an approximation of the P14 colour-colour plots with our own ULX sample (Fig.~\ref{fig:pintore}), although the energy bands directly available to us from the 3XMM-DR4 catalogue are not identical to those used in P14, therefore the plots are not precisely comparable. The high-hard/low-soft to high-soft/low-hard ULX sequence seen in P14 is still present in our plots in a rough sense, although there is a greater amount of scatter and a few objects that have similarly high values for both colours. The far left of the plot is mostly populated by the most highly-absorbed ULXs in spirals, as discussed earlier in this section, and bears similarity with the positioning of IC~342~X-1 in P14. However, the overall distribution cannot as easily be divided into two based on the soft colour as in P14. The general trend of where the various super-Eddington accretion regimes appear on the plot may well hold -- example sources from \citet{sutton13b} marked in Fig.~\ref{fig:pintore} fall roughly into the expected regions, with a soft ultraluminous source having a higher soft colour than a hard ultraluminous and a broadened disc source -- however these different accretion regimes form a spectrum across colour space rather than falling into easily divisible groups. There is also no discernible difference between the spiral and elliptical populations.

We can investigate the host galaxy dependence of ULX properties from a different angle by plotting the HRs against the detection luminosity for the elliptical and spiral populations (Fig.~\ref{fig:hrlum}). The majority of ULX detections have a luminosity below $10^{40}$\,erg\,s$^{-1}$. The harder subset found in spiral galaxies have a similar luminosity distribution to the rest of the ULXs, however there is a different subset of ULXs with luminosities greater than $10^{40}$\,erg\,s$^{-1}$ which are also only seen in spiral galaxies. We perform a similar statistical test to that performed on the high-HR2 population to determine whether it is significant that we only see sources of very high luminosity in spiral galaxies, and found a null-hypothesis probability of $p=0.006$. Therefore we are confident to $\sim2.8\sigma$ that the elliptical and spiral ULX populations have different luminosity distributions. While not enormously statistically significant, this result is consistent with indications in W11 that the luminosity function for ULXs sources in elliptical galaxies is steeper than that for spiral galaxies, which is also more firmly established as the case for lower luminosity sources \citep{grimm03,gilfanov04}.

\begin{figure}
\begin{center}
\includegraphics[width=80mm]{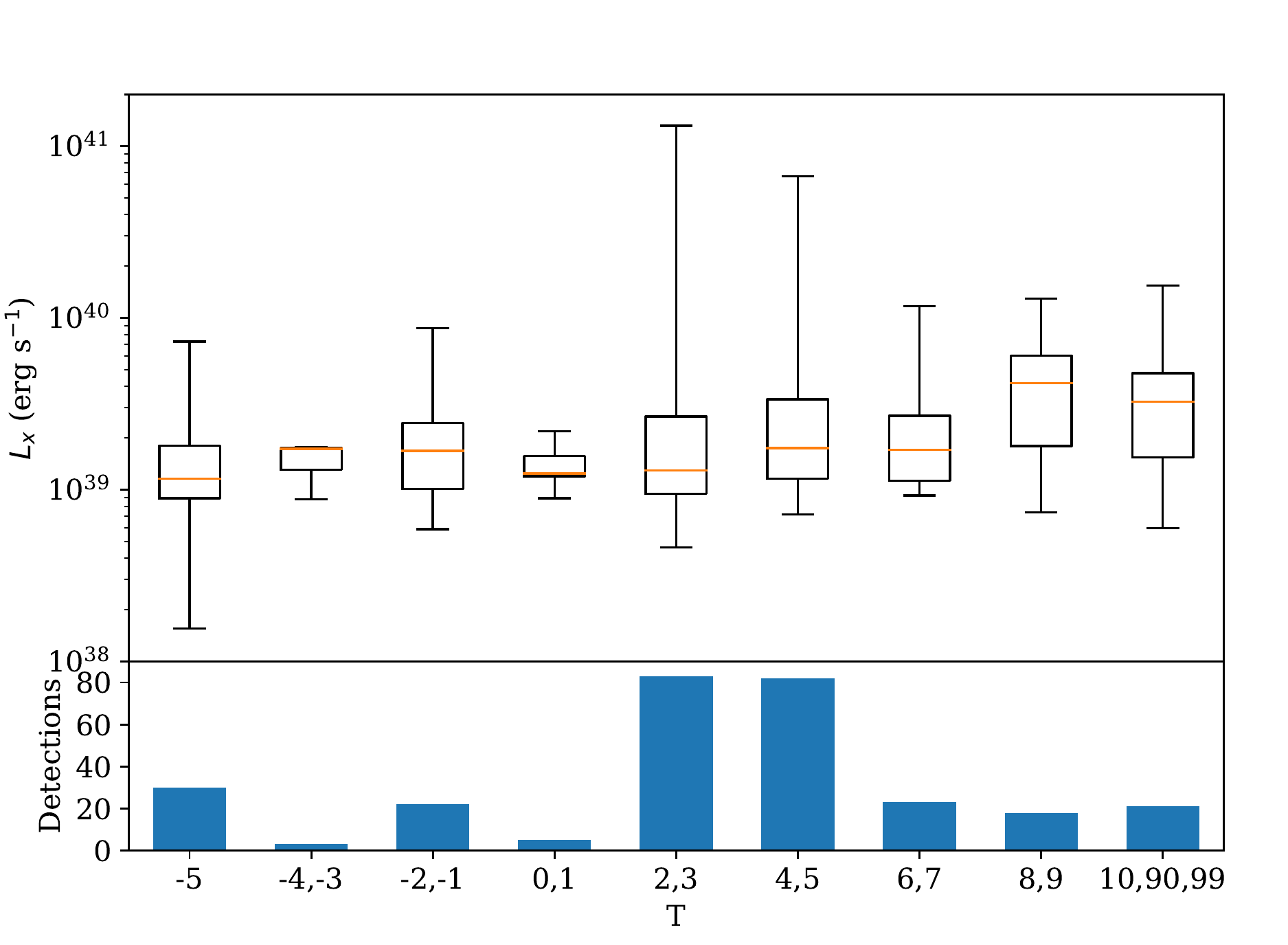}
\vspace{-2mm}
\includegraphics[width=80mm]{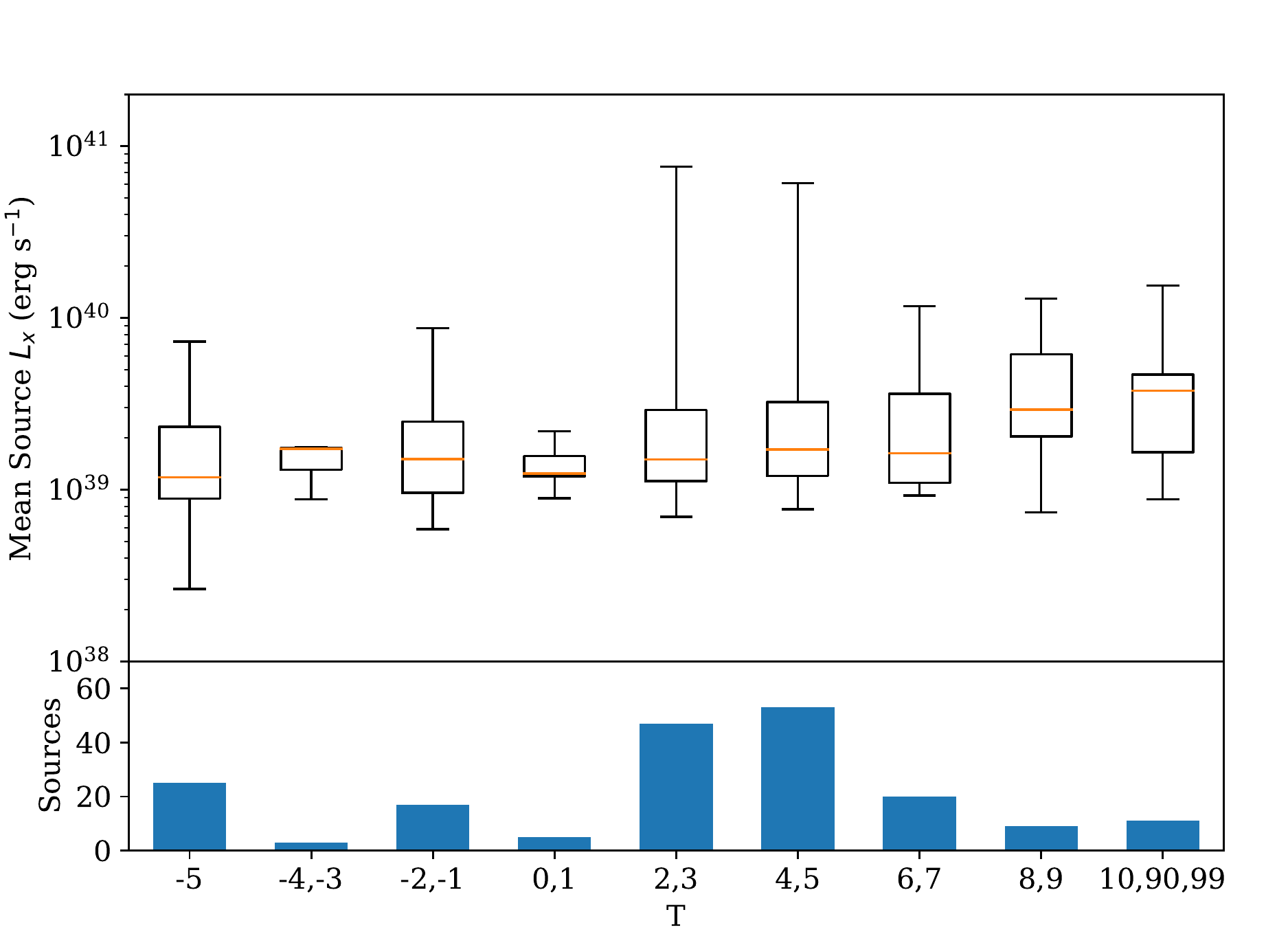}
\end{center}
\caption{Box plots of ULX luminosity for host galaxies with Hubble Stage $T$, for every detection of a source classed as a ULX ({\it top}) and for the average luminosity each individual ULX ({\it bottom}). Boxes span from the first to the third quartile, with the median marked in red and whiskers extending to the maximum and minimum values. A bar plot beneath each box plot shows the number of detections or sources in each $T$ bin.} 
\label{fig:lumbox}
\end{figure}

Previous population studies of ULXs have indicated that dwarf galaxies, which are less evolved and are metal-poor, tend produce more ULXs per unit mass than larger, higher-metallicity galaxies (e.g. \citealt{swartz08}). This is mainly due to higher amounts of star formation per unit mass than metallicity itself \citep{mapelli11,prestwich13}, although high-mass, metal-poor stars are able to leave more massive compact objects at the end of their lifespan \citep{heger03} which may lend themselves towards producing brighter ULXs. Therefore we expect to see a greater number and more luminous ULXs in star-forming (i.e. spiral) galaxies, but also possibly more higher-luminosity ULXs in the lowest-mass spirals and irregular galaxies ($T\geq8$).

To investigate this, we further divided the ULX groups into bins of their Hubble stage $T$, for which we produce a box plot of the luminosity (Fig.~\ref{fig:lumbox}), both for every detection of sources identified as ULXs and for the average luminosity of each individual source. Plotted in this way, we can see that although the most luminous ULXs occur in the moderately-wound spirals with $2\leq T\leq5$, these types of galaxy also have the most representation in our sample by both sources and detections, and the median luminosities for the more tightly-wound spirals remain fairly low. There is very tentative evidence to suggest that ULXs in the most diffuse, Magellanic spirals ($T=8,9$) and irregular galaxies ($T=10$--99) are brighter on average than those in tighter spirals, although the sample size is too small to draw any definite conclusions, and the luminosity distributions of ULXs between different spiral classifications are not found to be significantly different using a KS-test. 

\subsection{The most luminous ULXs}
\label{sec:bright}

One motivation for constructing samples of ULXs is the search for hyper-luminous X-ray sources (HLXs; $L_{\rm X}>10^{41}$\,erg\,s$^{-1}$), which are difficult to explain even with super-Eddington accretion onto a moderately-sized BH -- although one source at this luminosity, NGC~5907~ULX-1 \citep{israel17a}, has been discovered to be a highly super-Eddington NS -- and thus may represent the best candidates for IMBHs. The {\it XMM-Newton} Serendipitous Source Catalogue has previously been used in dedicated searches for HLXs (e.g. \citealt{sutton12,zolotukhin16}).

While there are no HLXs in our complete subsample, our wider sample contains 27 sources with at least one detection above $5\times10^{40}$\,erg\,s$^{-1}$. We investigated these sources on an individual basis to discover whether any viable new candidates for IMBHs had been identified. We list the sources in Tables~\ref{tab:bright} and~\ref{tab:best}, separated into sources which are either already well-studied or poor candidates for IMBHs, and the sources which are the most likely to be genuine IMBH candidates.

\begin{table*}
	\begin{minipage}{160mm}
		\caption{The 18 high-luminosity sources within the full (non-complete) version of our sample that have already been discovered or are otherwise poor IMBH candidates.} \label{tab:bright}
		\vspace{-4mm}
		\begin{center}
			\begin{tabular}{@{~}l@{~~}l@{~~}ccc@{~}l@{~}}
				\hline
				Name & Host Galaxy & $D^a$ & $C_{\rm X,max}^b$ & $L_{\rm X,max}^c$ & Note \\ 
				& & (Mpc) & (ct\,s$^{-1}$) & ($\times10^{40}$\,erg\,s$^{-1}$) & \\
				\hline
				3XMM J003937.5$+$005110 & NGC 201 & 58.9 & $0.078\pm0.002$ & $5.6\pm0.4$ & Optical counterpart in SDSS, redshift unknown \\
				3XMM J010746.7$-$173026* & IC 1623B & 81.2 & $0.128\pm0.004$ & $41\pm2$ & AGN of IC 1623B \\
				3XMM J011425.0$-$552349* & NGC 454 & 48.6 & $0.311\pm0.004$ & $64.4\pm0.9$ & AGN of NGC 45 \\
				3XMM J011942.7$+$032422 & NGC 470 & 31.7 & $0.48\pm0.01$ & $13.1\pm0.6$ & Examined in \citet{sutton12} \\ 
				3XMM J020937.6$+$354728 & UGC 1651 & 150.8 & $0.046\pm0.007$ & $23\pm3$ & Optical counterpart in DSS, redshift unknown \\
				3XMM J024025.6$-$082429 & NGC 1042 & 18.3 & $0.388\pm0.006$ & $6\pm2$ & Examined in \citet{sutton12} \\ 
				3XMM J072648.0$+$854550 & NGC 2276 & 32.2 & $0.123\pm0.006$ & $6.1\pm0.3$ & 3 blended ULXs \citep{sutton12} \\
				3XMM J080728.0$+$391135* & NGC 2528 & 52.4 & $0.034\pm0.007$ & $8\pm2$ & Background AGN at $z=0.13$ \citep{abazajian04} \\ 
				3XMM J090434.7$+$143539* & IC 2431 & 199.3 & $0.12\pm0.01$ & $80\pm20$ & AGN(s) of IC 2431 \\ 
				3XMM J091502.2$+$294314* & NGC 2789 & 84.6 & $0.030\pm0.003$ & $5\pm1$ & Background AGN at $z=0.32$ \citep{geller14} \\ 
				3XMM J110353.9$+$405100* & ARP 148 & 138.0 & $0.048\pm0.003$ & $14\pm1$ & AGN of ARP 148 \\
				3XMM J113355.4$+$490348 & IC 708 & 126.6 & $0.020\pm0.006$ & $7\pm3$ & Optical counterpart in SDSS, redshift unknown \\
				3XMM J121856.0$+$142419 & M99 & 32.1 & $0.177\pm0.006$ & $5.8\pm0.3$ & Examined in \citet{sutton12}, as NGC 4254 ULX. \\
				3XMM J134404.2$-$271410* & IC 4320 & 90.7 & $0.117\pm0.006$ & $26\pm2$ & Background AGN at $z\sim2.8$ \citep{sutton15} \\
				3XMM J151558.6$+$561810 & NGC 5907 & 16.4 & $0.583\pm0.007$ & $6.7\pm0.1$ & Well-studied \citep{sutton13a,walton15} \\ 
				3XMM J210741.3$+$035217* & UGC 11680 & 103.6 & $0.153\pm0.008$ & $49\pm5$ & AGN of UGC 11680 \\
				3XMM J223829.4$+$351947* & UGC 12127 & 110.3 & $0.160\pm0.003$ & $22.9\pm0.6$ & AGN of UGC 12127 \\ 
				3XMM J230457.6$+$122028 & NGC 7479 & 31.7 & $0.248\pm0.007$ & $6.5\pm0.3$ & Examined in \citet{sutton12} \\
				\hline
			\end{tabular}
		\end{center}
		$^a$The host galaxy distance in Mpc.\\
		$^b$The maximum count rate across all energy bands.\\
		$^c$The maximum source luminosity.\\
		*These sources were unambiguously identified as contaminants and thus removed from the final catalogue (see Section~\ref{sec:cont1}).
	\end{minipage}
\end{table*}

\begin{table*}
	\begin{minipage}{110mm}
		\caption{The nine high-luminosity sources from our sample that are the best new IMBH candidates.} \label{tab:best}
		\vspace{-4mm}
		\begin{center}
			\begin{tabular}{llccc}
				\hline
				Name & Host Galaxy & $D$ & $C_{\rm X,max}$ & $L_{\rm X,max}$ \\ 
				& & (Mpc) & (ct\,s$^{-1}$) & ($\times10^{40}$\,erg\,s$^{-1}$) \\
				\hline
				3XMM J022748.9$+$003023 & UGC 1934 & 163.7 & $0.006\pm0.001$ & $7\pm3$ \\ 
				3XMM J104414.4$+$064541 & NGC 3356 & 82.3 & $0.006\pm0.001$ & $7\pm3$ \\ 
				3XMM J120438.4$+$014716 & NGC 4077 & 94.8 & $0.072\pm0.005$ & $10\pm2$ \\ 
				3XMM J121117.8$+$392430 & UGC 7188 & 91.3 & $0.014\pm0.005$ & $13\pm7$ \\
				3XMM J125728.4$+$273015 & NGC 4839 & 98.2 & $0.031\pm0.007$ & $30\pm10$ \\
				3XMM J132727.7$-$271932 & IC 4252 & 181.5 & $0.03\pm0.01$ & $12\pm7$ \\ 
				3XMM J161604.0$-$223726 & IC 4596 & 100.9 & $0.021\pm0.001$ & $9\pm1$ \\ 
				3XMM J223843.9$+$353223 & NGC 7345 & 124.7 & $0.028\pm0.006$ & $9\pm5$ \\
				3XMM J233843.6$-$562849 & ESO 192-IG 011 & 135.7 & $0.012\pm0.003$ & $7\pm3$ \\
				\hline
			\end{tabular}
		\end{center}
		Columns as in Table~\ref{tab:bright}.
	\end{minipage}
\end{table*}

To produce these lists, we first ran these sources past previous studies of highly luminous ULXs to establish which had already been studied and which were potential new IMBH candidates. We also examined the optical data (from SDSS and {\it PanSTARRS} where available, and DSS where not) to determine whether there were optical counterparts to these sources, which may identify further background or foreground contaminants. 

Seven of these objects, located in NGC~470, NGC~1042, NGC~2276, M99, NGC~5907, IC~4320 and NGC~7479 are already well-studied sources, examined by \citet{sutton12} as some of the brightest sources in the W11 catalogue. Notable among these sources is NGC~2276~ULX, in fact a blended source of three separate ULXs which can only be resolved using the high spatial resolution of the {\it Chandra} telescope, one of which remains an IMBH candidate \citep{mezcua15}, and NGC~5907~ULX-1, subsequently found to be a NS ULX \citep{israel17a}. Another is the ULX in IC~4320, confirmed to be a background AGN at $z\sim2.8$ by \citet{sutton15} but missed before as it is not yet classified as such in NED. HLX-1 is not identified due to its host galaxy not appearing in RC3 or CNG (see Section~\ref{sec:limit}).

Of the remaining 20 high-luminosity sources, eleven have optical counterparts when examined with DSS, SDSS or {\it PanSTARRS} \citep{chambers16} data. Six of these, in IC~1623B, NGC~454, IC~2431, ARP~148, UGC~11680 and UGC~12127, are likely AGNs of their host galaxies. The AGNs in IC~1623B, NGC~454, ARP~148 and UGC~11680 are all halves of complex merger systems, for which the NED position of the galaxy is offset from the position of the AGN. IC~2431 is a system of four interacting galaxies, at such a distance that the PSF of the X-ray emission encompasses most of the optical extent of the system, therefore it is likely formed of contributions from the AGNs of several if not all of these galaxies. For UGC~12127, the NED position of the galaxy is slightly offset from the geometric centre of the optical data, potentially due to the presence of bright regions within the galaxy ellipse in addition to the galaxy core. 

For two of the other five sources with optical counterparts, in NGC~2528 and NGC~2789, the optical counterparts have redshifts associated with them, confirming them to be background AGNs. Therefore we removed these sources from the sample altogether, as well as the others above confirmed to be AGNs. For the three with optical counterparts but no recorded redshifts (in NGC~201, UGC~1651 and IC~708), we retained them in the overall X-ray source catalogue as it is not within the scope of this study to confirm them as background sources one way or another, however we note that, especially in the cases of NGC~201 and UGC~1651, the counterparts look like background galaxies upon visual inspection and so these sources are unlikely to be realistic IMBH candidates. 

\begin{figure*}
	\begin{center}
		\includegraphics[height=52mm]{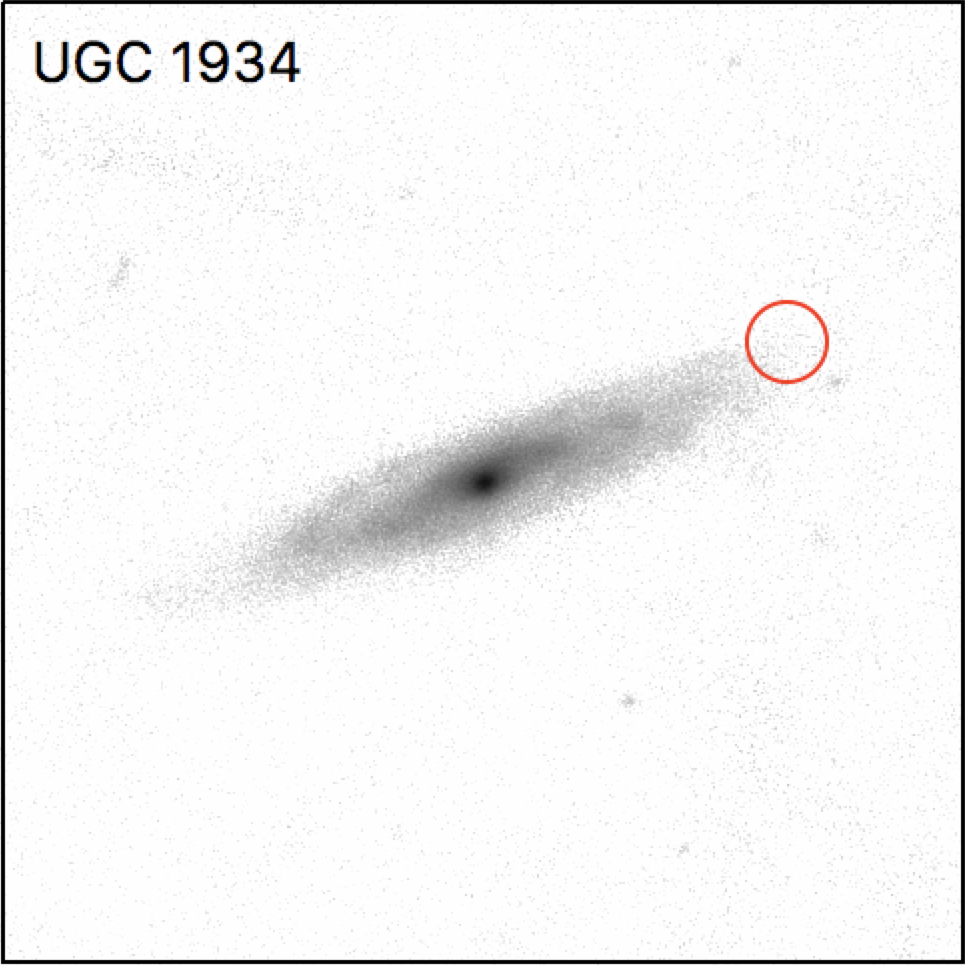}\hspace{1mm}
		\includegraphics[height=52mm]{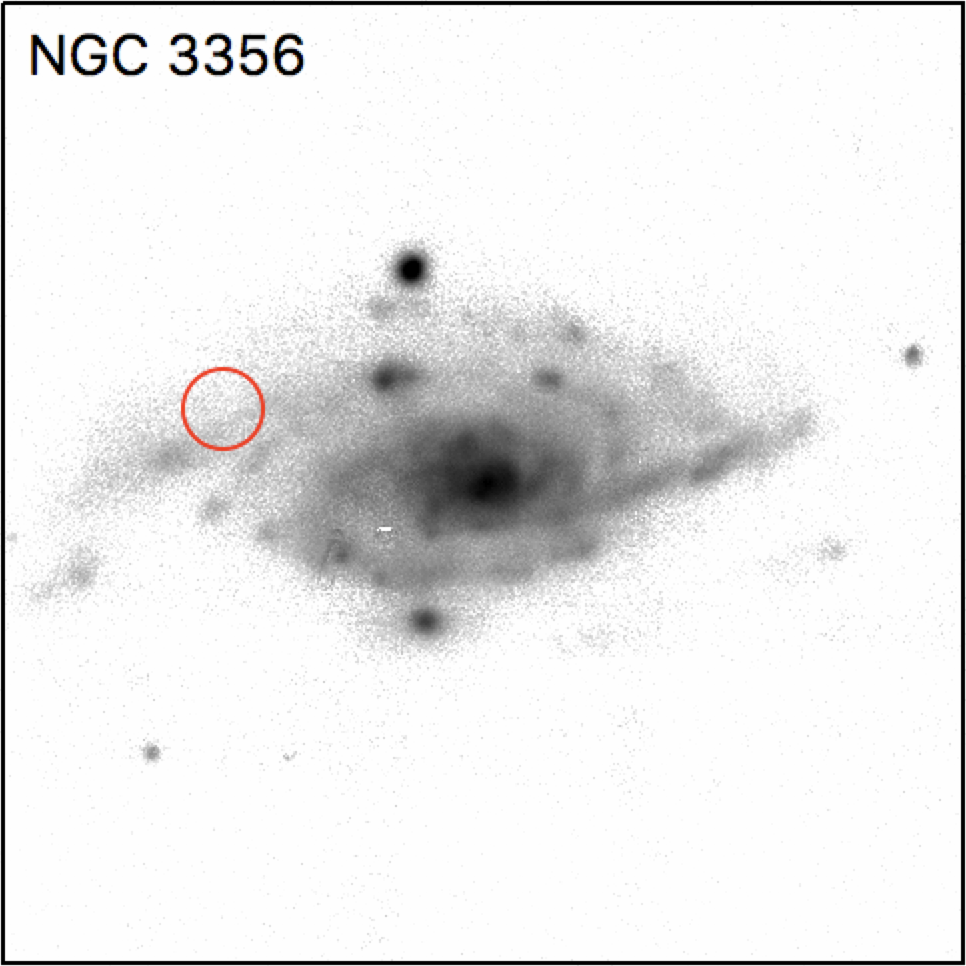}\hspace{1mm}
		\includegraphics[height=52mm]{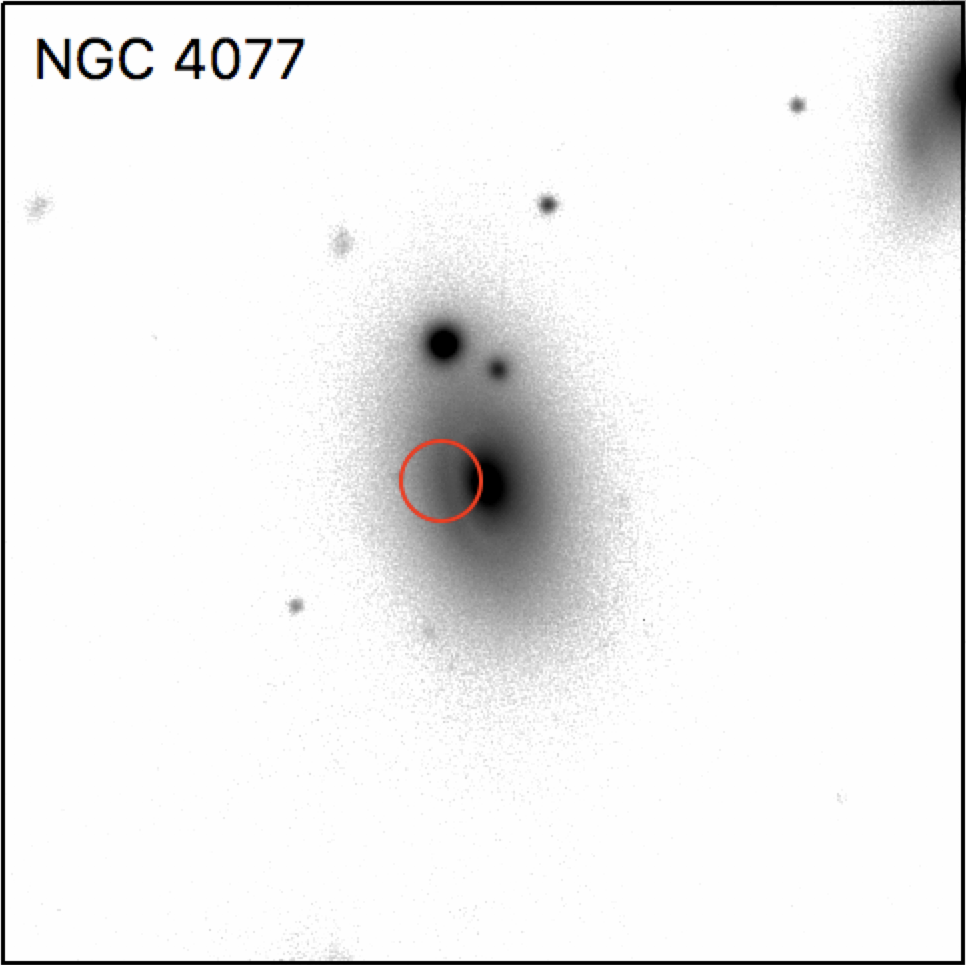}\\
		\vspace{1mm}
		\includegraphics[height=52mm]{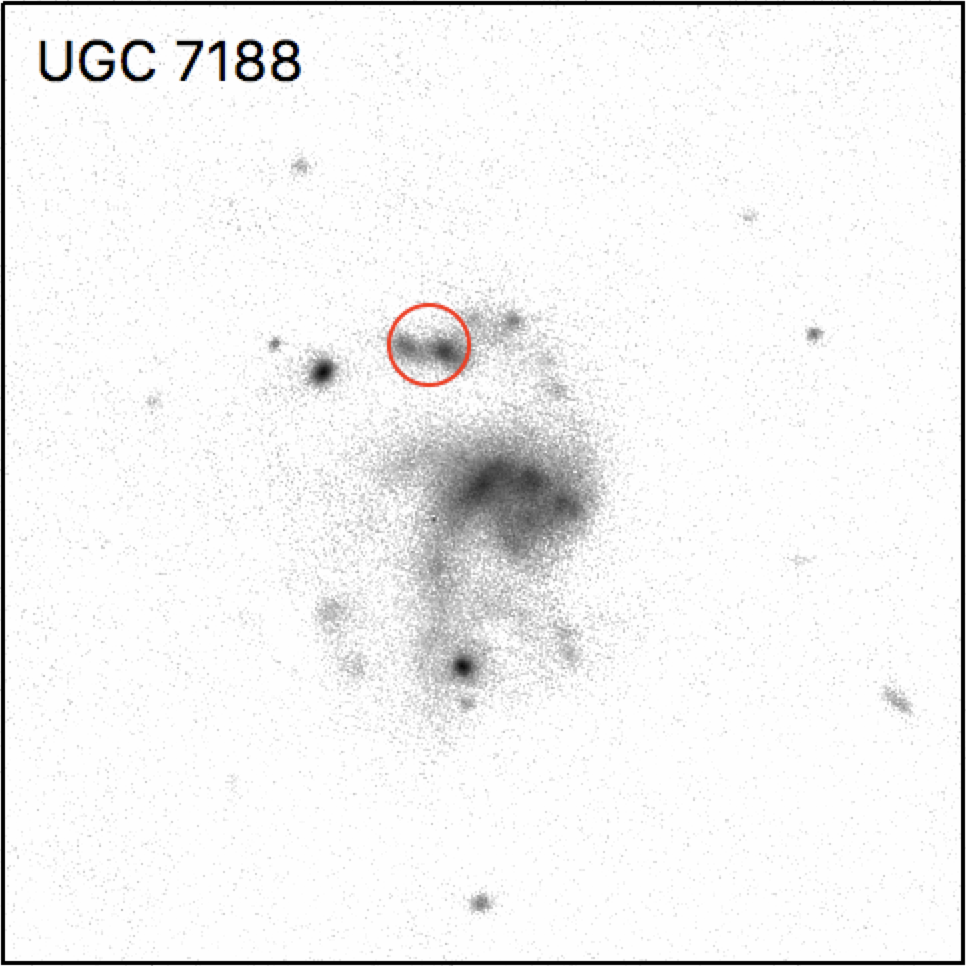}\hspace{1mm}
		\includegraphics[height=52mm]{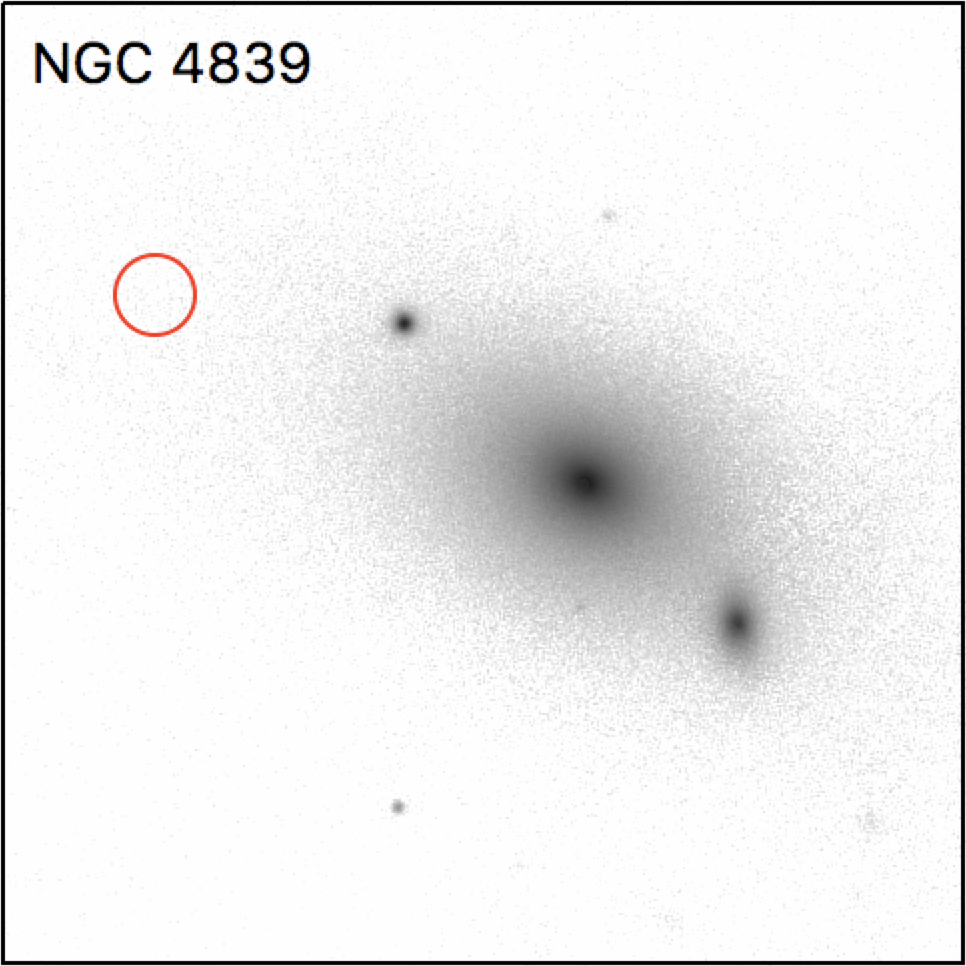}\hspace{1mm}
		\includegraphics[height=52mm]{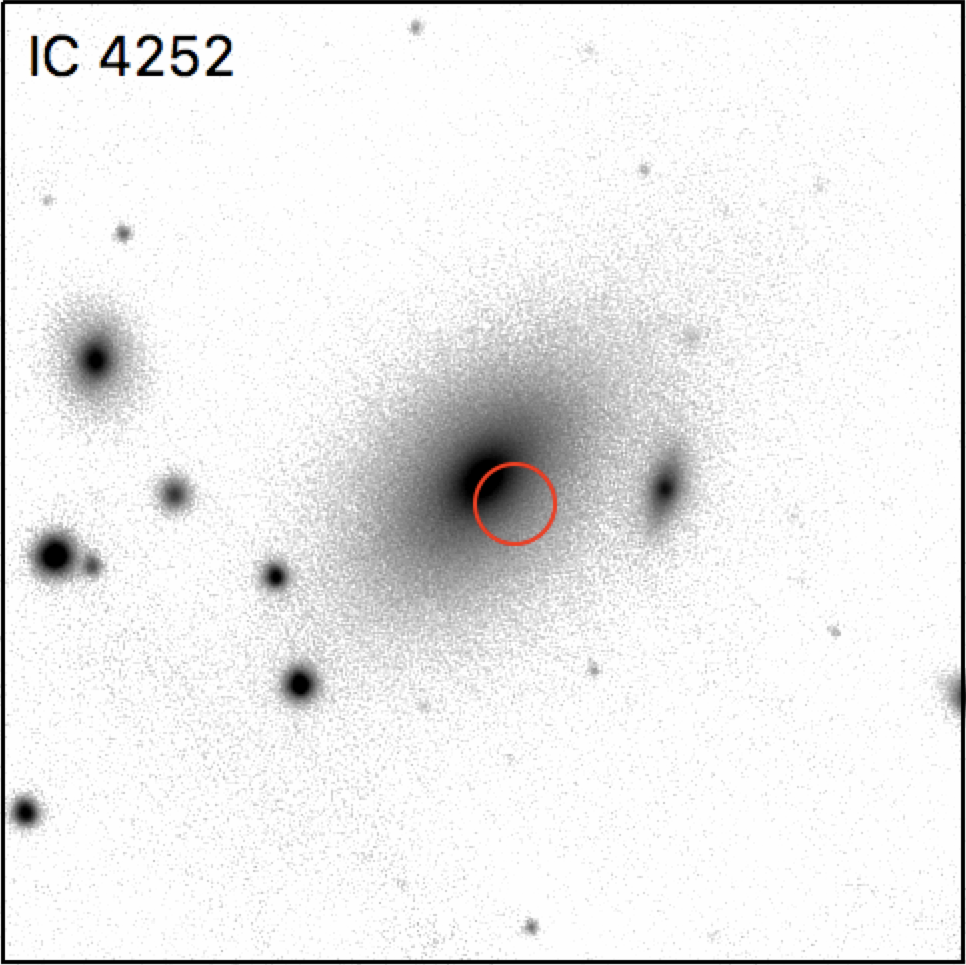}\\
		\vspace{1mm}
		\includegraphics[height=52mm]{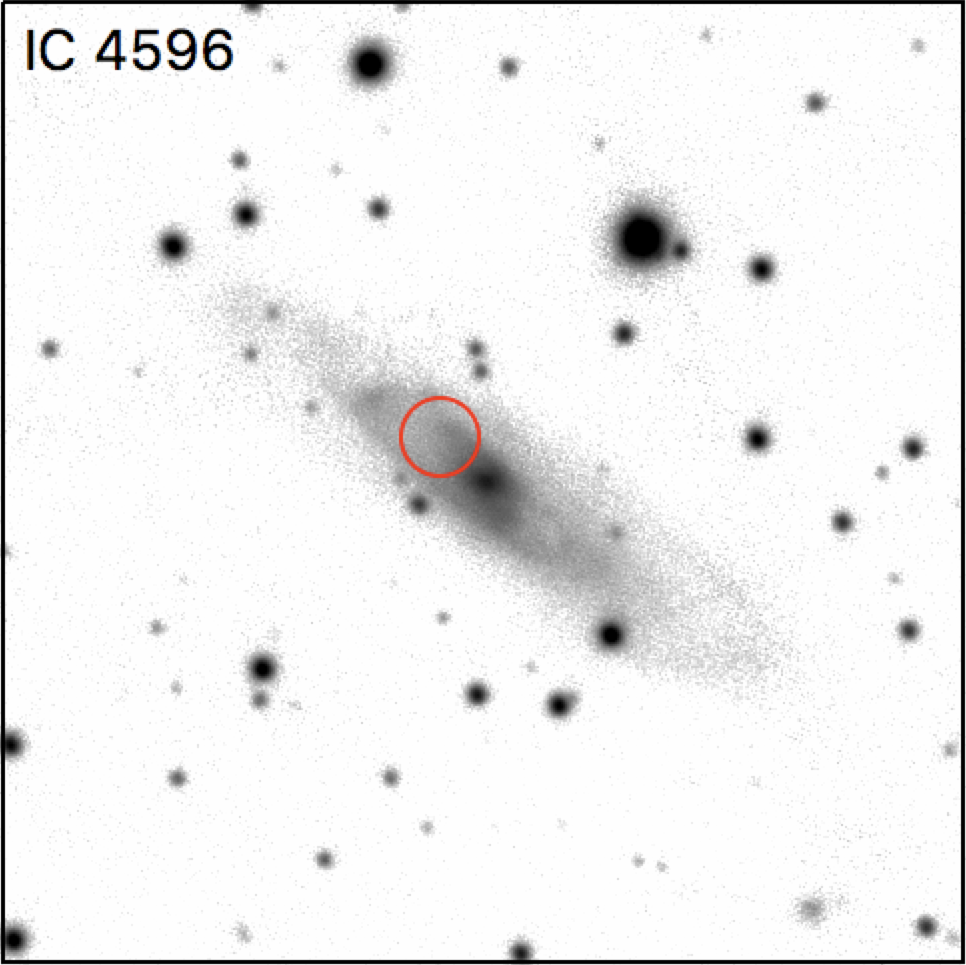}\hspace{1mm}
		\includegraphics[height=52mm]{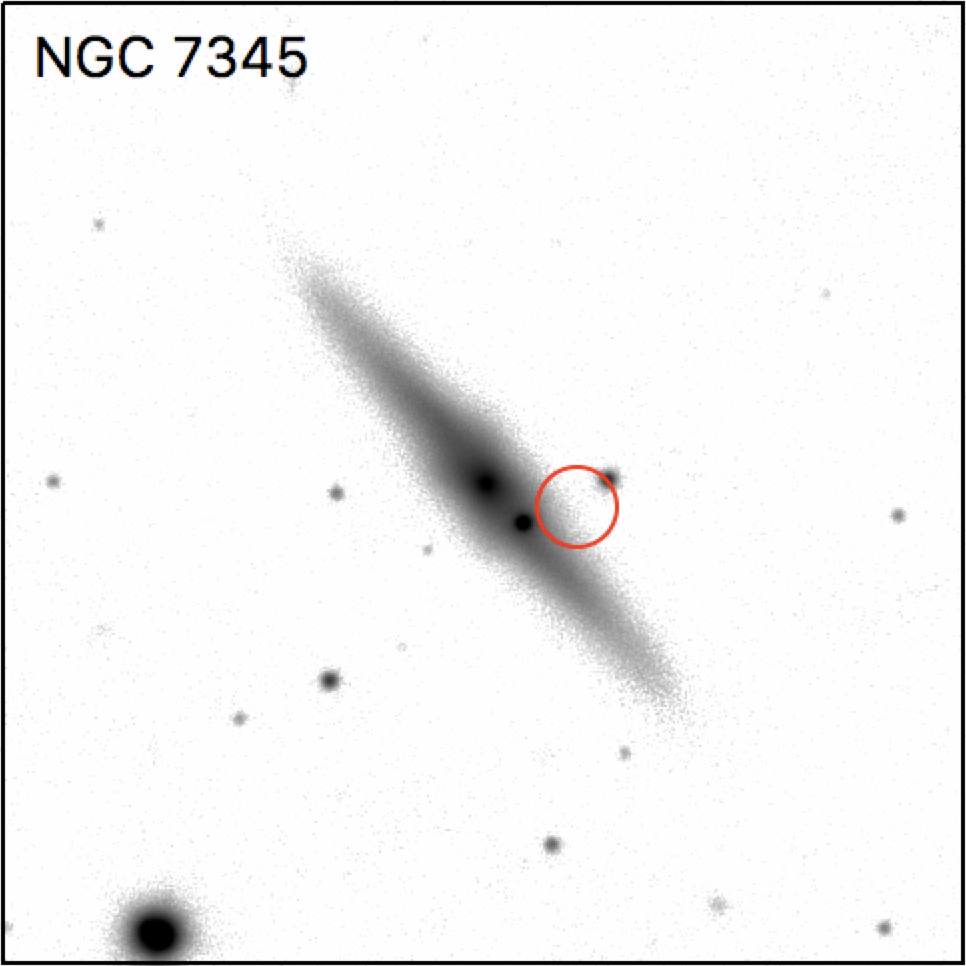}\hspace{1mm}
		\includegraphics[height=52mm]{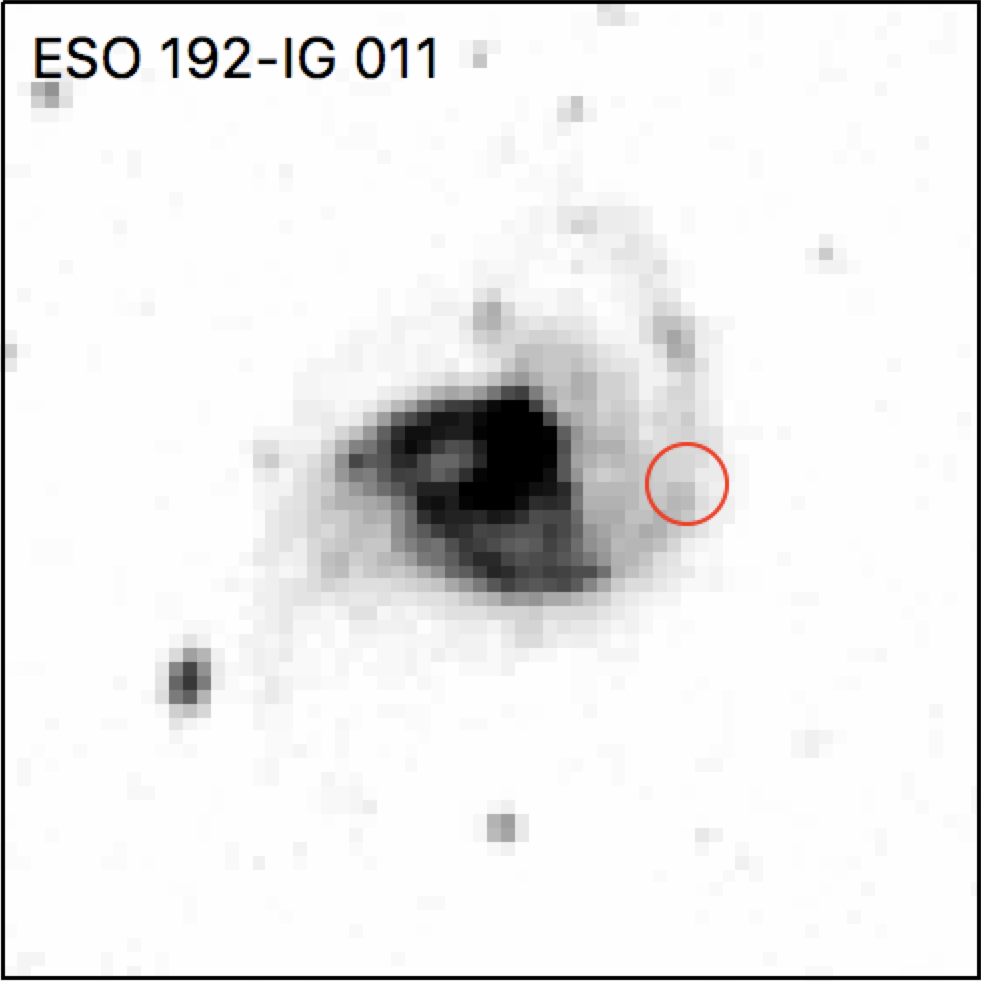}
	\end{center}
	\caption{The positions of nine potential IMBH candidates (see Table~\ref{tab:best}), marked on $2\times2$\,arcminute {\it PanSTARRS} g-band images of their host galaxies (except for ESO~192-IG~011, for which no {\it PanSTARRS} data is available and we show SAO-DSS data) by a 5\,arcsec red circle. {\it Top left}, UGC~1934, {\it centre top}, NGC~3356, {\it top right}, NGC~4077, {\it centre left}, UGC~7188, {\it centre}, NGC~4839, {\it centre right}, IC~4252, {\it bottom left} IC~4596, {\it bottom centre}, NGC~7345, {\it bottom right}, ESO~192-IG~011.}  
	\label{fig:highlum}
\end{figure*}

After these considerations, nine high-luminosity sources remain, none of which feature in previous ULX catalogues we have compared our sample to. All are at relatively large distances, with the nearest in NGC~3356, 82.3\,Mpc away. All also have low count rates -- distance is a factor in this, but also high off-axis angle in some cases such as the source in UGC~1934 -- with the highest rate recorded for the source in NGC~4077, $0.072\pm0.005$\,ct\,s$^{-1}$, requiring $\sim14$\,ks of good time to observe $\sim1000$ counts. Their low count rates do not necessarily put them beyond further study, but detailed investigations will require lengthy dedicated observing campaigns. We show the locations of these nine sources on {\it PanSTARRS} or DSS images in Fig.~\ref{fig:highlum}. Among these sources, the ULXs in NGC~4077, IC~4252, and IC~4596, while insufficiently close to their galaxy centre to be automatically removed as AGNs in our selection method, are close enough to potentially be offset AGNs. Observations with the high spatial resolution capabilities of {\it Chandra} would be sufficient to determine whether or not these sources are consistent with being the host galaxy AGN -- currently, only IC~4252 has been covered by a {\it Chandra} observation, which was too short for the source to be detected. 

The six remaining candidates which are not close to their host galaxy centre have less than 120 counts observed each. Therefore, while these sources represent our best new candidates for possible IMBHs, the collection of a significant amount of additional data is required before we can make any further claims as to their nature.

\subsection{The Eddington threshold}
\label{sec:threshold}

At the other end of the luminosity range of ULXs is the Eddington threshold, the luminosity regime $10^{38} < L_{\rm X} < 3\times10^{39}$\,erg\,s$^{-1}$. A sample of sources at these luminosities is valuable for a number of different reasons -- they provide a potential resource for studying sources accreting at or just below the Eddington limit or possibly transitioning between sub- and super-Eddington accretion, and also house objects related to ULXs such as ultraluminous supersoft sources (ULSs). They also have cosmological significance as a potential source of radiative and mechanical feedback at high redshifts (e.g. \citealt{power13,jeon14,artale15}). The differential X-ray luminosity function of star-forming galaxies -- that is, those galaxies dominated by HMXBs -- takes the form of a power-law with an index of $\sim1.6$ (\citealt{grimm03,mineo12}; W11), indicating that the bulk of radiative energy in the X-ray regime is emitted by the brightest sources. While this means that the highest proportion is produced by ULXs, the higher number of sources in the Eddington threshold luminosity regime and their presence in a larger proportion of galaxies (see Section~\ref{sec:prop}) makes their volume of influence larger than that of the brighter ULXs while still being bright and producing a large amount of energy. Since they are more numerous than ULXs and still at a comfortable luminosity for {\it XMM-Newton} to detect in a typical observation, they are also the most numerous luminosity group in our sample.

Our sample contains 608 sources with detections at luminosities in the Eddington threshold regime. Many of these sources are in nearby and well-studied galaxies, so while their fluxes are lower than ULXs at an equivalent distance, it is still possible to obtain good data on individual objects in this population as well as the population as a whole. Four Eddington threshold objects from our sample with particularly good data are examined in \citet{earnshaw17}, and reveal that the softest objects in this luminosity regime are a heterogeneous population that includes the highest-luminosity examples of sub-Eddington accretion as well as ultraluminous supersoft sources and potential intermediate objects that lie between the soft ultraluminous and the ultraluminous supersoft accretion regimes.

Given the generally low flux of these objects, it is difficult to obtain sufficient data for in-depth studies of individual sources for the majority of these objects. Therefore techniques such as spectral stacking, which we are employing in a new study of this population (Roberts et al. in prep), will also prove invaluable in further probing this population of extragalactic X-ray sources. 

\subsection{Variable ULXs}
\label{sec:var}

Another potentially interesting subset of our sample are variable ULXs. It is observed that strong variability on $\sim100$\,s timescales is more likely to be found in those sources dominated by soft emission (e.g. \citealt{sutton13b}). Under the models of super-Eddington accretion for sources in the ultraluminous regime, this is interpreted as the hard central emission being intermittently obscured by a soft, clumpy and fast-moving outflowing wind (e.g. \citealt{middleton15a}). 

We can perform a basic search for varying ULXs by finding sources variable enough to trigger being flagged as such ({\sc var\_flag} = True) in 3XMM-DR4. This occurs when the $\chi^2$ probability of a source having a constant brightness over the course of an observation has a value $p < 10^{-5}$ for at least one exposure in any of the cameras. This measure is conservative but it is easy to select on and provides a small list of highly-variable targets we can follow up on.

We find 16 sources in our sample flagged as variable in 3XMM-DR4 in at least one observation, eight of which are ULXs. Three of these ULXs are well-studied and known to be variable, NGC~1313~X\=/1 \citep{heil09}, NGC~5408~X\=/1 \citep{strohmayer07} and NGC~6946~X\=/1 \citep{hg15}. Two other well-studied ULXs are not particularly known for short-term variability but have one {\it XMM-Newton} observation that shows strong short-term variability, IC~342~X\=/1 \citep{middleton15a} and Ho~II~X\=/1 \citep{sutton13b}. Out of the remaining three, one particularly interesting variable ULX discovered in this way is M51~ULX\=/7, which not only is highly variable but has a hard spectrum, which goes against the prediction that variable ULXs will have predominantly soft spectra. An in-depth investigation into M51~ULX\=/7 revealed that it has spectral and timing properties similar to a source accreting in the sub-Eddington low/hard state, and may therefore be a candidate IMBH \citep{earnshaw16}.

By selecting on the {\it XMM-Newton} variability flags, we are selecting for strong variability on short timescales. Most ULXs however also show variability across long timescales between observations. A small number of ULXs show a difference in flux of over an order of magnitude between observations, which could potentially be an indication of a bimodal flux distribution caused by a NS entering and leaving a `propeller' mode (e.g. \citealt{tsygankov16}). We searched for such ULXs in a recent study, in which we found five ULXs with a high dynamic range in flux. One of these sources, M51~ULX\=/4, was found to possess a convincingly bimodal flux distribution, possibly indicating the presence of a NS accretor \citep{earnshaw18}.

\section{Conclusions}
\label{sec:conc}

Using an improved version of the method introduced in \citet{walton11}, we have produced a new, clean catalogue of 1,314 extragalactic non-nuclear X-ray sources from the 3XMM-DR4 data release of the {\it XMM-Newton} Serendipitous Source Catalogue, of which 384 are ULXs. This is one of the largest ULX samples to date and is a significant improvement to previous ULX catalogues in terms of cleanliness. We find that, within a subsample of galaxies complete to $10^{39}$\,erg\,s$^{-1}$, one in three galaxies contains a ULX. We also confirm previous findings that ULXs are more commonly found in spiral or irregular star-forming galaxies than in non-star-forming elliptical galaxies, and that this also applies to lower-luminosity objects.

By studying the hardness ratio properties of the complete subsample of ULXs, we have found that ULXs appear to have very similar colour properties to the extragalactic X-ray binary population as a whole, and also appear to have a different (although overlapping) distribution to AGNs, which are generally softer in HR2 and HR3 and extend to higher values of HR4. ULXs in spiral or irregular host galaxies and ULXs in elliptical host galaxies are mostly indistinguishable in hardness ratio space. Additionally, when producing a similar colour-colour plot to that introduced in \cite{pintore14}, we see that the ULX population cannot be easily divided into distinct groups of accretion regimes based on colour. However, we do find that our results are consistent with previous indications that the most luminous ULXs are found in star-forming host galaxies compared with non-star-forming galaxies.

Upon examination of the 27 ULXs in our catalogue with X-ray luminosity $L_{\rm X} > 5\times10^{40}$\,erg\,s$^{-1}$, we find four previously-identified extreme-luminosity ULXs, and nine new objects which are our best possible candidates for the discovery of more IMBHs. Our catalogue also possesses other subpopulations of interest, including 608 sources at the Eddington Threshold ($10^{38} < L_{\rm X} < 3\times10^{39}$\,erg\,s$^{-1}$), making it an ideal resource for expanding our exploration of accreting stellar remnants into the nearby Universe.

\section*{Acknowledgements}

We thank our anonymous referee for constructive comments on this paper. We thank Richard Saxton for his helpful assistance. We gratefully acknowledge support from the Science and Technology Facilities Council (HPE through grant ST/K501979/1, TPR through ST/P000541/1). HPE acknowledges support under NASA contract NNG08FD60C. MJM and DJW acknowledge support from STFC Ernest Rutherford fellowships. SM acknowledges financial support by the Spanish Ministry of Economy and Competitiveness through grant AYA2016-76730-P (MINECO/FEDER).

This research has made use of data obtained from the 3XMM {\it XMM-Newton} Serendipitous Source Catalogue, based on archival observations obtained with {\it XMM-Newton}, an ESA science mission with instruments and contributions directly funded by ESA Member States and NASA. We also make use of images obtained from the {\it PanSTARRS1} Surveys.

\bibliography{ulxcataloguepaper}
\bibliographystyle{../mn2e}
\bsp

\appendix

\section{Catalogue structure}
\label{app:catalogue}

\begin{table*}
	\begin{minipage}{168mm}
		\caption{Example data for the first thirty rows in our sample, including properties of the host galaxy and for each detection, along with their $1\sigma$ errors as given in (or in the case of $L_{\rm X}$, derived from) the {\it XMM-Newton} Serendipitous Source Catalogue, and whether the source is in W11 or in a galaxy in the complete subsample. The entire catalogue is available in digital format online.} \label{tab:catalogue}
		\vspace{-4mm}
		\begin{center}
			\begin{tabular}{@{}l@{~~}l@{~~}c@{~}c@{~}c@{~~~}c@{~~~}c@{~~~}c@{~~~}c@{~~~}c@{~~~}c@{}}
				\hline
				3XMM Name & Host Galaxy & $D^a$ & $L_{\rm X}^b$ & HR1 & HR2 & HR3 & HR4 & ULX? & In W11? & Complete to \\ 
				& & (Mpc) & ($\times10^{38}$\,erg\,s$^{-1}$) & & & & & & & $10^{39}$\,erg\,s$^{-1}$? \\
				\hline
				000155.3$-$152551 & WLM & 0.98 & $0.03\pm0.01$ & $<1$ & $0.52\pm0.29$ & $0.02\pm0.25$ & $-0.41\pm0.30$ & No & No & Yes \\
				000158.1$-$152758 & WLM & 0.98 & $0.08\pm0.01$ & $0.78\pm0.12$ & $0.31\pm0.10$ & $-0.31\pm0.10$ & $-0.57\pm0.15$ & No & No & Yes \\
				000201.4$-$153034 & WLM & 0.98 & $0.05\pm0.01$ & $-0.95\pm0.75$ & $0.99\pm0.08$ & $-0.27\pm0.14$ & $-0.38\pm0.19$ & No & No & Yes \\
				000205.0$-$152504 & WLM & 0.98 & $0.05\pm0.01$ & $0.57\pm0.17$ & $0.27\pm0.14$ & $-0.62\pm0.14$ & $-0.30\pm0.36$ & No & No & Yes \\
				002012.0$+$591756 & IC 10 & 0.76 & $0.005\pm0.002$ & $<1$ & $0.84\pm0.17$ & $-0.18\pm0.21$ & $-0.35\pm0.26$ & No & No & No \\
				002013.7$+$591626 & IC 10 & 0.76 & $0.006\pm0.001$ & $<1$ & $0.09\pm0.11$ & $-0.87\pm0.10$ & $>-1$ & No & No & No \\
				002014.8$+$591852 & IC 10 & 0.76 & $0.0061\pm0.0009$ & $0.82\pm0.08$ & $0.01\pm0.08$ & $-0.86\pm0.09$ & $-0.63\pm0.45$ & No & No & No \\
				002021.4$+$591901 & IC 10 & 0.76 & $0.007\pm0.002$ & $0.74\pm0.33$ & $0.53\pm0.18$ & $0.20\pm0.13$ & $-0.67\pm0.13$ & No & No & No \\
				002026.0$+$591844 & IC 10 & 0.76 & $0.006\pm0.001$ & $0.91\pm0.10$ & $0.17\pm0.13$ & $-0.35\pm0.14$ & $-0.41\pm0.21$ & No & No & No \\
				002029.0$+$591651 & IC 10 & 0.76 & $1.42\pm0.01$ & $0.85\pm0.01$ & $0.63\pm0.01$ & $-0.20\pm0.01$ & $-0.72\pm0.01$ & No & No & No \\
				002121.2$-$483827 & NGC 88 & 45.9 & $14\pm4$ & $0.46\pm0.16$ & $-0.14\pm0.15$ & $-0.56\pm0.18$ & $-0.96\pm0.43$ & Yes & Yes & No \\
				003251.7$+$482556 & NGC 147 & 0.71 & $0.005\pm0.003$ & $-0.37\pm0.38$ & $0.73\pm0.25$ & $-0.39\pm0.33$ & $0.16\pm0.27$ & No & No & Yes \\
				003257.9$+$483101 & NGC 147 & 0.71 & $0.007\pm0.003$ & $0.95\pm0.24$ & $-0.07\pm0.31$ & $0.16\pm0.32$ & $-0.36\pm0.33$ & No & No & Yes \\
				003259.5$+$482700 & NGC 147 & 0.71 & $0.022\pm0.005$ & $0.24\pm0.83$ & $0.86\pm0.20$ & $0.33\pm0.14$ & $-0.51\pm0.16$ & No & No & Yes \\
				003302.5$+$483227 & NGC 147 & 0.71 & $0.004\pm0.002$ & $0.60\pm0.34$ & $-0.14\pm0.28$ & $0.25\pm0.27$ & $-0.60\pm0.29$ & No & No & Yes \\
				003306.0$+$483134 & NGC 147 & 0.71 & $0.005\pm0.002$ & $0.58\pm0.34$ & $0.39\pm0.22$ & $-0.55\pm0.24$ & $-0.76\pm0.51$ & No & No & Yes \\
				003307.9$+$482504 & NGC 147 & 0.71 & $0.012\pm0.005$ & $0.81\pm0.19$ & $0.00\pm0.20$ & $-0.21\pm0.23$ & $0.04\pm0.28$ & No & No & Yes \\
				003312.9$+$482950 & NGC 147 & 0.71 & $0.004\pm0.002$ & $0.94\pm0.15$ & $-0.09\pm0.23$ & $-0.98\pm0.20$ & $0.55\pm0.92$ & No & No & Yes \\
				003313.3$+$482611 & NGC 147 & 0.71 & $0.005\pm0.002$ & $0.61\pm0.29$ & $0.13\pm0.24$ & $-0.23\pm0.23$ & $-0.98\pm0.35$ & No & No & Yes \\
				003322.0$+$483327 & NGC 147 & 0.71 & $0.010\pm0.003$ & $0.37\pm0.27$ & $0.34\pm0.25$ & $-0.05\pm0.21$ & $-0.67\pm0.18$ & No & No & Yes \\
				003324.6$+$482751 & NGC 147 & 0.71 & $0.005\pm0.002$ & $0.96\pm0.08$ & $-0.13\pm0.17$ & $-0.82\pm0.15$ & $>-1$ & No & No & Yes \\
				003330.2$+$483049 & NGC 147 & 0.71 & $0.003\pm0.002$ & $0.09\pm0.73$ & $0.80\pm0.27$ & $-0.04\pm0.27$ & $-0.93\pm0.26$ & No & No & Yes \\
				003338.7$+$483340 & NGC 147 & 0.71 & $0.003\pm0.002$ & $0.62\pm0.43$ & $0.52\pm0.25$ & $-0.85\pm0.22$ & $-0.65\pm0.70$ & No & No & Yes \\
				003739.2$-$334249 & ESO 350-G 040 & 120.7 & $222\pm58$ & $0.72\pm1.30$ & $0.98\pm0.16$ & $0.08\pm0.23$ & $-0.20\pm0.21$ & Yes & No & No \\
				003742.4$-$334249 & ESO 350-G 040 & 120.7 & $40\pm18$ & $0.74\pm0.15$ & $-0.17\pm0.20$ & $-0.86\pm0.26$ & $0.32\pm0.73$ & Yes & Yes & No \\
				003742.4$-$334249 & ESO 350-G 040 & 120.7 & $52\pm26$ & $0.66\pm0.17$ & $-0.69\pm0.17$ & $0.01\pm0.46$ & $-0.71\pm0.49$ & Yes & Yes & No \\
				003830.1$+$481928 & NGC 185 & 0.65 & $0.006\pm0.001$ & $>-1$ & $<1$ & $0.89\pm0.14$ & $0.05\pm0.15$ & No & No & Yes \\
				003840.5$+$482231 & NGC 185 & 0.65 & $0.0020\pm0.0007$ & $0.60\pm0.26$ & $0.38\pm0.16$ & $-0.26\pm0.15$ & $-0.65\pm0.28$ & No & No & Yes \\
				003841.9$+$482031 & NGC 185 & 0.65 & $0.003\pm0.001$ & $0.73\pm0.44$ & $0.10\pm0.37$ & $0.56\pm0.21$ & $-0.62\pm0.18$ & No & No & Yes \\
				003842.6$+$481751 & NGC 185 & 0.65 & $0.0054\pm0.0003$ & $0.02\pm0.38$ & $0.72\pm0.16$ & $-0.85\pm0.15$ & $0.00\pm0.63$ & No & No & Yes \\
				\hline
			\end{tabular}
		\end{center}
		$^a$The host galaxy distance.\\
		$^b$The detection luminosity.
	\end{minipage}
\end{table*}

Our catalogue, which we include in a digital format alongside this paper, is structured as follows.

There is one entry per {\it XMM-Newton} detection (identified with the column DETID) of each source (identified with SRCID and the IAU name) included in the catalogue. The first 318 columns are taken from 3XMM-DR4 and include fluxes, count rates, counts and detection likelihoods for each energy band, observation details, hardness ratios, quality flags, and average source properties. 

The next 66 columns are taken from RC3 and contain the host galaxy properties including names, dimensions, brightness and colour -- they also contain the galaxy position and distance, which we retain for completeness but note that the more accurate NED values were those used in analysis. Properties of galaxies taken from CNG were inserted into the equivalent RC3 columns. The following 8 columns are taken from the best NED match to the RC3/CNG galaxy, 10 further columns from the best NBG match, and 4 further columns from the NED-D distance database. 

The remaining 17 columns are defined by the authors. DISTANCE\_BEST and DISTANCE\_BEST\_SRC are the best distance to the host galaxy in Mpc, and the catalogue from which this best distance was taken: NED-D, NBG, NED or RC3/LV in that order of precedence (for more details on the calculation of the distance in each case, see Section~\ref{sec:samp}). The BT\_ABS\_MAG column is the absolute magnitude of the host galaxy, derived from BT\_MAG\_CORR in RC3, and the EP\_8\_LUMINOSITY and EP\_8\_LUMINOSITY\_ERR columns are the full-band detection luminosity and error derived from the EP\_8\_FLUX and EP\_8\_FLUX\_ERR columns in 3XMM-DR4. DISTANCE\_BEST is the distance used in the calculation in each of these cases. EP\_8\_LUMINOSITY\_MAX is the sum of EP\_8\_LUMINOSITY and EP\_8\_LUMINOSITY\_ERR, and at least one detection with EP\_8\_LUMINOSITY\_MAX $> 10^{39}$\,erg\,s$^{-1}$ is the basis upon which we class a source as a ULX. SEPARATION is the sky distance in arcsec between the detection position and the NED galaxy position, and RMIN is this value minus three times the detection position error POSERR (and may be negative) -- the filtering out of central sources is done on the basis of RMIN at the contaminant removal stage. Finally, there are six boolean flag columns indicating whether the source appears in one of the six other ULX catalogues we compare with in Section~\ref{sec:prop}, and three boolean flags indicating whether the source is in any of the three complete subsamples -- that is, it lies in a galaxy for which we are confident that all sources above $10^{38}$, $10^{39}$ and $10^{40}$\,erg\,s$^{-1}$ respectively have been detected (see Section~\ref{sec:comp}).

Altogether, the catalogue table contains 2,139 rows and 423 columns.

\label{lastpage}

\end{document}